\documentclass[epjH]{svjour}
\usepackage{graphicx}
\begin{document}

\title{JETS AND QCD: A Historical Review of the Discovery of the Quark and Gluon Jets and
 its Impact on QCD\footnote{\it In Memoriam: Hans Joos}}
\author{A.Ali\inst{1} \and G.Kramer\inst{2} 
        }
\institute{DESY, D-22603 Hamburg (Germany) \and Universit\"at Hamburg,
           D-22761 Hamburg (Germany)}  

\abstract{
The observation of quark and gluon jets has played a crucial role in establishing
Quantum Chromodynamics [QCD] as the theory of the strong interactions
within the Standard Model of particle physics. The jets, narrowly 
collimated bundles of hadrons, reflect configurations of quarks and gluons 
at short distances. Thus, by analysing energy
and angular distributions of the jets experimentally, the properties 
of the basic constituents of matter and the strong forces 
acting between them can be explored. In this review, which is primarily 
a description of the discovery of the  quark and gluon jets and the impact 
of their observation on Quantum Chromodynamics, we elaborate, in particular, the role
of the gluons as the carriers of the strong force. 
Focusing on these basic points, jets in $e^+ e^-$ collisions 
will be in the foreground of the discussion and we will concentrate on the
theory that was contemporary with the relevant experiments at the electron-positron colliders.
 In addition we will delineate the role 
of jets as tools for exploring other particle aspects in $ep$ and $pp/p\bar{p}$ 
collisions - quark and gluon densities in protons, measurements of the QCD coupling,
fundamental 2-2 quark/gluon scattering processes, 
but also the impact of jet decays of top quarks, and $W^\pm, Z$ bosons 
on the electroweak sector. The presentation to a large extent 
is formulated in a non-technical language with the intent to recall the significant
steps historically and  convey the significance
of this field also to communities beyond high energy physics.
}
\maketitle
\section{Introduction}
\label{intro}
Quantum Chromodynamics [QCD], the theory of the strong interactions within 
the Standard Model of particle physics~\cite{QCD,QCD-2,asyfree,asyfree-2}, describes 
the building blocks of strongly interacting particles, like proton, neutron and many others, 
and the forces acting between them. The fundamental building blocks of these
particles are the spin-1/2 quarks $q$~\cite{quarks,quarks-2}, which come in three families. 
Their masses cover a large range~\cite{Nakamura:2010zzi}. The three lightest quarks $u,d,s$
 weigh only a small 
fraction of the proton mass, the charm quark $c$ just about the proton mass 
while the two heavy quarks $b,t$ weigh more than 5 and 180 times the proton mass,
respectively. Baryons, like proton and neutron, are composed of three quarks $qqq$,
while mesons, like pions, are bound states $q \bar{q}$ of quark-antiquark pairs. 

Since the spin and the spatial $S$-wave functions of the lightest baryons are
symmetric under the exchange of quarks, the Pauli principle demands the quarks to be labelled 
by new charges, called colours, which discriminate between the three components
of the baryons~\cite{color,color-2,color-3}. Rephrased within the SU(3)$_C$ symmetry group for three
colour degrees of freedom, the colour wave function is antisymmetric. This
threefold antisymmetric combination of colours renders baryons non-coloured, {\it i.e.} 
they are white SU(3)$_C$ singlets; summing up the quark colours symmetrically in mesons, 
these hadrons are white too. 
By reducing the lifetime of the neutral pion by a factor $3^2 = 9$, the three-colour
 extension reconciles the prediction of the quark model with the experimental measurement,
 a crucial point in establishing the colour charges.  

Equivalently to the electric charges in electrodynamics, 
the colour charges of the quarks can serve 
as sources for force fields, which bind the quarks within the mesons and
baryons~\cite{Nambu}. Eight such gluon fields $g$ are predicted by
the non-abelian gauge group SU(3)$_C$. Like the photon field they are vector 
fields with spin = 1, but in contrast to the photon they carry colour charges  
themselves, mediating colour flips of quarks by absorption or emission.
This theory, Quantum Chromodynamics, is theoretically described 
by a non-abelian Yang-Mills gauge 
theory \cite{YangMills}. Gluons couple to each other, giving rise to 
three- and four-gluon interactions. These self-interactions of the
gluons have profound consequences for the QCD coupling. While virtual fermionic 
quarks render the vacuum colour-diamagnetic, the large contribution 
of virtual bosonic gluons renders the vacuum finally colour-paramagnetic.
Thus, in contrast to the electric coupling, the QCD coupling decreases    
with decreasing distance and the quarks and gluons become asymptotically free~\cite{asyfree,asyfree-2}.
 Quarks and gluons therefore interact weakly at short
distances while the strength of their interactions grows with increasing
distance, suggesting the permanent confinement of particles carrying 
non-zero colour charges \cite{Wilson}.  

Quarks can be {\it seen} in the scattering of electrons or neutrinos
off nucleons. The final-state pattern of these processes reveals 
that the leptons scatter off point-like, nearly massless spin-1/2 constituents 
of the nucleons which carry the electric and weak charges of the quarks. 
Gluons inside nucleons, which do not carry electric nor weak charges, 
manifest themselves only indirectly. Half of the momentum 
of fast moving nucleons cannot be accounted for by the spectrum 
of the quarks alone, and it must be attributed to gluons as flavour-neutral 
constituents \cite{ChLlSm}. In addition, the quark spectrum is modified by gluon 
bremsstrahlung if the energy of the impinging leptons is raised from low
to high values \cite{Gross}. 

However, QCD suggests another method to unravel its basic constituents.
As a result of asymptotic freedom, quarks and gluons move as
quasi-free particles, called partons~\cite{Feyn}, at short distances. When these
 coloured objects 
are accelerated in scattering processes, they develop bremsstrahlung cascades 
of narrowly collimated gluons and quark-antiquark pairs, which finally transform
to equally well collimated hadrons at distances at the colour 
confinement radius of about 1 fm [$10^{-13}$ cm]. 
Thus, the quarks and gluons at small distances map themselves into jets of hadrons 
at large distances. Since the quanta associated with the confinement forces are soft,
their impact on the energies and momenta of the jets is small
so that the configurations of high-energy quarks and gluons at short
distances are truly reflected in the energy and angular distributions 
of the jets. Since these jets can be observed experimentally, the properties 
of quarks and gluons can be determined experimentally by jet analyses, such as
their spins, flavour and colour charges, and their basic interactions.  

It should be stressed here that the field of jet physics and QCD owes a great deal of
gratitude to the development and successful operations of high energy colliders, in
particular, electron-positron colliders. Starting from SPEAR at SLAC, which started the
physics runs in 1972 and had a maximum beam energy of 4 GeV, the subsequently built
$e^+e^-$ colliders DORIS (physics start 1973; maximum beam energy 5.6 GeV) and PETRA
(physics start 1978; maximum beam energy 23.4 GeV) at DESY, PEP (physics start 1980; 
maximum beam energy 15 GeV) and SLC (physics start 1989; maximum beam energy 50 GeV)
at SLAC, TRISTAN at KEK (physics start 1987; maximum beam energy 32 GeV), and LEP
(physics start 1989; maximum beam energy 104.6 GeV) at CERN, saw the main jet activity
and detailed tests of QCD. The results from these machines are the primary focus of
this review and are discussed in the first six chapters. However, in a long epilogue,
described in chapter 7, entitled jets as tools, we have discussed some selected results
related to QCD and jets,
which have come out from the electron-proton collider HERA (physics start 1992; maximum
$e^+/e^-$-beam energy 27.6 GeV and maximum proton energy 920 GeV) at DESY, and the hadron colliders
Tevatron (physics start 1987; maximum $p$ and $\bar{p}$ energy 980 GeV) at Fermilab and
finally the LHC (physics start 2010; maximum proton beam energy so far 3.5 TeV). It is
not our mandate to discuss the technical aspects of these machines, which will take us
too far afield from the main focus, namely historical development of jets and QCD
from the theoretical and experimental points of view. For the interested readership
of this review, the high energy machine related aspects are  summarised concisely in
 Reviews of Particle Physics by the Particle Data Group
(PDG). Many of these colliders, in fact all the $e^+e^-$ colliders, are no longer working
in particle physics, and for these we refer to the 1996
 PDG review~\cite{Barnett:1996hr}, while for the others to the 2010
 PDG review~\cite{Nakamura:2010zzi}.      

\begin{figure}
\center{
\resizebox{1.0\columnwidth}{!}{
\includegraphics{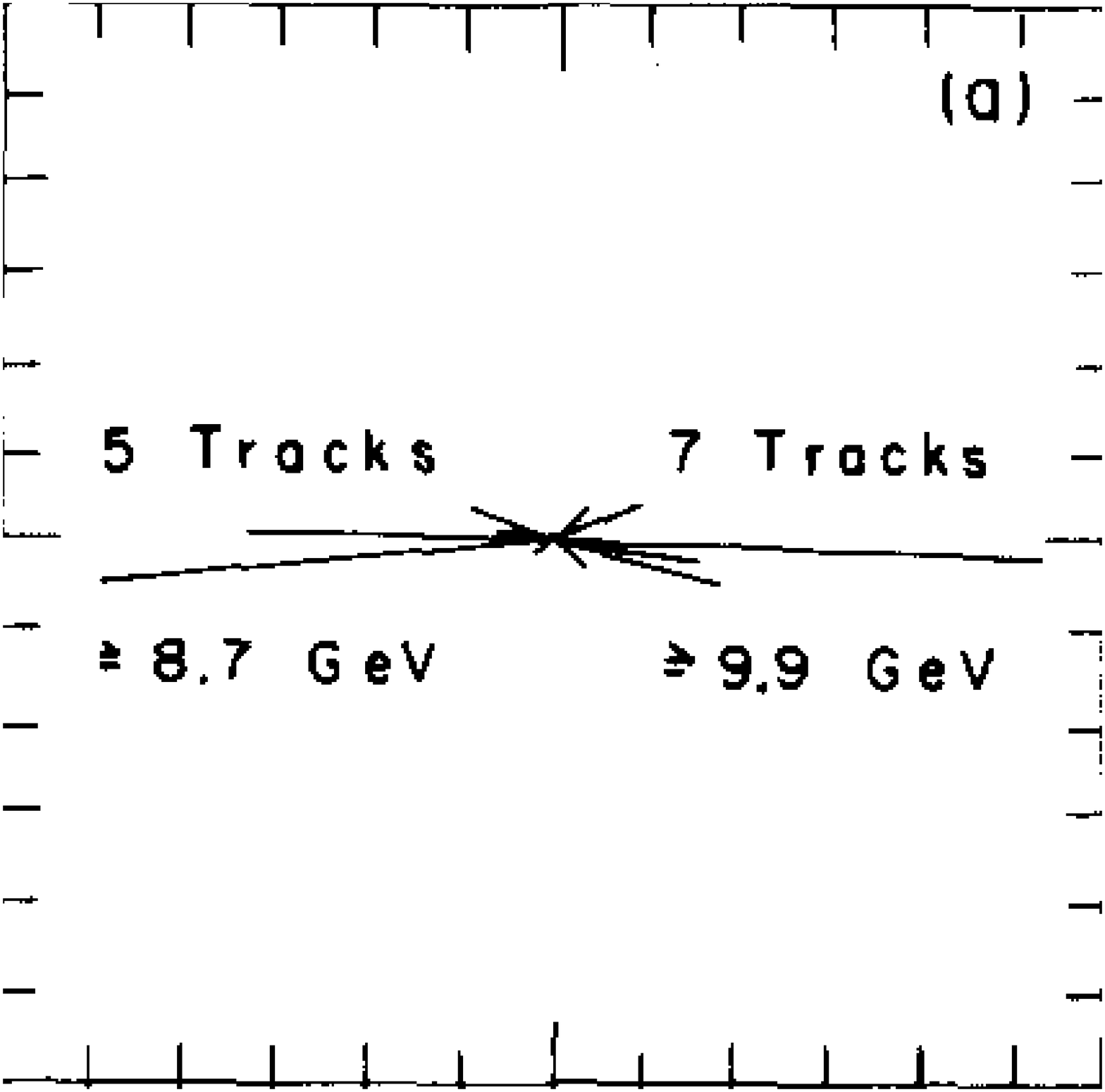} $\;\;\;\;\;\;\;\;\;\;\;$ \includegraphics{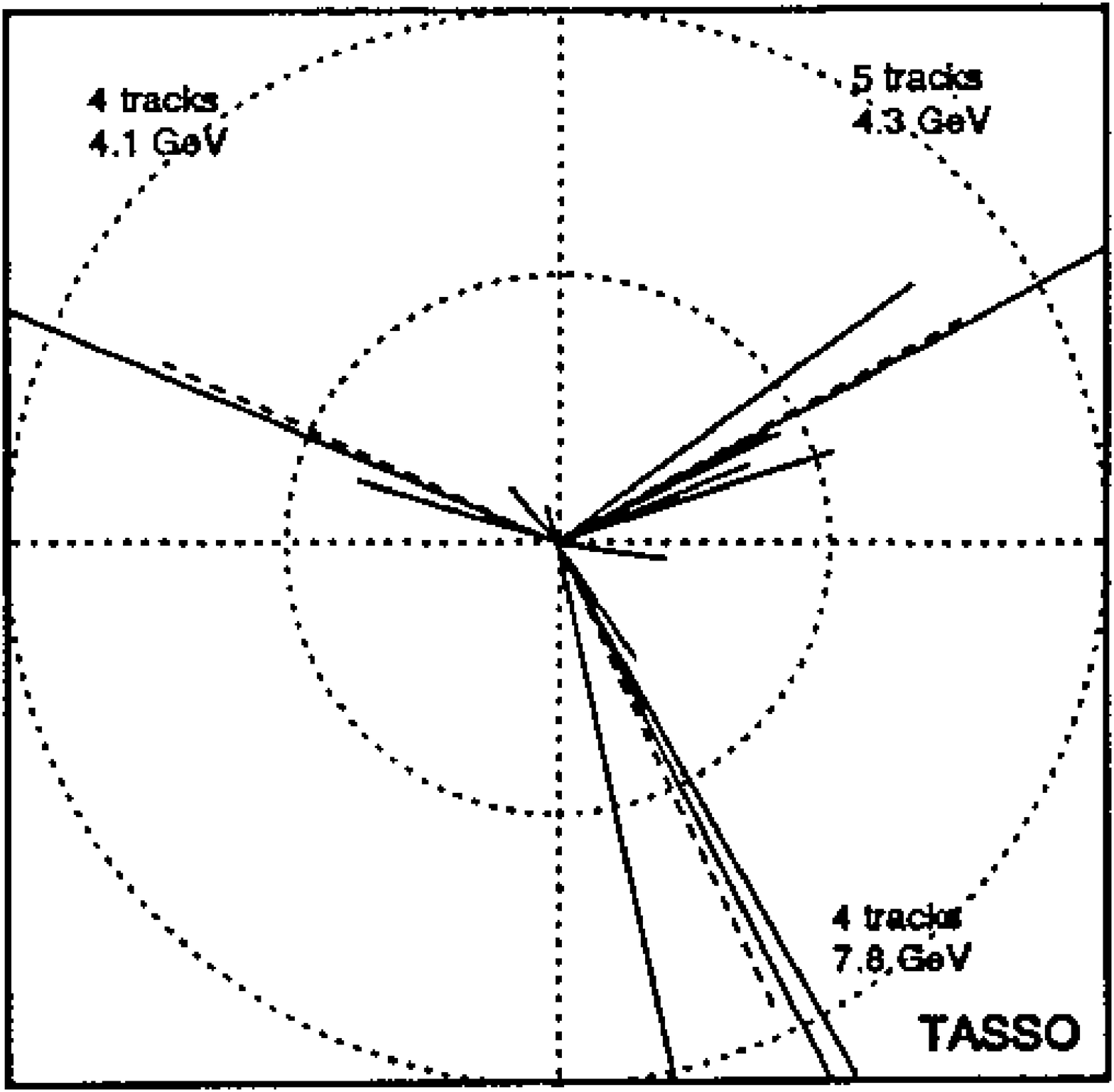}}}
\caption{Observation of (a) 2-jet final states in 
         electron-positron annihilation to hadrons: $e^+ e^- \to q  
         \bar{q} \to 2 \, jets$ (TASSO~\cite{EXP4gluon-TASSO}); and (b) 3-jet final states
         in gluon bremsstrahlung off quarks in $e^+ e^-$ annihilation: 
         $e^+ e^- \to q  \bar{q}  g \to 3 \, jets$ in the TASSO detector~\cite{EXPgluonjet}.}
\label{fig:1.1ab}  
\end{figure}

Quite early, the final states in $e^+ e^-$ annihilation 
to hadrons had been predicted to consist [primarily] of two jets evolving
from a quark-antiquark pair produced in the collision process \cite{THquark,THquark-2,THquark-3}:
\begin{equation}
e^+ e^- \to q  \bar{q} \to 2 \, jets \,.
\end{equation}
Experimental evidence for these quark jets was first provided at the 
$e^+ e^-$ collider SPEAR \cite{EXPquark,EXPquark-2} by demonstrating that the
hadrons in the final states were not isotropically distributed but 
accumulated near the central event axis specified by the momenta
of the quarks \cite{PHENquark}. At PETRA energies ($12 \leq \sqrt{s} \leq 46.6$ GeV)
the two jets could be recognised without any sophisticated analysis,
{\it cf.} Fig.{\ref{fig:1.1ab}} (left-side frame). Angular distributions and charge analyses 
finally proved the jets to be associated with spin-1/2 quarks indeed.      

First indirect evidence of gluons was provided by the PLUTO collaboration at the $e^+ e^-$ collider
DORIS~\cite{Berger79} from the decay $\Upsilon(1S) \to ggg$. 
However, as $\Upsilon(1S)$ has a mass of 9.46 GeV, significant non-perturbative contributions 
 had to be taken into account.
PLUTO used their 2-jet data below the resonance to extract the $q^* \to ~{\rm hadrons}$
fragmentation and used this to estimate also the fragmentation $g^* \to ~{\rm hadrons}$.
With this, their analysis was in agreement with the expectations from
the underlying process  $\Upsilon(1S) \to ggg$. 
Gluon jets were later discovered unambiguously at the $e^+ e^-$ collider PETRA
\cite{EXP4gluon-TASSO,EXP4gluon-MARK-J,EXP4gluon-PLUTO,EXP4gluon-JADE}
 running at higher energy (typically 30 GeV). A 3-jet event from the very
early PETRA data~\cite{EXPgluonjet} is shown in Fig.~\ref{fig:1.1ab} (right-side frame).
Such events had been predicted theoretically \cite{THgluon}
for configurations in which the quark pair produced in $e^+e^-$ annihilation
radiates a hard non-collinear gluon:
\begin{equation}
e^+ e^- \to q  \bar{q}  g \to 3 \, jets \,.
\end{equation}
This bremsstrahlung mechanism is characteristic for gauge theories 
and it is familiar from electrodynamics where charges accelerated in 
collision processes emit photons, as in electron-positron scattering
$e^+ e^- \to e^+ e^- \gamma$, for example.   
Bremsstrahlung gluons in QCD which transform to hadron jets generate
characteristic patterns in the final states which allow one to prove 
the existence of gluons: With increasing energy the primary
quark jets become narrower; the events become flat and
 ``Mercedes-Star-like'' ($Y$-shaped); and finally three well separated jets emerge.
Detailed comparison of the event structure with the underlying theory (QCD) required
apart from the perturbative (hard) processes also modeling of the non-perturbative (soft)
features of the quark and gluon fragmentation. Hence,  event generators,
incorporating the perturbative and non-perturbative aspects of QCD 
were necessary to relate the emerging jet distributions to the predictions derived
from gluon bremsstrahlung 
in QCD. References~\cite{PHENgluon1,PHENgluon2} illustrate the early use of such
event generators, taking the form of
Monte Carlo simulations to match the theretical calculations with the
experimental measurements, which were state-of-the-art tools at that time, and which helped
in establishing the properties of the quark and gluon jets. To avoid any confusion,
Monte Carlo in the present context is a numerical computational technique to calculate
multi-dimensional distributions of a process in which events are generated randomly but
weighted to reflect the underlying dynamics.
Dedicated experiments at PETRA and PEP and theoretical progress in the
80's  greatly consolidated jet physics and led to quantitative tests of
QCD. A more modern view of the use of Monte Carlo programs, in particular, their role as tools
in hard hadronic collisions can be found in recent reviews, for example~\cite{Mangano:2005dj}.

The program to establish QCD in studying quark and gluon jets 
was naturally continued at the $e^+ e^-$ collider LEP, see, for example, 
\cite{LEP}, where the increased energy could be exploited 
to measure the gluon self-interactions in multijet events,
\begin{equation}
e^+ e^- \to q  \bar{q}  q^\prime  {\bar{q}}^\prime, \;\;
            q  \bar{q}  g  g
        \to 4  \, jets \,, 
\end{equation}
with the production amplitudes dominated by the $q\bar{q} gg$ states, which
included the virtual gluon splitting, {\it e.g.} $g^\ast \to gg$. By measuring energy and angular
distributions of these 4jet-events the colour charge 
of gluons could be determined, the crucial element for generating 
asymptotic freedom in QCD. Correspondingly, the variation of the
quark/gluon coupling could be examined for a large range of energies,
though experiments at PEP, PETRA and TRISTAN had already confirmed the
running of $\alpha_s(Q^2)$ in agreement with the renormalisation group (RG)
equation.

The quark/gluon jet phenomena were also indicated at the $pp$ collider
ISR~\cite{Breakstone:1983pb}, before high-energy jets were unambiguously isolated at the 
$Sp\bar{p}S$~\cite{Scott:1985sr}. Since then, jets in hadronic collisions have become
precision tools in not only testing QCD and the electroweak physics at the highest available
energies (such as at the Tevatron and the LHC), but also in searching for phenomena
 beyond-the-Standard-Model (BSM),
such as dijet resonances and quark substructure. By the same token, jet phenomena observed
at hadron-hadron and lepton-hadron colliders have provided fundamental information on the
quark and gluon densities (parton distribution functions) of the proton. We shall review this
towards the end of this paper, but for now concentrate on the general development of
jet physics and QCD which took place in the context of $e^+e^-$ colliders.
 
Quark and gluon processes at short distances can be treated, due to 
asymptotic freedom of QCD, in perturbative expansions for the weakly 
interacting fields. Therefore the basic short-distance processes
as well as the evolution of the quark/gluon cascades are well
controlled theoretically. However, the final transition from the
quark/gluon configurations to hadrons is a long-distance 
process governed by large values of the QCD coupling which
cannot be treated in perturbation theory and which, so far, cannot 
be analysed rigorously. Instead, QCD-inspired models have been 
developed which parametrise the transition phenomenologically.
Two alternative approaches have been worked out in detail.
In the first picture a quark moving out of the short-distance
regime generates a string-like gluonic flux tube which breaks
up repeatedly by spontaneous quark-antiquark creation when 
its length approaches about 1 fm. This mechanism generates
a jet of collimated hadrons with energy and direction corresponding 
to the initial high-energy quark \cite{FieldF}. Gluons had
been treated analogously \cite{PHENgluon1,PHENgluon2}, 
or they were assumed to generate kinks, local depositions
of energy and momentum
in the strings stretched between quarks and antiquarks \cite{Lund}.   
Alternatively in cluster fragmentation, after splitting 
all final gluons in a quark/gluon cascade to quarks and 
antiquarks, $q \bar{q}$ pairs with low invariant masses 
transform to hadronic resonances which eventually 
decay to the standard low-mass mesons and baryons 
\cite{Herwig1}. 

After the important work of Ref.~\cite{WuZo} (see, also~\cite{Lanius:1980nv,Lanius:1980mz,Daum:1980rp}),
numerous methods have been proposed, with steadily increasing refinement, 
to reconstruct the jets experimentally. One class
consists of algorithms based on sequential jet recombination.
Particles are sequentially combined if their distance in
momentum space falls below a pre-set minimum. Typical examples are 
the JADE algorithm \cite{JADE}, where the distance is defined 
by the invariant mass of pairs, developed later to algorithms
based on transverse momenta $k_t$. A second class is 
built by cone algorithms in which particles belonging to 
pre-defined cones are grouped 
into jets. Originally introduced to regulate singularities in infrared 
and collinear quark-gluon configurations \cite{SterW,Sterman:1979uw}, they have
been developed to a standard method in hadron collider analyses.  

The original jet analyses at PETRA were based on independent-jet
fragmentation \cite{PHENgluon1,PHENgluon2}, providing a valid tool 
for reconstructing the quark/gluon configurations at small distances
in $e^+ e^-$ annihilation. Subtle effects observed later in the
hadron distributions between jets, were interpreted as string 
effect~\cite{Lund,Andersson:1983ia}, or explained alternatively by additional
soft gluon radiation with angular ordering~\cite{Azimov:1986sf}. 
PYTHIA \cite{Pythia,Sjostrand:2007gs}, HERWIG \cite{Herwig} and SHERPA \cite{Sherpa} 
are modern versions of Monte Carlo programs which are used in
present jet analyses.  

The connection of jets with QCD has been extensively treated in
the literature under theoretical and experimental aspects, see
{\it e.g.}~\cite{Kramer}-\cite{EllisK}. This review 
will summarise the basic concepts of jet physics, intended 
to describe how jet physics has been exploited to establish QCD 
as the non-abelian quark/gluon gauge field theory of the strong 
interactions. Addressing also communities outside high energy 
physics, the review is presented mostly in a non-technical language, 
giving a qualitative account of theoretical and experimental 
developments which have dramatically changed the earlier picture 
of the strong forces in particle physics. In doing this, we have
included some landmark measurements in a chronological order as they
were reported. The same remark applies to the discussion of the theoretical aspects,
and we have emphasized only works which were contemporary with the discoveries.
The picture now is based on a few fundamental principles summarised succinctly in Quantum
Chromodynamics. 

 The topics on which we concentrate are the non-perturbative and 
perturbative elements of quark/gluon jets, 
including experimental and phenomenological 
methods to define the jets. Early evidence and 
indirect indications of quark and gluon jets in $e^+ e^-$ 
annihilation to quarks at SPEAR and $\Upsilon$ decays to gluons 
at DORIS will be reviewed. In the central core of this 
paper, we will describe the theoretical basis of the discovery 
of gluons in the three-jet events at PETRA and the measurement 
of their properties. The picture will be completed with LEP. Finally 
we will demonstrate in a few examples how jets can be used 
as tools for measuring other parameters and fundamental processes of QCD, the 
gluon content of nucleons, QCD Rutherford scattering, {\it etc.},
but also how to exploit jets for identifying electroweak $W,Z$ and Higgs
bosons, top-quark physics, and search for new phenomena, in particular possible substructure of
partons. Such problems have been addressed at HERA and the Tevatron, 
and they will play an important role at the LHC. However, despite discussing some of the
most recent measurements in jets and QCD, this is not a review of the up-to-date theoretical
advances. We have included some of these topics to introduce the readers to the
vast areas of particle physics research in which jets and QCD have branched out,
but emphasize that this article aims primarily at providing a historical perspective.

 This paper is organised in 8 sections and the main topics discussed
are as follows: Fragmentation properties of quarks and gluons (section 2),
discovery of quark jets at SPEAR and the first application of perturbative QCD to
derive the 2-jet cross section in $e^+e^-$ annihilation (section 3),
gluon jets in $\Upsilon$ decays and the basic partonic process $\Upsilon \to ggg$
(section 4), jets in QCD and at PETRA
and PEP (section 5), jets and QCD studies at LEP (section 6), jets as tools, with 
applications in Deep Inelastic Scattering Processes, $\gamma \gamma$ collisions, and hard hadronic
collisions at the Tevatron and the LHC  (section 7). A brief summary (section 8) will conclude
this review.  

\section{The fragmentation of quarks and gluons}
\label{sec:fragmentation}
\begin{figure}
\center{
\resizebox{0.95\columnwidth}{!}{
\includegraphics{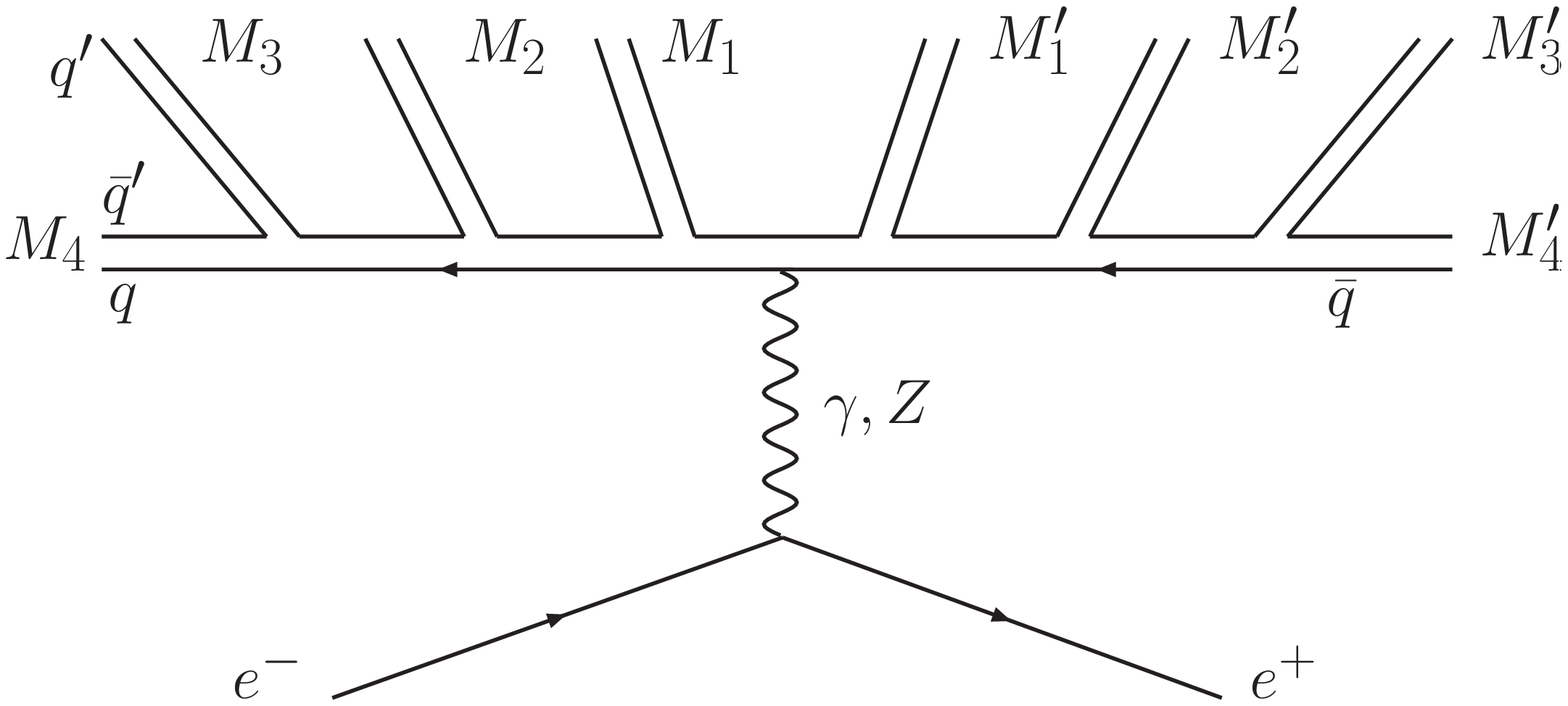} \hspace*{2cm} \includegraphics{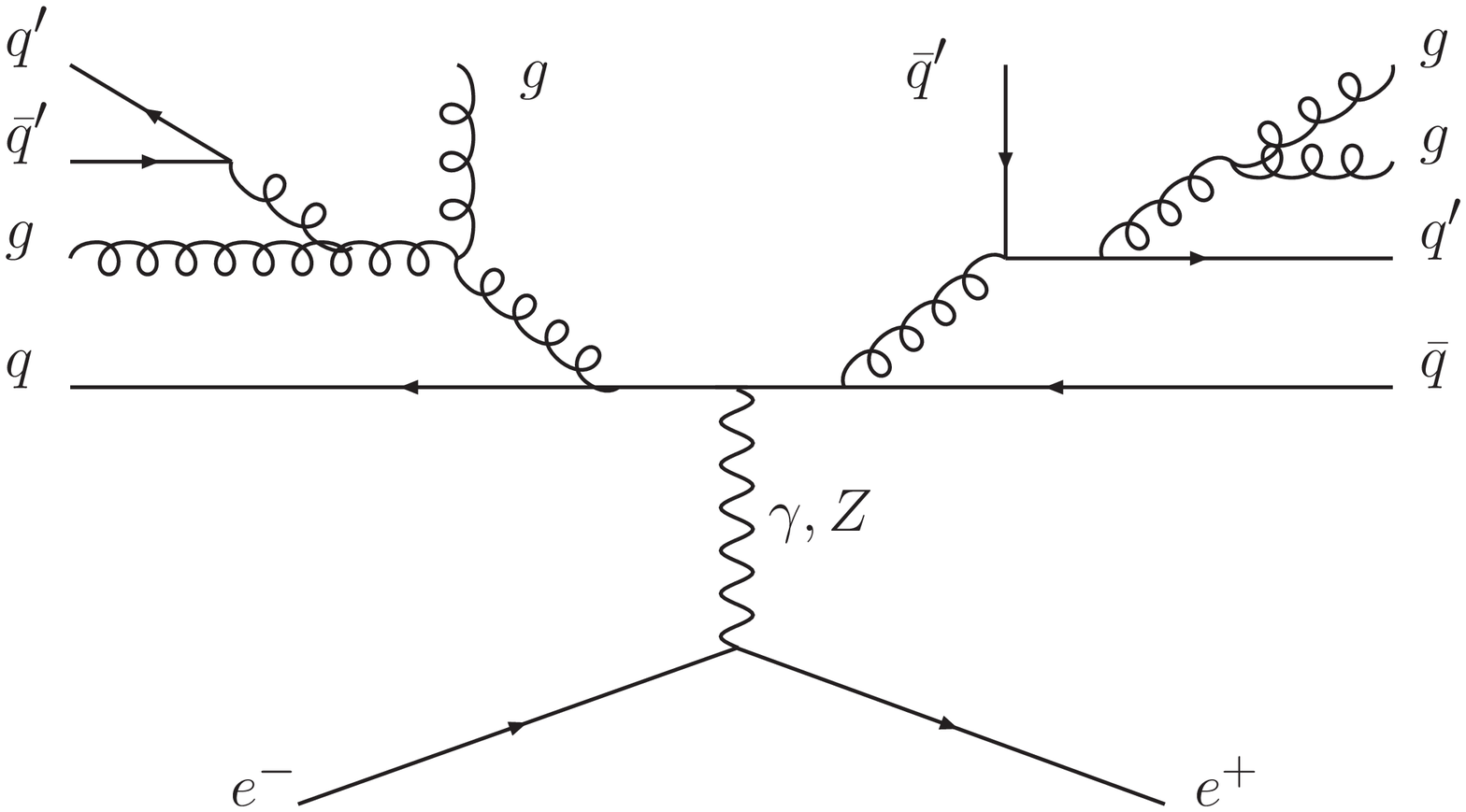}}}
\caption{(a) Quark fragmentation to hadrons induced by confinement forces;
         (b) Quark/gluon cascades at high energies in QCD.
        }
\label{fig:fragm.ab}
\end{figure}

Quarks and gluons move, due to asymptotic freedom of QCD, as quasi-free
particles at short distances of the order of $10^{-15}$ cm ($10^{-2}$ fm) in the 
femto-universe. When these coloured objects separate to more than 
of the order of 
1 fm, confinement forces become effective 
which bind the quarks and gluons in hadrons. The hadronisation proceeds 
through the formation of jets in high energy processes which is driven 
by two dynamical mechanisms. These mechanisms can be explicated most easily 
in $e^+ e^-$ annihilation to hadrons, $e^+ e^- \to q \bar{q}, q \bar{q} g, 
...$, {\it cf.} Fig.{\ref{fig:fragm.ab}}. {\it (i)} Quarks 
which are suddenly accelerated in the production 
process at short distance and time of the order of $1/E \ll 1$ fm, 
will radiate gluons preferentially into a cone of small aperture, 
$dN/d\Theta^2 \sim 1/\Theta^2$. Subsequently the gluons may split 
into two gluons or quark-antiquark pairs, and, repeatedly, quarks and gluons 
again into quark and gluon pairs, so that the original quark fragments finally 
into a quark/gluon cascade within a narrow cone. {\it (ii)} When 
the coloured quarks on the way out of the femto-universe to large distances 
separate to more than 1 fm, a gluonic flux tube of narrow transverse dimensions 
builds up which fragments into ordinary hadrons. Similar mechanisms
lead to the hadronisation of gluons. In total, the perturbative quark/gluon 
cascade as well as the partons fragmenting non-perturbatively into hadrons
generate jets of particles preserving, in momentum and energy, the original
kinematic characteristics of their parent partons. 

\subsection{Quark fragmentation}
When quarks and antiquarks in high energetic processes, like $e^+ e^- \to 
q \bar{q}$, separate from each other, a linear gluonic flux tube is expected 
to build up, with energy density of about 1 GeV/fm and small transverse size,
which will confine the two coloured objects. For sufficiently large separations 
$R \sim 1$ fm, enough energy will be accumulated in the flux tube so that new
quark-antiquark pairs can be created spontaneously and the flux tube breaks up. 
This expectation is borne out by lattice analyses \cite{Bali} which, in static 
approximation, support this picture qualitatively. Beyond the short-distance 
regime, the potential between heavy quarks rises linearly with distance, 
$V(R) = \sigma R$ with $\sigma \approx 1$ GeV/fm. However, when the distance 
between the heavy quarks exceeds a value of about 1.2 fm, light quark pairs 
$q \bar{q}$ are created and the heavy-quark $[Q \bar{Q}]$ system breaks up 
into two separate mesons $[Q \bar{q}]$ and $[\bar{Q} q]$. 

Repeating this break-up process, adjacent quarks and antiquarks coalesce to
hadrons with small transverse momenta of the order of 350 MeV so that narrow jets 
of collimated hadrons are generated~\cite{FieldF,Lund}. If the partition 
of energies in $q \to h_{[q{\bar{q}}']} + q'$ scales with the energy 
of the primary quark, the number density of hadrons $D(z)$ observed 
with energy $z = E^h/E_q$ obeys, for a single species, the recursion formula~\cite{FieldF}
\begin{equation}
D(z)= f(1-z) + \int_z^1 f(\eta) F(z/\eta) \frac{d\eta}{\eta}~,
\end{equation}
with $f(\zeta)$ denoting the break-up probability 
for fractional energy $\zeta$. This equation states that the primary meson might be
the first in rank primary meson, with probability $f(1-z) dz$, or if not, then the first-rank
 primary meson has left a momentum fraction $\eta$ with probability $f(\eta) d\eta$, and in this
remaining cascade the probability to find $z$ in $dz$ is $F(z/\eta) dz/\eta$.
Dividing out by $dz$ gives the above equation. The
 probability $f(\zeta)$, which 
must be determined experimentally, is generally parametrised as a polynomial 
$\sim (1-\zeta )^\beta$ or as an exponential $\sim \zeta^{-1} (1-\zeta)^\beta 
\exp[-\alpha /\zeta]$ in string fragmentation. 

From this picture two important consequences can be derived. 

{\it (i)} Solving the evolution equation generates a pole in $z \to 0$,
most easily seen for polynomial probabilities, 
\begin{equation}
D(z) \to \frac{const}{z} \;\; {\rm for} \;\; z \to 0 \,.
\end{equation}
Thus the fragmentation mechanism predicts a large number of soft (low energy) hadrons 
in the jets, {\it i.e.} a long constant plateau in the rapidity $y = \log z^{-1}$.

{\it (ii)} Summing up the hadron charges in the jets reflects the charge 
of the parent quark. If $u,d,s$ quark pairs were created in the flux tube 
spontaneously with equal probabilities, the sum would measure the charge 
exactly. However, since $s$-quarks weigh a little more than $u$- and $d$-quarks, 
the probabilities for spontaneous quark-pair creation deviate from 1/3 
by a small amount and a small fraction of the charge leaks into the 
plateau:
\begin{equation}
\langle Q_q \rangle = \sum_{h \in jet} e^h = e_q - \gamma \,.
\end{equation}
In the parton model language~\cite{FieldF} one finds $\gamma \approx 0.067$, {\it i.e.} 
$\langle Q_u \rangle = 0.60$, $\langle Q_d \rangle = \langle Q_s \rangle 
= -0.40$. The close relation to the ideal values +2/3 and -1/3 therefore 
allows the efficient tagging of the parent quark charges in the jets.  In practise, things are
more involved in perturbative QCD. Jet-charge studies have been undertaken extensively at
 LEP~\cite{Abbiendi:1999ry},
where a lot of data are available at the $Z$ peak and flavour-tagged results~\cite{Albino:2005me},
 distinguishing between the light-quark, charm and bottom contributions, have been obtained.

The light quark fragmentation to mesons can effectively be described by the
fragmentation function
\begin{equation}
D(z) = (1+\beta) \frac{1}{z} (1-z)^\beta 
                                   \;\; {\rm with} \;\; \beta \sim 0.2~,
\end{equation}
for small jet energies $\sim 7$ GeV. For higher energies QCD predicts 
a stronger fall-off of the spectrum. [For a detailed overview 
of quark fragmentation to various types of mesons and baryons
 see~\cite{Saxon,Albino:2008gy}.]  

The fragmentation function of the heavy $c,b$-quarks behaves rather
differently. It was recognized very early~\cite{heavyq,heavyq-2} that a heavy flavoured meson
(containing a charm or bottom quark and a light antiquark) retains a good fraction of the
momentum of the primordial heavy quark. Thus, 
 due to the inertia of the heavy quarks, the fragmentation function of a heavy quark
should be much harder than that of a light hadron. Estimating the 
size of the transition amplitude by the energy transfer in the 
break-up process $Q \to h_{[Q \bar{q}]} + q$, the 
fragmentation function behaves, for example, as \cite{Pet}
\begin{equation}
D_Q(z) \sim \frac{1}{z \left[ 1 - \frac{1}{z} 
                                - \frac{\epsilon_Q}{1-z} \right] ^2}
       \;\; {\rm with} \;\; 
       \epsilon_Q \sim \Lambda^2 / M^2_Q  \,,
\end{equation}
with $\Lambda \sim$ 200 MeV denoting the strong interaction scale. 
The spectrum develops a narrow maximum near $z_{0} \sim 
1 - \sqrt{\epsilon_Q}$. This form describes the essential characteristics of the hard spectra 
of $Q$-flavoured mesons in the heavy $c,b$ jets with $M_c \sim$ 1.5 GeV 
and $M_b \sim$ 4.5 GeV. For more recent works on the heavy quark fragmentation,
see~\cite{Mele:1990cw,Ma:1997yq,Qfrag,Cacciari:2005uk,Kneesch:2007ey,Kniehl:2008zza}.

It should be pointed out that at higher energies, where the heavy quarks are produced
with momenta much larger than the heavy quark mass, one
expects important perturbative QCD corrections, enhanced by the powers of the logarithms of
the transverse momenta to the heavy quark mass, which modify the shape of the fragmentation
functions. These effects can be implemented using the framework of an evolution equation which
is discussed later in this review. They have to be incorporated in the analysis of data. 

After the discovery of quark jets in 1975 in $e^+ e^- \to q \bar{q}$ at SLAC,
detailed studies in understanding the hadronisation process, and hence  the
 energy-momentum profiles of the quark jets,
were initiated in 1977 by Feynman and Field~\cite{FieldF}. In their approach,
the initial quarks
and antiquarks produced in $e^+e^-$ annihilation fragmented independently
in a cascade process,
 $q \to q+ (\bar{q}^\prime q^\prime) \to h_{(q \bar{q}^\prime)} +q^\prime$,
conserving the charge and other flavour quantum numbers at
 each step of this cascade. To determine the energy-momentum profile,
 light-cone variables $p= (p_+,p_-,\vec{p}_T)$ were used with $p_- \ll p_+$,
where $p_\pm = E \pm p_\parallel $. The fragmentation
 $q \to h+q^\prime$ is then affected through a primordial
fragmentation function
\begin{equation}
f_q^h(z)= 1- a + 3a (1-z)^2, \;\;\;\; z=\frac{(E+ p_\parallel)_{\rm h}  }
{(E+ p_\parallel)_{\rm q}}~, 
\end{equation}
with $a$ an adjustable energy-independent parameter, fixed by data.
As already discussed,  this gives rise to an scale-invariant longitudinal
energy
distribution of hadrons in a jet. Heavy quark fragmentation (for example of
 a charm quark into a
$D$ meson) is encoded by a different primordial $c \to D$ fragmentation
function, as already discussed in this section above. 
The $\vec{p}_T$-distribution ($\vec{p}_T$ is the transverse momentum
 measured with respect to the jet-axis, which can be identified with the
direction of the fragmenting quark, for the time being) was 
implemented in terms of a Gaussian function, characterised by
$\sigma_q\simeq 0.35$ GeV, also determined phenomenologically:
 $g(p_T^2)= (2 \sigma_q^2)^{-1} {\rm e}^{-p_T^2/2\sigma_q^2}$. Like the flavour
quantum numbers, $\vec{p}_T$ is locally compensated, implying that 
 an $r^{\rm th}$-rank primary meson has a momentum
$\vec{k}_T(r)$, with $\vec{k}_T(r)= \vec{q}_{T r} - \vec{q}_{T(r-1)}$.
The striking feature of the Feynman-Field
jet is its simple algorithm with  the phenomenological profile determined
in terms of a few parameters, which  provided an adequate description of the
non-perturbative aspects of jets initiated by quarks. 
\subsection{Gluon fragmentation}
The fragmentation of gluon jets follows rules similar to quarks. 
Two paths had been chosen in the analysis of PETRA jets. 
The properties of $g$-jets may be described either as a nearly
flavour-neutral average \cite{PHENgluon1} of $u,d$ and, with 
reduced probability, $s$-quark jets, or, alternatively, gluon jets 
may be reinterpreted as a superposition of quark and antiquark 
jet pairs \cite{PHENgluon2} with the spectra derived from the 
$g \to q \bar{q}$ splitting function. In any case, the transition
from gluons to quarks $g \to q \bar{q}$ in creating the leading 
particle will soften the gluon fragmentation compared with the
quark fragmentation, accounted for effectively by raising 
the power fall-off towards $z \to 1$ of the fragmentation function of the order 
of $\sim 1.5$.   
\subsection{Hadronisation Models}
Quark and gluon configurations created at small distances must 
transform to bundles of hadrons due to confinement at large distances. 
The transformation requires non-perturbative mechanisms and, therefore,
cannot be treated rigorously at the present time. Instead, models
have been developed which describe this step at various levels of
sophistication.

\subsubsection{Independent jet fragmentation}
Gluonic flux tubes, built up when coloured objects separate, may hadronize
into a jet of collimated hadrons as argued earlier. While the basic picture
had first been described for quark jets \cite{FieldF}, gluon jets can be
analysed similarly when the gluons are either treated globally as partons
\cite{PHENgluon1} or split into quark-antiquark pairs, fragmented incoherently
again \cite{PHENgluon2}. Implementing energy-momentum
conservation in the overall event was an unsatisfactory element of the model.
Nevertheless, independent 
jet fragmentation has a simple and transparent structure including a small number 
of parameters. The picture could account quite successfully for the essential 
properties of two- and three-jet events in $e^+ e^-$ annihilation 
at PETRA and PEP. Thus, it had initially been the right theoretical tool 
for proving experimentally the gluonic nature of the third jet 
in three-jet events.

\subsubsection{String model}
Apart from the different choice of the primordial splitting function
$f(\zeta)$, motivated by covariance and side-independence of the 
beginning of the break-up sequence, quark jets in the string model
\cite{Lund} are not very different from independent fragmentation
schemes. However, gluons are incorporated quite differently. They 
generate kinks which locally transfer energy and momentum to the strings 
stretched between quarks and antiquarks. A small number of hadrons 
is boosted from the segment between quark and antiquark jets to the
segments between quark or antiquark and gluon jets. This string effect
has been observed experimentally as reshuffling of hadrons
between jets, discussed later.

\subsubsection{Cluster hadronisation}
Quite a different hadronisation mechanism is based on colour pre-confinement
\cite{preconf}. Neighbouring coloured partons in cascades arrange themselves 
in colour-neutral islands with preferentially small invariant masses. In
practise, the quark/gluon partons in cascades are evolved down to low invariant masses 
of the order of several $\Lambda_{\rm QCD}=O(200~{\rm MeV})$, where $\Lambda_{\rm QCD}$
is the scale parameter specific to QCD and appears in the argument of 
$\alpha_s(Q^2)$.
 Splitting the gluons in the final step 
into $q \bar{q}$ pairs, neighbouring quarks and antiquarks form
the colourless clusters which may finally decay to standard hadrons 
according to phase space rules \cite{Herwig1}. The reduced number
of radiated gluons off heavy quarks and the small number of
large invariant masses in the tail of the distribution 
of the colour-neutral clusters can be covered by non-perturbative 
string-type compliments to the cluster hadronisation scheme.

Based on these schemes QCD event generators have been constructed which
describe hadron spectra at a high level of accuracy. While
the prototypes had originally been developed for hadron production 
in $e^+ e^-$ annihilation, the event generators 
have been expanded to proton-(anti)proton collisions and complimented 
by programs for lepton-nucleon collisions. The modern versions of 
PYTHIA \cite{Pythia,Sjostrand:2007gs}, HERWIG \cite{Herwig}, SHERPA \cite{Sherpa} and
others involve the cascading of quarks/gluons in the final and the $p/\bar{p}$
initial states, and string or cluster hadronisation in the final states. 
For multijet final states, frequently produced at high energies 
in colliders, elaborate techniques have been developed, based 
on the relation~\cite{Sudakov:1954sw}
 ${\rm PS}(Q^2) = {\rm ME}(Q^2) \times {\rm Sudakov \; factor} \, [Q^2_{max} \to Q^2]$,
 to accomplish smooth transitions between quark/gluon parton 
showers (${\rm PS}$) and well separated multijet final states described by
fixed-order perturbation theory matrix elements (${\rm ME}$) squared.
\subsection{Inclusive jet measures}
\begin{figure}
\center{
\resizebox{0.75\columnwidth}{!}{
\includegraphics{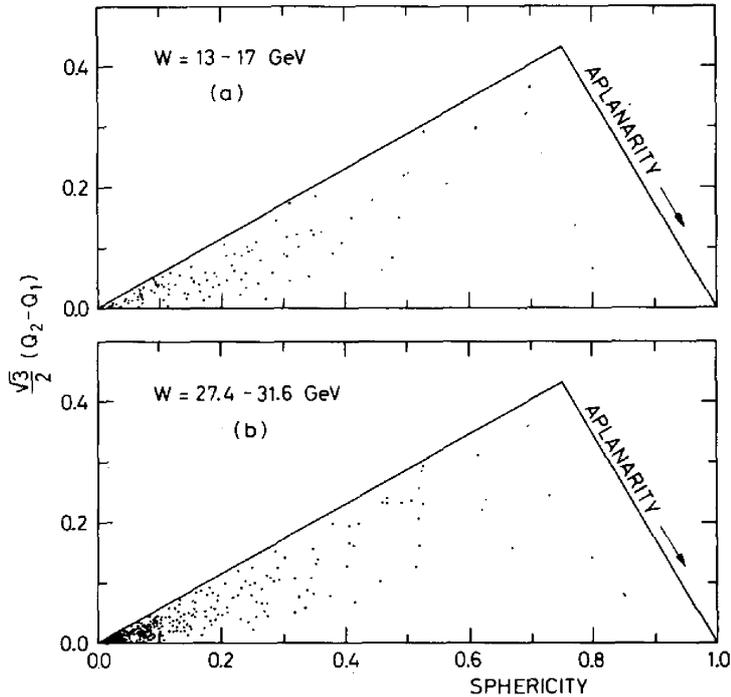}}}
\caption{Distributions of the $e^+e^- \to$ hadron events as a function of aplanarity
  and sphericity defined in the text for the low (a) and high (b)
energy PETRA data (TASSO~\cite{EXP4gluon-TASSO}).
        }
\label{fig:PLB86-5}
\end{figure}

 In this section we discuss some inclusive jet measures which have played
an important role in the quantitative tests of QCD. The first of these is
the observable called sphericity which played a central role in the discovery of
quarks jets at SPEAR. In its tensorial form it is
defined as follows~\cite{PHENquark}:
\begin{equation}
S_{\alpha \beta}=\frac{\sum_i p_{i\alpha} p_{i\beta}}{\sum_{i}\vec{p_i}^2}~,
\label{eq:Salfabeta}
\end{equation}
which can be diagonalised obtaining the 
principal axes $\vec{n_1},\vec{n_2}$ and $\vec{n_3}$ with
corresponding eigenvalues $Q_1,Q_2$ and $Q_3$. The $Q_i$ can be
ordered $Q_1<Q_2<Q_3$ and normalised so that $Q_1+Q_2+Q_3=1$.
Then the squares of the transverse momenta are minimal with respect to the
axis $\vec{n_3}$ and the sphericity $S$ is given by
\begin{equation}
 S= \frac{3}{2}(1-Q_3)=\frac{3}{2}(Q_1+Q_2)~,
\end{equation}
with the sphericity axis equal to $\vec{n_3}$. For events with two particles
with equal and opposite momenta (ideal two-jet event) we have $S=0$ and
$S \to 1$ for completely isotropic events. Because of the normalisation
$Q_1+Q_2+Q_3=1$ only two of the eigenvalues are needed to characterise an event.
For example one can take in addition to S the so-called aplanarity $Ap$, which
is
\begin{equation}
Ap = \frac{3}{2}Q_1=
\frac{3}{2} {\rm min} \frac{\sum_{i}|\vec{p_{iT,out}}|^2}{\sum_{i}\vec{p_i}^2}~.
\end{equation}
The aplanarity $Ap$ minimises the transverse momenta with respect to a
plane. Events with small $Ap$ are almost planar. The jet variables 
of an event, $Q_1,Q_2$ and $Q_3$ can be plotted inside a triangle as shown
in Fig.~\ref{fig:PLB86-5}, in which events obtained by the
 TASSO Collaboration at PETRA
are shown. In this plot planar events are found in the strip with small
$Ap$, 2-jet events have in addition also small $S$.
Alternatively, sphericity can be defined
 as~\cite{PHENquark}
\begin{equation}
 S=\frac{3}{2}\, {\rm min}\, \frac{\sum_{i} |\vec{p_{iT}}|^2}{\sum _{i}|\vec{p_i}|^2}~.
\end{equation}
Here,  $\vec{p_{iT}}$ are the transverse momenta of all produced
hadrons in the final state event relative to an axis chosen such that the
numerator is minimised.

 The method based on
the sphericity  tensor, first applied  to the
analysis of the 3-gluon decay of the $\Upsilon$ resonance~\cite{Alexander} and 
to the analysis of $q\bar{q}g$ events~\cite{WuZo}, has the 
advantage that the eigenvalues $Q_i$ and the principal axes $\vec{n_i}$ and 
from this $S$ and $Ap$ can be calculated quite easily. Since in these jet 
measures the momenta enter quadratically, the high momentum particles are 
weighted stronger in the calculation of $S$ and $Ap$. Also, these variables 
are not invariant against clustering of particles and depend strongly on 
details of the fragmentation of quarks (and gluons) into hadrons. This has, for
example, the effect, that the sphericity changes if a particle momentum
splits up by decay, as for example, $\rho^0 \to \pi^+\pi^-$ or by 
fragmentation, for instance $q \to q' + meson$ into two or more momenta. 
Therefore these variables are also sensitive to the emission of soft or 
collinear gluons. 

There exist other variables  
which are infrared safe (this term is used for observables which are free of 
divergences when calculated in perturbation theory in the limiting case of low energy radiation)
 and which depend on linear sums of 
momenta. Known examples are thrust $T$ and acoplanarity $A$ which are defined 
by
\begin{equation}
  T = max \frac{\sum_{i}|\vec{p_{iL}}|}{\sum_{i}|\vec{p_i}|}~,
\end{equation}
\begin{equation}
  A= 4min \left(\frac{\sum_{i}|\vec{p_{iT,out}|}}{\sum_{i}|\vec{p_i}|}\right)^2~.
\end{equation}
For thrust $T$, which was introduced in \cite{Brandt,Farhi}, 
the thrust axis $\vec{n}$ is obtained by maximising the
longitudinal momenta $\vec{p_{iL}}$  along it. T varies 
in the range $0.5<T<1.0$, where the lower limit corresponds to isotropic events
and the upper limit to completely collinear configurations. In a similar
way for spherocity $S^\prime$, defined as~\cite{Brandt79}   
\begin{equation}
  S^\prime = (\frac{4}{\pi})^2 {\rm min} 
\left(\frac{\sum_{i}|\vec{p_{iT}}|}{\sum_{i}|\vec{p_i}|}\right)^2~,
\end{equation}
the $|\vec{p_{iT}}|$ is minimised with respect to a unit vector $\vec n $. It lies between 0
and 1 for configurations from collinear to fully isotropic events. Similar to 
$Ap$ the acoplanarity is obtained in such a way that the $\vec{p_{iT,out}}$ is 
minimal with respect to a plane. Planar event must have small $A$ values. For
massless particles $A$ varies between 0 and 2/3.

Various other jet measures have been proposed: For example a generalisation
of thrust to three clusters instead of two, called triplicity 
\cite{Brandt79} or jettiness \cite{WuZo}. A variable 
introduced for the analysis to verify the existence of four-jet events is the 
variable tripodity $D_3$ \cite{Nachtmann}. These and other jet
variables will be defined explicitly when they are used to interpret
specific final state data in $e^+e^-$ annihilation in later sections.

\subsection{Jet algorithms}
Classifying multi-particle final states qualitatively in jets with high 
energies is straightforward for a coarse picture. However, when the picture 
is refined to a high level of precision, algorithms must be employed 
which define the jets in a rigorous way. In addition, when experimental 
measurements are compared with theoretical predictions, the algorithms 
used in the experimental analyses must conform with the algorithms
adopted in the theoretical analyses.

A multitude of algorithms  \cite{Salam:2009jx} has been developed to describe 
jets at high energies. A few representative examples should characterise
the two classes, sequential recombination and cone algorithms.
Recombination algorithms have been introduced originally in $e^+ e^-$ 
annihilation, while cone algorithms have been used frequently at 
hadron colliders so far. We shall discuss some of these algorithms
later while discussing jets in hadronic collisions.

\subsubsection{Sequential recombination algorithms}
The JADE algorithm \cite{JADE} is a prominent representative for recombination
algorithms applied in $e^+ e^-$ annihilation. Particles are clustered 
in a jet iteratively as long as their distance remains less than a 
prescribed cut-off value. The distance of two particles is defined by
the invariant mass of the pair:  
\begin{equation}
y_{ij} = 2 E_i E_j (1-\cos\theta_{ij}) / E^2_{cm}  \,.
\label{eq:yij}
\end{equation}
If the criterion $y_{ij} \le y_{cut}$ is fulfilled, the particles $i$ and 
$j$ are combined to a compound by adding up energy and momentum,
for instance, and the recombination continues by pairing the compound
with the next particle $k$. The procedure stops after all particles are 
associated with jets. The cut-off value $y_{cut}$ is generally chosen 
in the range from $10^{-1}$ down to $10^{-3}$. 

The presence of $E_iE_j$ in the numerator of $y_{ij}$ means that two soft particles moving
in {\it opposite directions} can get combined into a single particle in the early stages of
clustering, which is counter-intuitive to the idea of a jet as consisting of particles
restricted in the angular dimension. Apart from this, JADE algorithm leads to a structure
in higher orders of pertrurbation theory which does not allow itself to be expressed in
a compact resummed form. Technically, this means that the double logarithms of the type
$\alpha_s^n \ln^{2n}y_{\rm cut}$ ($n=1,2,... $), which arise in higher orders of perturbation theory
and which should be resummed to have the correct perturbative form in a limited kinematic region,
 named after Sudakov, are either not present or not discernible easily. 
To rectify both of these shortcomings, the JADE algorithm has been improved
by substituting $E_i E_j \to \min[E^2_i,E^2_j]$ in the DURHAM algorithm~\cite{Catani:1991hj}.
This amounts to defining the distance by the minimal transverse momentum 
$k_t$ of the particles in nearly collinear pairs. 

The concept has been transcribed to hadron colliders~\cite{Ellis,Catani-algo}, where the total 
sub-energy is experimentally not well defined, by switching to
un-normalised measures and replacing the angles between particles by
the differences of the rapidities $y = 1/2 \log(E+p_z)/(E-p_z)$ 
along the beam axis and the 
azimuthal angles $\phi$ in the plane transverse to the beam axis,
\begin{equation}
d_{ij} = \min[p^{2p}_{ti},p^{2p}_{tj}] \, 
         [(y_i - y_j)^2 + (\phi_i - \phi_j)^2]/R^2 \,,            
\end{equation}
with $p_{ti(j)}$ denoting the transverse momenta of the particles
with regard to the beam axis. 
The jet parameter $R$ is chosen of the order of one. Since the individual quantities 
$(y_i-y_j)$, $(\phi_i-\phi_j)$, $p_{ti}$ and $p_{tj}$ are all invariant under longitudinal boosts,
the distance  measure $d_{ij}$ is also longitudinally invariant.
 Recombination with the beam jets
is controlled by the observable $d_{iB} = p^{2p}_{ti}$, included parallel
to the measure $d_{ij}$ when recombining all the particles to jets with non-zero 
transverse momenta and beam jets. Originally, the power parameter $p$ had 
been chosen 1 in the $k_t$ algorithm~\cite{Catani:1991hj} and 0 in the Cambridge/Aachen
 algorithm~\cite{Wobisch:1998wt}. 
However, clustering involving hard particles are favoured by choosing $p = -1$ 
in the $anti$-$k_t$ algorithm~\cite{Cacciari:2008gp}. This algorithm, which makes jets 
grow outwards from hard seeds as intuitively expected, is the preferred tool 
for LHC analyses.

\subsubsection{Cone algorithms}
Cone algorithms had been introduced in QED to tame infrared and collinear
singularities encountered in photon radiation off charged particles. The concept has been
translated to QCD in formulating the Sterman-Weinberg jets \cite{SterW}. Defining 2-jet events
as hadron configurations in which all but a fraction $\epsilon$ of the
total energy is contained in cones of half-angle $\delta$ around the
central event axis, the ratio
\begin{equation}
\frac{\sigma_2}{\sigma} = 1 - \frac{32}{3} \, \frac{\alpha_s}{2 \pi} \,
                          \log\frac{1}{\delta} \, \log\frac{1}{\epsilon} \,.
\end{equation}
describes the 2-jet fraction of hadronic events in $e^+ e^-$ annihilation
in the leading logarithmic approximation.
 
The transition to hadron collisions has been formulated again by adopting
the definition of distances based on rapidities and azimuthal angles.
The clustering requires a seed location; the 4-momentum of the cluster is
determined by summing the 4-momenta of all the objects within a distance
$R= \sqrt{ (y-y_c)^2 + (\phi - \phi_c)^2}$ from the seed $(y_c,\phi_c)$. In one variant,
 used in the analysis of the Run II
Tevatron data, the 4-momenta are summed using the so-called
 E-scheme~\cite{Blazey:2000qt} (this should not be confused with the E-scheme in the
analysis of the jets in $e^+e^-$ annihilation),
$(E,p_x,p_y,p_z)=\sum (E,p_x,p_y,p_z)_i$, and the various variables are defined as
\begin{equation}
p_T=\sqrt{p_x^2 + p_y^2}, ~~~y=\frac{1}{2} \ln \left( \frac{E+p_z}{E-p_z}\right)~,
~~\phi=\tan^{-1} (p_y/p_x)~.
\end{equation}
This differs from the Snowmass clustering algorithm~\cite{Huth:1990mi}, used in the
analysis of the Tevatron I data, in which the clustering centroid was defined as the 
$E_T$-weighted averages of the rapidity $y$ and the azimuthal angle $\phi$. 
The cones are either centred around seed particles (defined as those particles setting
the initial direction and one sums the momenta of all the other particles around these
seed particles within a specified jet measure), an approach which is not
infrared safe, or they are defined by grouping all particles into
subsets such that the subsets correspond exactly to pre-defined cones. 
For further discussion of these and related issues, we refer to the works of
Seymour~\cite{Seymour:1997kj} and the comprehensive review of jet measures by
 Salam~\cite{Salam:2009jx}.
\section{Discovering quark jets}
\label{sec:3}
\subsection{Quark jets at SPEAR}
\label{sec:3.1}
\begin{figure}
\center{
\resizebox{0.50\columnwidth}{!}{
\includegraphics{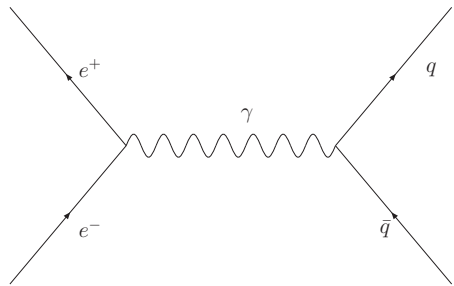}}}
\caption{Born diagram for $e^+ e^- \to \gamma \to q \bar{q}$.}
        \label{fig:Born}
\end{figure}

 The notion of jets in 
$e^+e^-$ annihilation is closely connected with the discovery of Bjorken 
scaling in deep-inelastic electron-nucleon scattering in 1968 at SLAC. As 
mentioned in the introduction,  inelastic 
electron scattering on protons and neutrons at large spacelike ($q^2<0$) 
momentum transfer and large inelasticity $\nu$ can very well be described 
in terms of an interaction of  the spacelike virtual 
photon  with the pointlike constituents of the
nucleon, the partons, identified as the $u$ and $d$ quarks inside the proton
and neutron \cite {Feynman}.
The analogous process with a timelike ($q^2>0$) virtual photon is 
$e^+e^-$ annihilation into a quark-antiquark pair, $e^+e^- \to q\bar{q}$, as 
shown in Fig.~\ref{fig:Born}, where $q$ stands for the quarks $u, d, s, c, b$.
 As explained 
already in the introduction, in this simple model the virtual photon from the
annihilating electron and positron creates a quasi-free quark and
antiquark. The occurrence of real
quark and antiquark particles in the final state is prevented by the the fact that
they carry non-trivial colour quantum numbers. The quarks and antiquarks transform
themselves into hadrons with unit probability under the
confinement forces, which act at much later times $t \simeq 1~GeV^{-1}$. 
These hadrons should appear in the final state aligned roughly along the momentum
direction of the original $q$ and $\bar{q}$, so that two distinct hadron
jets with oppositely directed momenta appear in experiments. This discussion mirrors
the early ``outside-in'' picture of jet formation, which was used in the formative
stages of jet physics. This approach was later replaced by the so-called ``inside-out''
description where quark-antiquark pairs were created out of the vacuum before the
step of hadronisation. We will discuss the salient feature of this development later.

\begin{figure}
\center{
\resizebox{0.75\columnwidth}{!}{
\includegraphics{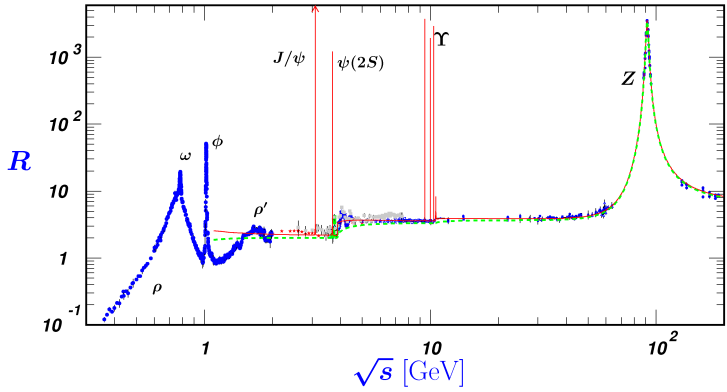}}}
\caption{Measurements of the ratio $R$  
as a function of the $e^+e^-$ centre-of-mass energy $\sqrt{s}$ [From
  PDG~\cite{Nakamura:2010zzi}].}
        \label{fig:Ree}
\end{figure}

 This simple quark model~\cite{THquark,THquark-2,THquark-3} 
was supported by the fact that the total annihilation cross section for hadron 
production $\sigma(e^+e^- \to hadrons)$ is given by the square of the
quark charges $Q_f$ multiplied with the number of colours $N_C$ of each quark 
$q$
\begin{equation}
\sigma(e^+e^- \to hadrons) \equiv \sigma_0=\frac{4\pi\alpha^2}{3s} N_C \sum_{f} Q_f^2~,
\label{eq:sigma-0}
\end{equation}
where the sum over $'f'$ is over all active flavours which can be produced at a
given center-of-mass energy $\sqrt{s}$; 
 $\alpha$ is the fine structure constant $\alpha \simeq
e^2/137$. Dividing by the cross section for the production of a
$\mu^+\mu^-$ pair, $\sigma(e^+e^- \to \mu^+\mu^-)$, one 
obtains the famous Drell-ratio $R$, defined as
\begin{equation}
R\equiv\frac{\sigma(e^+e^- \to {\rm hadrons})}{\sigma(e^+e^- \to \mu^+\mu^-)} = 
N_C\sum_{f} Q_f^2~,
\end{equation}
which has the numerical value 2  (for $f=u,d,s$), and assumes the values
$10/3$ (for $f=u,d,s,c$) and $11/3$ (for $,f=u,d,s,c,b$), as the threshold for
the processes $e^+ e^- \to c\bar{c}$ and $e^+ e^- \to b\bar{b}$ are crossed.
A recent compilation of the hadronic cross section 
$\sigma(e^+e^- \to {\rm hadrons})$ and the corresponding ratio $R$
is shown in Fig.~\ref{fig:Ree}
 (taken from the Particle Data Group~\cite{Nakamura:2010zzi})
where the various resonances ($\rho, \omega, \phi, J/\psi,...)$ encountered in $e^+e^-$
annihilation and the  transition regions are clearly visible. Away from the resonances, the
ratio $R$ is almost flat, increasing as a new quark-antiquark threshold is crossed in
agreement with the values quoted above.  
Note that the $t\bar{t}$ threshold (at around
350 GeV) lies beyond the energies of the $e^+e^-$ collider rings operated so far.

\begin{figure}
\center{
\resizebox{0.75\columnwidth}{!}{
\includegraphics{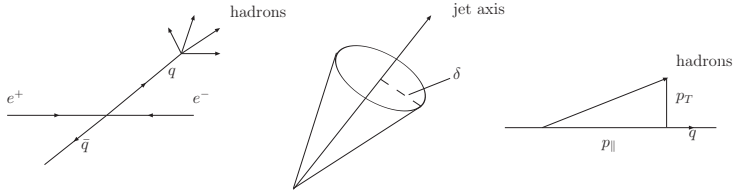}}}
\caption{The process $e^+ e^- \to q \bar{q} \to$ jets, with the jets defined using the
 jet-cone angle $\delta$, arising from limited $p_T$ of the hadrons.}
\label{fig:2jets}
\end{figure}

The production of hadron jets in $e^+e^-$ annihilation as a signature of
the process $e^+e^- \to q\bar{q}$ was suggested by Bjorken and Brodsky already 
in 1970 \cite{PHENquark}.
However, it was not until 1975 that they were discovered
experimentally at SLAC's $e^+e^-$ storage ring SPEAR
 by the SLAC-LBL Collaboration using the
MARK I detector~\cite{EXPquark} when high enough centre-of-mass energies $\sqrt{s}$ became
available. At low energies, for
example at the ADONE ring at Frascati or the original DORIS ring at DESY, it 
was not possible to see jets because the jet cones were too broad.
This is easy to understand if we assume that the transverse momentum
$p_T$ with respect to the jet direction (which, theoretically is the
momentum direction of the original quark $q$ or antiquark $\bar{q}$ ), which 
are emitted back-to back in the c. m. system,
is limited and that the hadron multiplicity $<n>$ increases only modestly
with $\sqrt{s}$. The jet cone becomes narrower with
increasing $\sqrt{s}$, characterised by the mean half angle
$<\delta>$ of the jet cone (see Fig.~\ref{fig:2jets}).
 At $\sqrt{s} = 4~GeV$ the particle multiplicity is about 6, so that
with $<p_T> \simeq 0.35~GeV$ the half-angle of the jet-cone
 $<\delta> \simeq <p_T> <n>/\sqrt{s} \simeq 0.53 
\simeq 30^{\circ}$. This shows that at this energy each of the two 
jets is broader than $60^{\circ}$ on average.

\begin{figure}
\center{
\resizebox{0.70\columnwidth}{!}{
\includegraphics{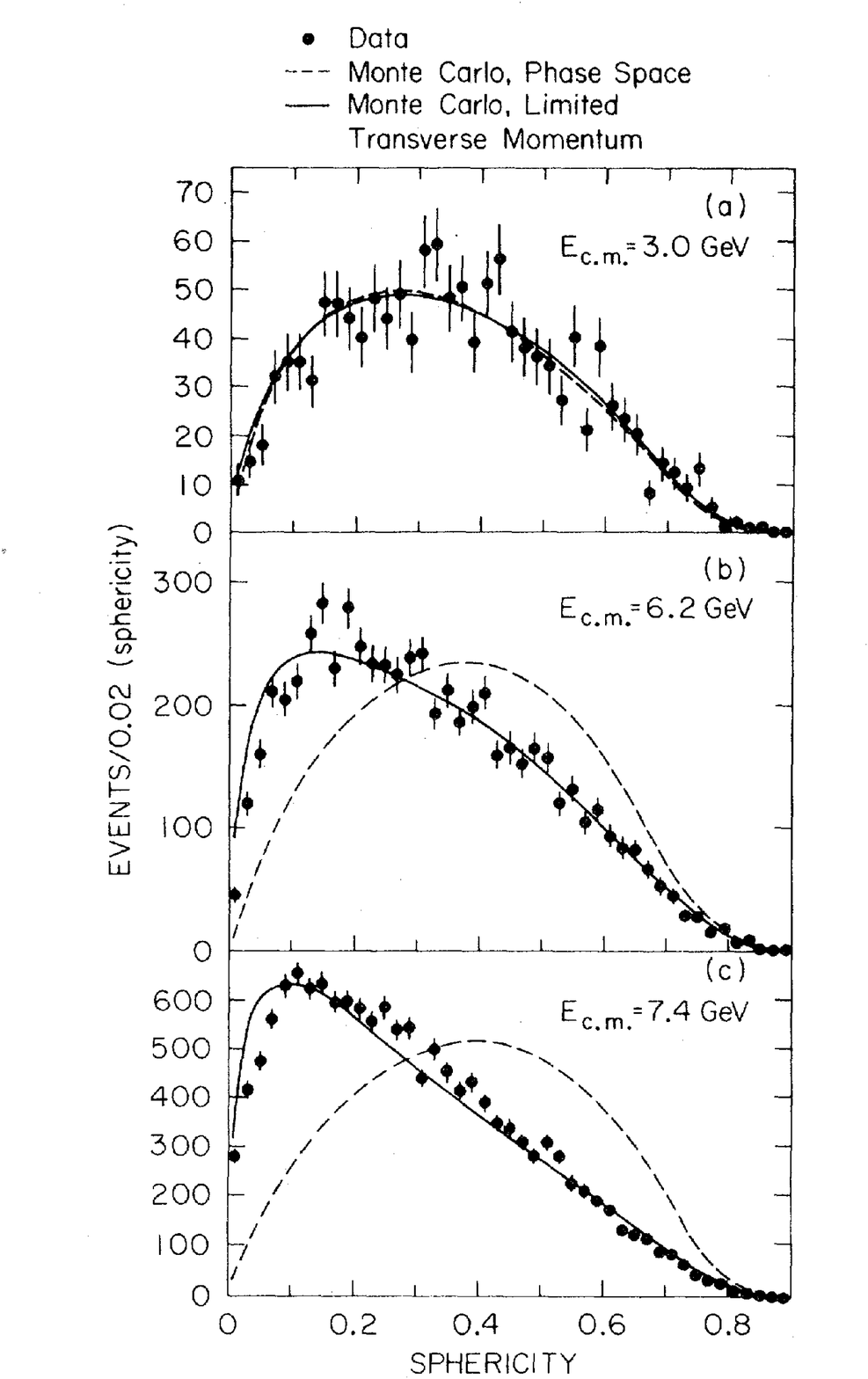}}}
\caption{Observed sphericity distributions for data from MARK I detector, jet model
 (solid curves), and phase-space model (dashed curves) for (a) $E_{c.m.}=3.0$ GeV, (b) 
$E_{c.m.}=6.2$ GeV, and (c) $E_{c.m.}=7.4$ GeV. (From Ref.~\cite{EXPquark}.)
        }
\label{fig:PRD26-3}
\end{figure}

 For establishing the jets it is necessary to 
determine the jet axis along which the transverse momenta of the produced
hadrons are minimised,  In the early work of the SLAC-LBL collaboration, the jet axis was defined 
in terms of the sphericity variable defined earlier.  In the SLAC-LBL experiment
 the mean sphericity was found to be  approximately constant
as a function of the total $e^+e^-$-energy $E_{c.m.}=\sqrt{s}$ up to $4~GeV$ 
and then it decreases with increasing $E_{c.m.}$. 
A detailed comparison is shown in Fig.~\ref{fig:PRD26-3}, in which the measured
sphericity distributions $d\sigma/dS$ at $E_{c.m.}= 3.0$, $6.2$ and $7.4~GeV$ are
compared with the calculated distributions based on  a two-jet model and the phase-space.
At $3.0~GeV$ there is no distinction between the two models and the
data agree with both.  At $6.2$ and 
$7.4~GeV$ the $S$ distributions are peaked towards low $S$ favouring the jet 
model. But the $S$ distributions are still very broad. This comparison shows 
quite clearly that (i) the $E_{c.m.}$ must be high enough to see the 
production of jets in $e^+e^-$ annihilation, and (ii) that even at the 
higher $E_{c.m.}$ energy range of the SPEAR storage ring, the jet 
structure is visible only through a detailed comparison with the prediction of an 
appropriate jet model. Observing the jet structure was easier at  PETRA energies, where 
most of the events have a two-jet topology, which, because of the higher energy
had much narrower angular jet-cones.
 An example of such an event measured by the TASSO collaboration  at 
$E_{c.m.}= 31.6~GeV$, is shown in Fig.~\ref{fig:1.1ab} (left-hand frame).

\begin{figure}
\center{
\resizebox{0.75\columnwidth}{!}{
\includegraphics{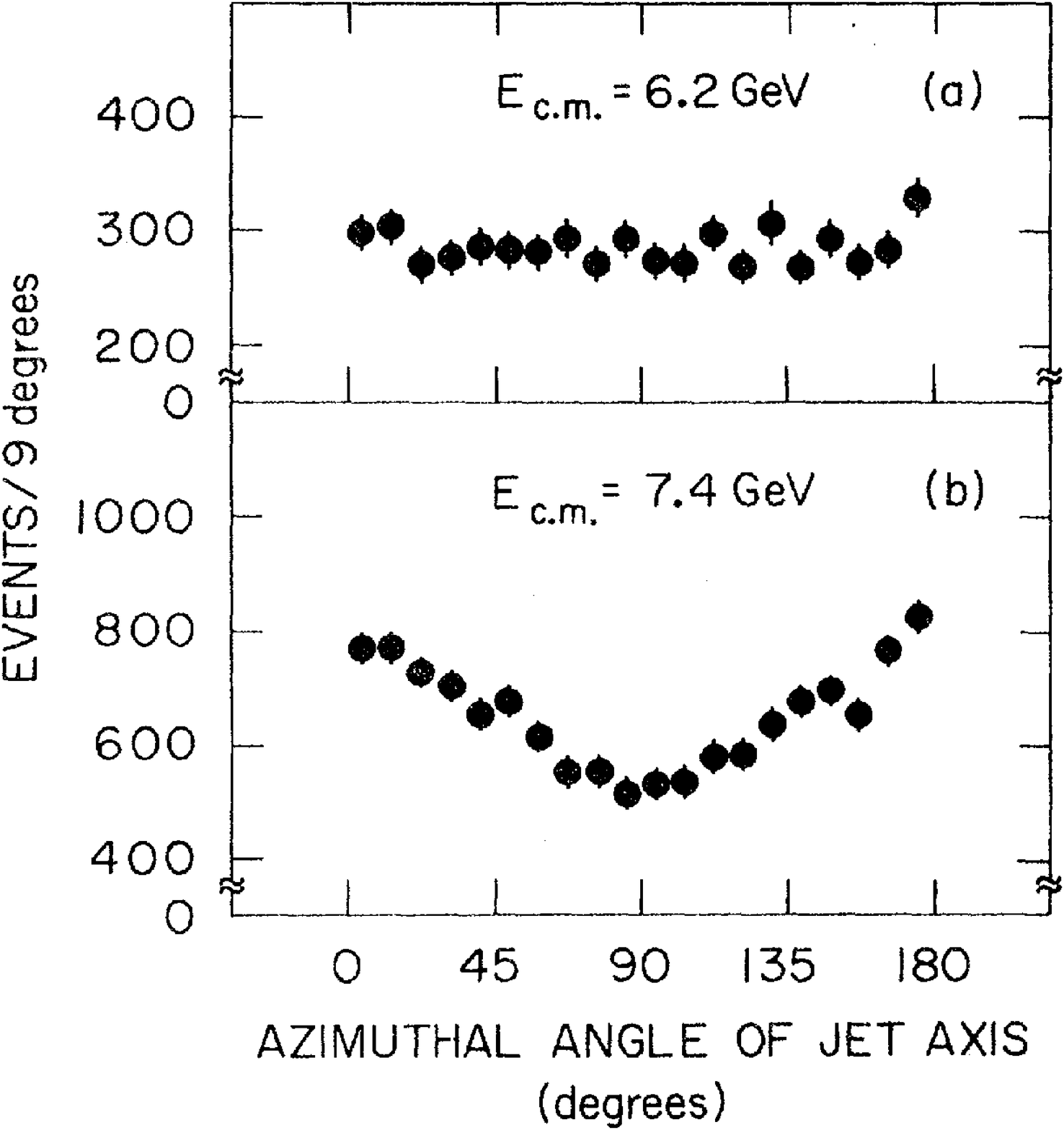}}}
\caption{Observed distributions of jet-axis azimuthal angles from the plane of the
storage rings for jet axis with $|\cos \theta | \leq 0.6$ for (a) $E_{c.m.}=6.2$ GeV and
 (b)  $E_{c.m.}=7.4$ GeV. (From Ref.~\cite{EXPquark}.)
        }
\label{fig:PRD26-11}
\end{figure}

 Further tests of the underlying
quark structure of the jets in $e^+e^-$ annihilation were undertaken at
SPEAR. One such test is the
measurement of the angular distribution $d\sigma/d\cos\theta$ of the jet axis 
with respect to the beam direction. This distribution for the production of
massless spin $1/2$ particles is \cite{Gatto}
\begin{equation}
 \frac{d\sigma}{d\cos\theta} \sim 1+\cos^2 \theta~.
\end{equation}
The first data came from the SLAC-LBL Collaboration at SPEAR. They did the
measurement with transversely polarised $e^+$ and $e^-$ beams available 
at the lower c.m. energies of the SPEAR ring.
 With transversely polarised beams the angular
distribution has the following form
\begin{equation}
\frac{d\sigma}{d\Omega} \sim 1+\alpha\cos^2\theta +\alpha P_+P_-\sin^2\theta~
  \cos2\phi~, 
\end{equation}
where $\phi$ is the azimuthal angle of the jet axis with respect to the
storage ring plane and $P_+$ and $P_-$ are the polarisations of the $e^+$ and
$e^-$ beams, respectively.
The measured $\phi$ distributions (averaged over $\theta$) for 6.2 and
7.4 GeV are seen in Fig.~\ref{fig:PRD26-11}. At 6.2 GeV the beam polarisations are $P_+=P_-=0$
and therefore the $\phi$ distribution is isotropic. At 7.4 GeV, where 
$P_+P_-=0.5$ the characteristic $\cos2\phi$ behaviour is observed. From this 
measurement at SPEAR, the value $\alpha=0.97\pm0.14$ \cite{EXPquark,EXPquark-2} is in 
agreement with the expectation for spin $1/2$ quarks, $\alpha=1$.

 Similar, but  less accurate, results  were obtained by the PLUTO Collaboration at DORIS for 
$E_{c.m.} =  7.7$ and $9.4~GeV$ \cite{Berger78}. Measurements of the angular
distribution of jets at higher energies were also performed at 
the $e^+e^-$ storage rings PEP and PETRA.
 Although the beam energies were
much higher, yielding a much better defined jet axis, the result 
$\alpha =1.04\pm0.16$~\cite{Elsen} does not have a better 
accuracy than the SPEAR measurement, which had the benefit of polarised beams.
This test of the spin $1/2$ nature of the quarks produced in $e^+e^-$
annihilation is very much the same as the verification of the Callan-Gross
relation \cite{Callan} in deep inelastic lepton-nucleon
scattering: $F_2(x)=2xF_1(x)$, which is also very well satisfied 
experimentally.

\subsection{Sterman-Weinberg Jets}
\label{sec:SW-jets}
The existence proof of jets in QCD was provided by Sterman and
Weinberg~\cite{SterW}. As already noted, they calculated the cross section
 $\sigma_{\rm 2-jet}(\epsilon, \delta)$
 for the process $e^+ e^- \to 2-{\rm jets}$, where the jets are defined by two
cones in opposite hemispheres  with half-angular size $\delta$, having all but
a fraction $\epsilon$ of the total c.m. energy. In the general field
theory context, jets were anticipated due to the Lee-Nauenberg theorem~\cite{Lee:1964is},
 which states
that the transition probability in a theory involving massless particles is finite
provided summation over degenerate states is performed.

\begin{figure}
\center{
\resizebox{0.60\columnwidth}{!}{
\includegraphics{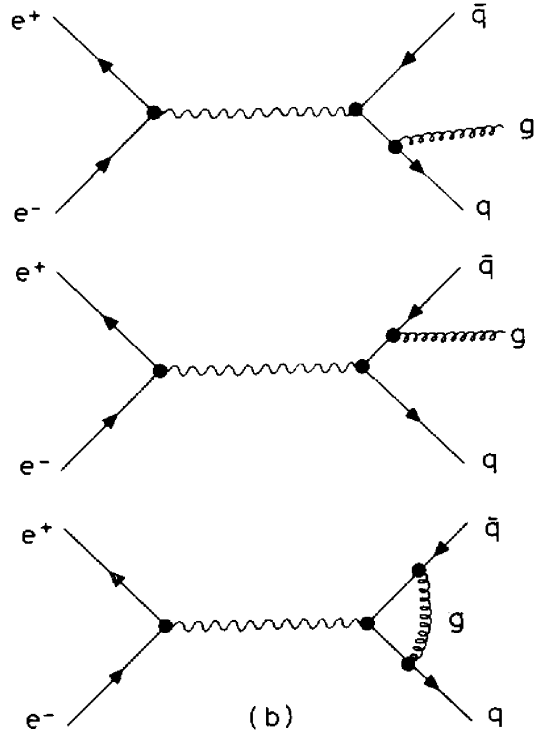}}}
\caption{Lowest order Feynman diagrams contributing to 
$e^+e^- \to q\bar{q} g$ (upper two diagrams)  and vertex corrections in
 $e^+e^- \to q\bar{q}$ (b).}
\label{fig:qqg}
\end{figure}

 The Feynman diagrams which
contribute in order $\alpha_s(Q^2)$ are shown in Fig.~\ref{fig:qqg}. 
 For small $\epsilon$ and
 $\delta$, and to leading order in $\alpha_s(Q^2)$ one has 
\begin{equation}
\sigma_{\rm 2-jet}(\epsilon, \delta)=\sigma_0 \left[ 1 + C_F \frac{\alpha_s(Q^2)}
{\pi} \left(-4 \ln 2\epsilon \ln \delta -3 \ln \delta -\frac{\pi^2}{3}
+ \frac{5}{2} + O(\epsilon) + O(\delta) \right)\right]~,
\label{eq:sw-2jets}
\end{equation}
where $\sigma_0$ is the lowest order cross section given in
 Eq.~(\ref{eq:sigma-0}), $C_F=4/3$ and $\alpha_s(Q^2)$ is the QCD coupling
constant defined in the lowest order
\begin{equation}
\alpha_s(Q^2)=\frac{12 \pi}{(33-2n_f)\ln\frac{Q^2}{\Lambda^2}}~,
\label{eq:alphas-0}
\end{equation}
where $n_f$ is the number of quark flavours.
 The terms $O(\epsilon)$, $O(\delta)$ neglected by Sterman and Weinberg
are all finite, essentially proportional to the phase space and have
been subsequently worked out~\cite{Stevenson:1978td}.
Here $\Lambda$ is a scale parameter, to be determined experimentally, typically of
 $O(200)$ MeV. As $Q^2 \to \Lambda^2$, $\alpha_s(Q^2) \to \infty$, signaling the breakdown of
perturbation theory. The above expression for 
$\alpha_s(Q^2)$ also states that $\alpha_s(Q^2) \to 0$ as $Q^2 \to \infty$,
implying that QCD is an asymptotically free field theory. In the range of $Q^2$ where
 $\alpha_s(Q^2)/\pi \ll 1$, one has a controlled perturbative region.

 Since, in
leading order in $\alpha_s(Q^2)$, the inclusive hadronic cross section
for $e^+ e^- \to \gamma \to {\rm hadrons}$ is
\begin{equation}
\sigma(e^+ e^- \to \gamma \to {\rm hadrons})=\sigma_0(1 + 
\frac{\alpha_s(Q^2)}{\pi})~,
\label{eq:sigma1}
\end{equation} 
the complement of $\sigma_{\rm 2-jet}(\epsilon, \delta)$ is the 3-jet
cross section 
\begin{equation}
\sigma_{\rm 3-jet}(\epsilon, \delta)= \sigma_0 \frac{\alpha_s(Q^2)}
{\pi} C_F\left(4\ln 2\epsilon \ln\delta + 3 \ln \delta +
 \frac{\pi^2}{3}-\frac{7}{4} + O(\epsilon) + O(\delta)\right)~.
\label{eq:sw3jets}
\end{equation}
This implies that for typical jet resolutions, a small fraction of hadronic
 events should consist of three-jets. They were found subsequently in
$e^+e^-$ annihilation at PETRA and we shall discuss them later quantitatively.

Another example of a jet measure which can be used with ease to
characterise jets is in terms of the invariant mass of the partons $y_{ij}$
emerging from a hard process, defined in Eq.~(\ref{eq:yij}).
 Requiring $y_{ij} > y_{\rm min} > 0$, one avoids both
infrared and collinear singularities in a perturbative calculation.
The first of such $y_{\rm min}$-dependent 2-jet cross-section was calculated
in \cite{Kramer}, yielding ($y_{\rm min}=y$)
\begin{equation}
\sigma_{2-jet}= \sigma_0\left[1 + C_F \frac{\alpha_s(Q^2)}{2\pi}
\left( -2 \ln^2 y -3 \ln y + 4y\ln y -1 + \frac{\pi^2}{3} +O(y)\right)
\right]~.
\end{equation} 
The $O(y)$ terms have been derived in~\cite{Kramer:1986sg}.

The two prescriptions just discussed have been used  in the experimental analysis
of data concerning jets. Thus, for example, the JADE
algorithm~\cite{Bartel:1986ua}, used mostly in the analysis of the $e^+e^-$ data at
PETRA and PEP, is based on the $y_{\rm min}$-prescription, which can be used to classify
also muli-jet events. This was subsequently replaced by the $k_t$-jet algorithm, as
discussed in the preceding section. The modified form of the
the cone-prescription is widely used in the analysis of jets in
hadroproduction processes.

\section{Gluon jets in Upsilon decays}
\noindent
The $\Upsilon$ meson first produced in proton-nucleus collisions at FERMILAB
\cite{Herb,Innes} and identified by the
$\Upsilon \to \mu^+\mu^-$decay was later observed as a narrow resonance with 
mass $m_{\Upsilon} = 9.46~GeV$ and width 
$\Gamma_{\Upsilon}=(40^{+13}_{-8})$ keV
in the process $e^+e^- \to \Upsilon \to {\rm hadrons}$
\cite {Berger76,Berger76-2,Darden-78,Bienlein-78,Andrews-80,Bohringer-80}.
This resonance is a $b\bar{b}$ bound state and has the quantum numbers
$J^{PC}=1^{--}$. As the $B\bar{B}$ threshold lies above  $m_\Upsilon$,
the $\Upsilon(9.46)$  state is
predicted to decay mainly into 3 gluons ($g$) in QCD, the massless colour-octet vector
 particles~\cite{Appelquist,Appelquist-2,Koller,Koller-2,Koller-3,deGrand-77,deRujula-78} in 
complete analogy with the decay of orthopositronium into 3 photons 
\cite{Ore}. While $\Upsilon(9.46) \to ggg$ is the dominant decay mode, 
with $3\%$ probability it can decay also into a photon 
and 2 gluons, $\Upsilon(9.46) \to \gamma gg$.
%Verification of the  $\Upsilon(9.46) \to ggg$ decay in terms of  
%three well-separated topological jets  was hindered due to mass of the $\Upsilon(9.46)$.
Average energies of the three partons
were measured  as $\langle E_1\rangle \simeq 4.1$ GeV for the most energetic of 
the three gluons, with the other two having energies  $\langle E_2\rangle \simeq 3.4$ GeV
and $\langle E_3\rangle \simeq 2.0$ GeV, respectively~\cite{Berger81}.
in approximate accord with the lowest order QCD matrix elements. However, only the fastest of
the three partons yielded a collimated jet of hadrons and its detailed phenomenological profile
was studied by PLUTO~\cite{Berger81,Stella:2010ne}. Phenomenological models, which included
the lowest order QCD matrix elements and the fragmentation of the partons (quarks and gluons),
were found to be in conformity with a number of inclusive measurements undertaken by
PLUTO.
\begin{figure}
\center{
\resizebox{0.75\columnwidth}{!}{
\includegraphics{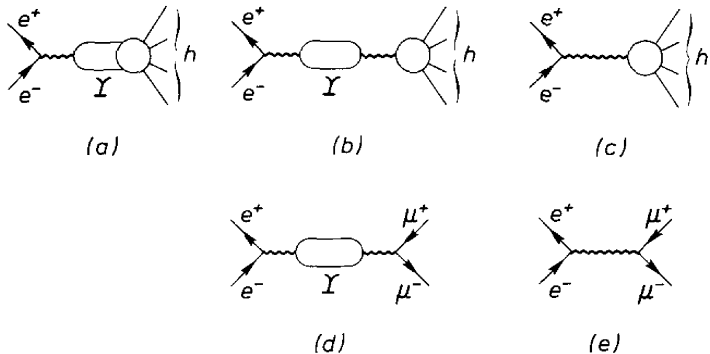}}}
\caption{Processes contributing to hadronic final states in the $\Upsilon(9.46)$ resonance
region: (a) from direct decays of the $\Upsilon$, (b) from the $\Upsilon$ vacuum polarization
 and (c) from the non-resonant continuum. The $\mu^+\mu^-$ final state can be produced
from the $\Upsilon$ vacuum polarization (d), and from the continuum (e).
 (from Ref.~~\cite{Berger81}).}
\label{fig:ZfP-C8-3}
\end{figure}

\begin{figure}
\center{
\resizebox{0.75\columnwidth}{!}{
\includegraphics{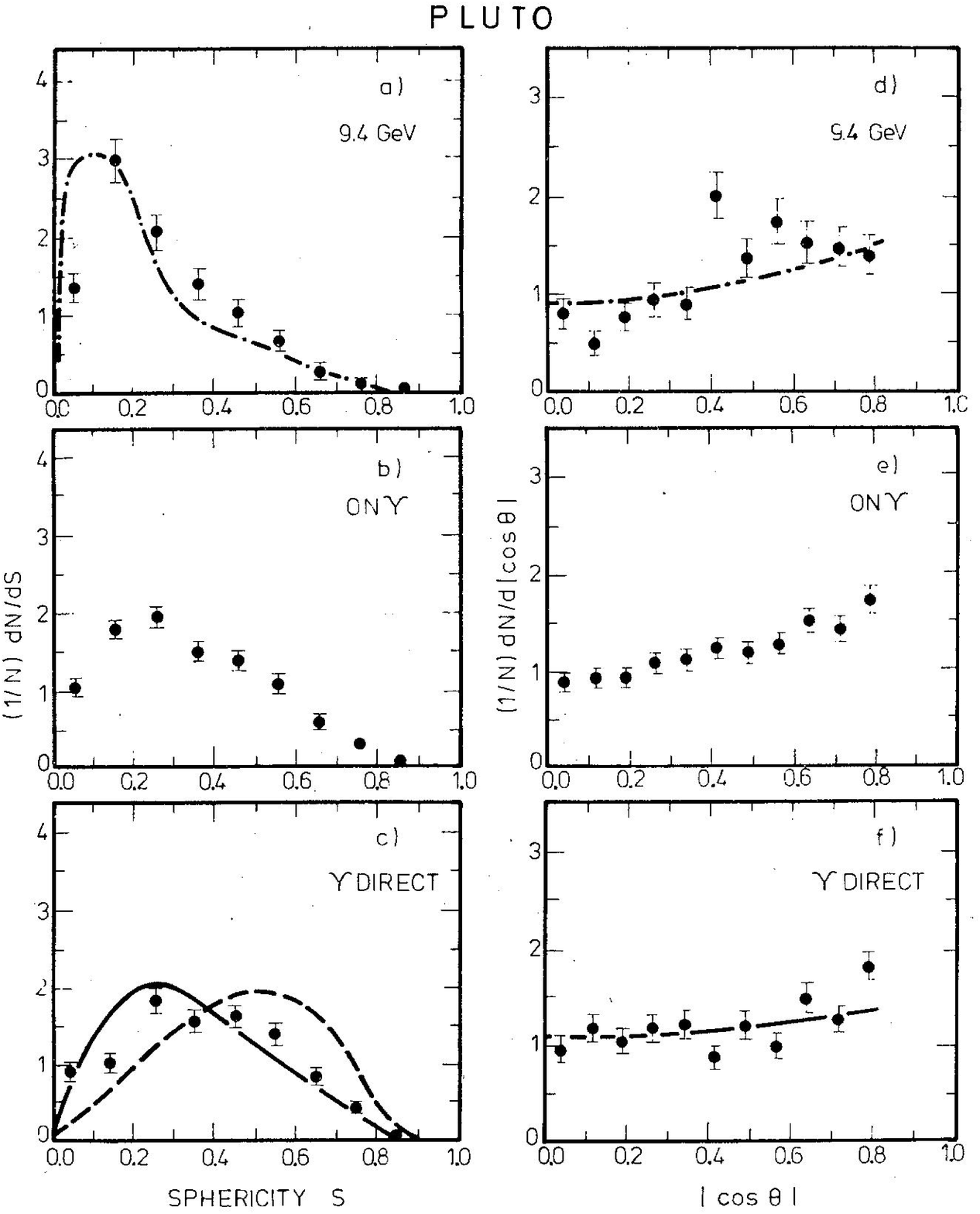}}}
\caption{Differential sphericity distributions and the sphericity angular distributions.
The dash-dotted line in (a) represents the two-jet model. The dashed and solid lines in
 (c) represent respectively phase-space and the three-gluon decay models. The dash-dotted
line in (d) is proportional to $1 + \cos^2\theta$ and the solid line in (f) to
$1+0.39 \cos^2\theta$ (from Ref.~~\cite{Berger79}).
        }
\label{fig:PL82B-1}
\end{figure}

The contributions to the multi-hadron events from the $\Upsilon$ mass region
originates from three sources~\cite{Berger81}, as shown in Fig.~\ref{fig:ZfP-C8-3}. The first row
in this figure shows the direct decay of the $\Upsilon$ (a), decay through the vacuum polarisation
 (one-photon decay) (b), and the non-resonating continuum (c).
Denoting the corresponding cross sections as $\sigma_{\rm on}$ (cross-section in the $\Upsilon(9.46)$
energy range), $\sigma^{\rm vp}$ (cross-section for the $\Upsilon$  vacuum polarisation), and
$\sigma^{\rm off}$ (cross-section for the non-resonant continuum), the cross-section for the
$\Upsilon(9.46)$-production with direct decay is
 $\sigma^{\rm dir}= \sigma^{\rm on}-\sigma^{\rm off} - \sigma^{\rm vp}$.
 Since for $\mu^+\mu^-$ final states,
a 'direct' production does not exist, the term $ \sigma^{\rm vp}$ can be obtained by scaling the 
$\mu$-pair production on and off-resonance to the hadron production level. This yields:
\begin{eqnarray}
\sigma^{\rm dir}=\sigma^{\rm on}-\sigma^{\rm off}-\sigma^{\rm vp}=
\sigma^{\rm on}-\sigma^{\rm off}
-\sigma^{\rm off}\frac{\sigma^{\rm on}_{\mu\mu}-\sigma^{\rm off}_{\mu\mu}}
{\sigma^{off}_{\mu\mu}}~.
\end{eqnarray}
 Using 
$(\sigma^{\rm on}_{\mu\mu} -\sigma^{\rm off}_{\mu\mu})/\sigma^{\rm off}_{\mu\mu}
=0.24\pm0.22$
\cite{Berger79} and the number of events in the two energy
regions, the $\Upsilon$ direct decay cross section is obtained. This is
evaluated as a function of sphericity $S$.
The results are shown in Fig.~\ref{fig:PL82B-1} a, b, c,
separately for the off-resonance data, the data at the $\Upsilon$ 
resonance and the subtracted distribution for the $\Upsilon$ direct decay. These
experimental results are compared to the two-jet model based on the Feynman-
Field model, already discussed (dash-dotted line in  Fig.~\ref{fig:PL82B-1} a)
 and the predictions of the
phase-space model in Fig.~\ref{fig:PL82B-1} c (dashed line) together with the
 three-gluon decay model (solid line).
The off-resonance data are well described by the two-jet model in agreement
with the earlier findings at SPEAR. The distribution for the 
direct decay is shifted to larger sphericity values and is best reproduced
by the three-gluon decay model.

The $\Upsilon$ meson is produced at rest. Therefore, the scaled momenta
$\vec{x_i}=2\vec{k_i}/m_{\Upsilon}$ obey the 
relations ($x_i=|\vec{x_i}|$): $\vec{x_1} + \vec{x_2} + \vec{x_3} = 0~,~x_1 + x_2 + x_3 = 2$.

 Another possibility to describe 
the configuration of the final state uses the angles between the gluons.
For massless gluons the relation between the $x_i$ and the
angles $\theta_i$ is:  $x_i =\frac{2\sin \theta_i}{\sum_{i} \sin \theta_i}$.

This relation allows one to characterise the final gluon configuration in the
corners of the Dalitz plot ($x_1=x_2=x_3=2/3$ is the
 "Mercedes-Star"-like configuration, $x_1=x_2=1,x_3=0$ is a two-jet configuration 
 with the third gluon perpendicular to the direction of the first two 
and $x_1=1,x_2=x_3=0.5$ is the configuration, where the fastest jet recoils 
against the two others with $x_2=x_3$). The momentum distribution of the gluon 
as calculated in leading-order (LO) QCD is~\cite{Koller,Koller-2,Koller-3,deGrand-77,deRujula-78}
\begin{eqnarray}
\frac{1}{\sigma}\frac{d^2\sigma}{dx_1dx_2}=\frac{6}{(\pi^2-9)x_1^2x_2^2x_3^2}
\left(x_1^2(1-x_1)^2+x_2^2(1-x_2)^2+x_3^2(1-x_3)^2 \right)~.
\label{eq:vecglue}
\end{eqnarray}
The above cross
section formula is the basis for Monte-Carlo model calculations mentioned
above. In these models the hard cross section for $\Upsilon \to 3g$ 
must be supplemented with the additional fragmentation of the 3 gluons into
hadrons.

To compare the decay  $\Upsilon \to 3g$ with vector gluons as 
follows from QCD, also models with scalar gluons  have 
been considered. The momentum distribution corresponding to scalar gluons
was derived in~\cite{Krasemann} leading to the result that they 
peak at the corners of the Dalitz plot and have zeros in the middle of
each boundary. In contrast, vector gluons populate nearly uniformly the Dalitz plot.
As the majority of the events have one gluon with very low
energy, a 2-jet structure is expected for scalar gluon theories ~\cite{WalshZerwas}.

\begin{figure}
\center{
\resizebox{0.75\columnwidth}{!}{
\includegraphics{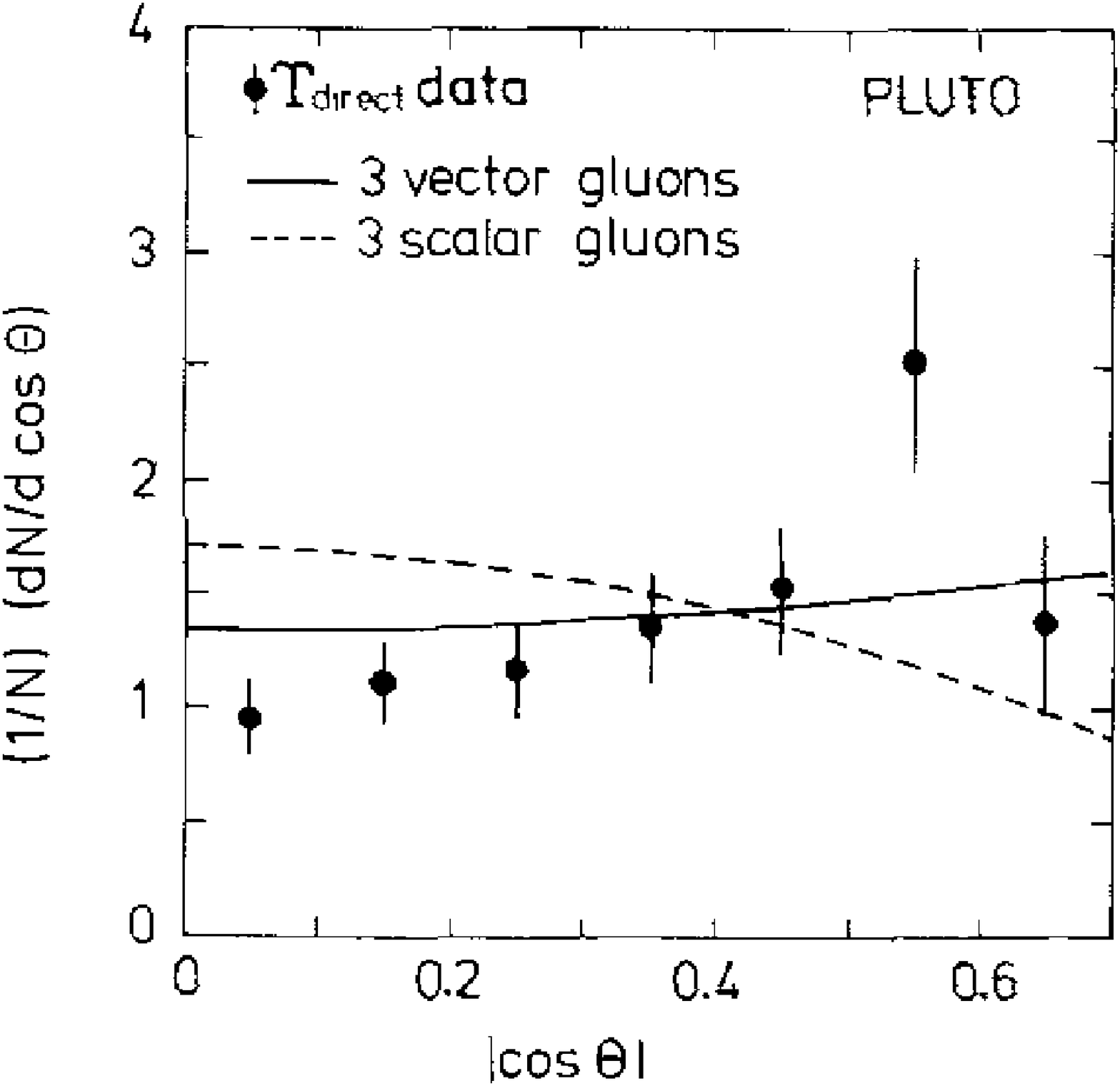}}}
\caption{Corrected experimental distribution in $|\cos \theta|$
from the decays $\Upsilon \to$ hadrons by the PLUTO collaboration, where $\theta$ is
the angle between the thrust axis and the beam axis. The curves are theoretical
distributions for $\Upsilon$-decay into vector (solid curve) and scalar (dashed curve)
gluons, respectively (from Ref.~~\cite{Berger81}).
        }
\label{fig:PLB88-3}
\end{figure}

The distribution of the gluon direction in space is essentially
determined by the QCD theory~\cite{Koller,Koller-2,Koller-3,deGrand-77,deRujula-78}.
 For vector gluons QCD predicts
\begin{equation}
 W(\cos\theta) \sim 1+0.39 \cos^2\theta~,
\end{equation}
where $\theta$ is the angle between the most energetic gluon and the
momentum of the incoming initial electron (see Fig.~\ref{fig:PL82B-1}). Scalar gluons
 would give rise to the angular distribution~\cite{Krasemann}
\begin{equation}
  W(\cos\theta) \sim 1-\cos^2\theta~.
\end{equation}
The PLUTO collaboration \cite{Berger81} has 
measured a number of observables to strengthen the 3-gluon interpretation of the 
hadronic $\Upsilon$ decay. 
The test of vector gluons versus scalar gluon has been done using the
angular distribution in $\cos \theta$, where $\theta$
is the angle between the thrust axis and the beam momentum. Theoretical distributions
predicted for $\Upsilon$ decay into vector and scalar gluons, respectively,
are  shown in Fig.~\ref{fig:PLB88-3} compared with the PLUTO measurements  
\cite{Berger81}. The data clearly prefer the vector gluon decay. 
Similar conclusions have been reached by the LENA collaboration on 
the basis of their measurements \cite{Niczyporuk}.

\begin{figure}
\center{
\resizebox{0.75\columnwidth}{!}{
\includegraphics{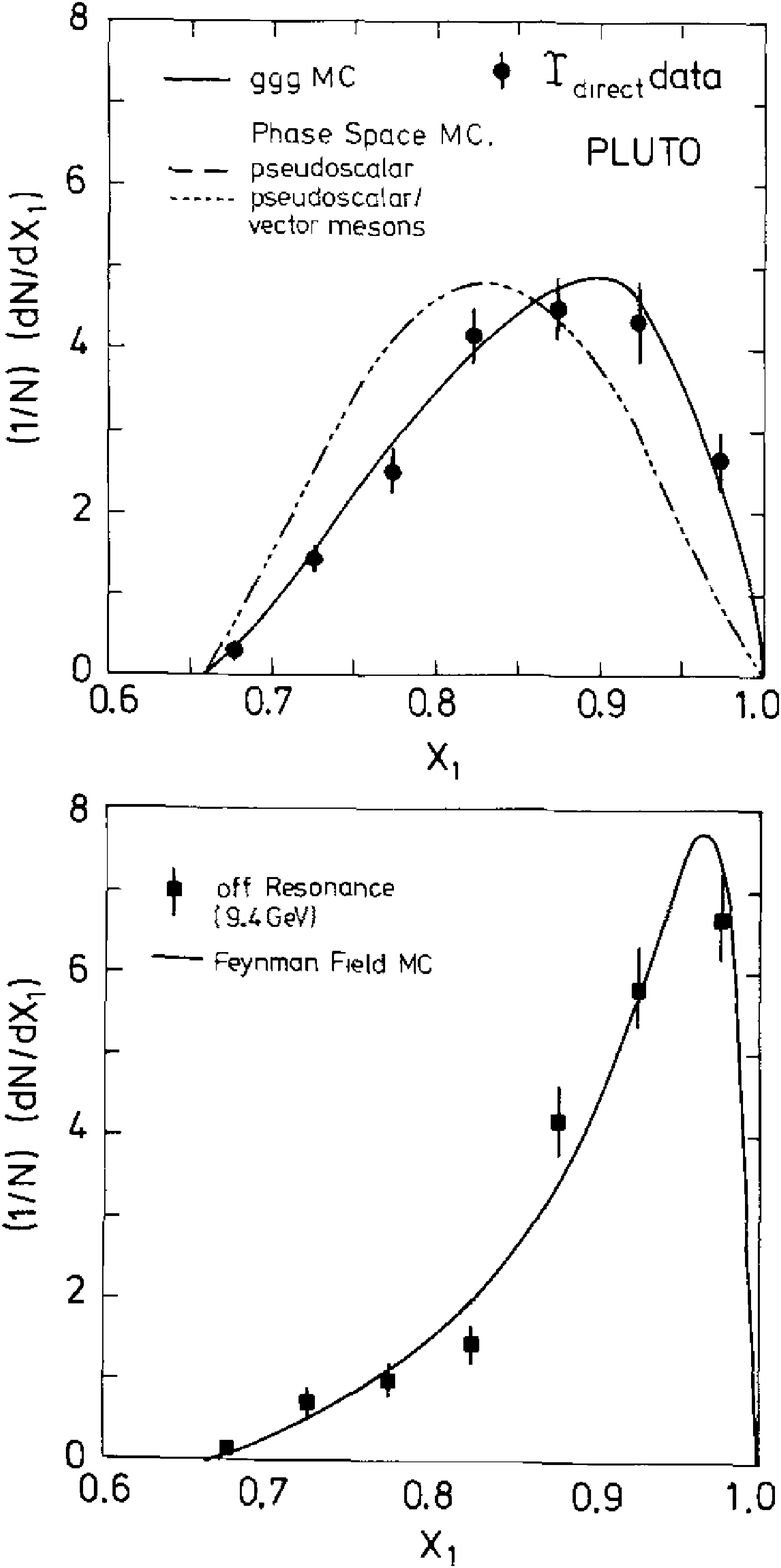}}}
\caption{Experimental distribution of the reconstructed reduced energy $x_1$ of the fastest triplicity jet for the $\Upsilon$-direct and off-resonance data taken by the PLUTO collaboration
 compared to Monte Carlo
 calculations for various models (from Ref.~\cite{Berger81}).
        }
\label{fig:ZPC8-11}
\end{figure}

Further comparisons with the 3-gluon model presented in \cite{Berger81} 
consist of topological tests with the jet variable thrust, defined
earlier, and the variable triplicity $T_3$ \cite{Brandt79}, an
extension of thrust to 3-jet configurations, where the final state
particles are grouped into 3 classes with total momentum $\vec{P}(C_l)$,
$l=1,2,3$. The values of triplicity vary between $T_3=1$ for a perfect 3-jet
event and $T_3=3\sqrt{3}/8$ for completely spherical events.  PLUTO data  
have been analysed also in terms of the fractional energies $x_i$ 
of the jets, where the jet axes are obtained from the triplicity analysis. If 
the three jets would be completely separated in space, the fractional energies
would be independent of the fragmentation of the gluons and would depend, in the lowest
order perturbation theory,
only on the QCD matrix element $W(x_1,x_2,x_3)$ given above (see Eq.~\ref{eq:vecglue}).
In Fig.~\ref{fig:ZPC8-11}
 the projections of the two-dimensional
histograms, spanned by the axes $x_3$ and $(x_1-x_2)/\sqrt{3}$, on the $x_1$ 
axis is shown (this is also the distribution of the most energetic triplicity jet). 
The prediction of the 3-gluon Monte Carlo model is compared to the data and impressive
agreement is obtained, whereas two versions of the phase-space Monte Carlo
model fail to do so. (See~\cite{Stella:2010ne}
for a recent reappraisal of the PLUTO experimental analysis).
 All this
information taken together demonstrates that the $\Upsilon$-direct decay data are
very well reproduced by the 3-gluon model while all the other models
disagree with $\Upsilon$-direct data.
These findings were further backed up later by the CLEO collaboration at the
CESR storage ring at Cornell \cite{Berkelman} and the ARGUS
collaboration at the DORIS ring at DESY \cite{Albrecht}.

%%%%%%%%%%%%%%%%%%%%%%%%%%%%%%%%%%%%%%%%%%%%%%%%%%%%%%%%%%%%%%%%
\section{Jets in QCD and at PETRA and PEP}
\label{sec:PETRA}
To put the contents of this chapter in historical perspective, we would like to introduce the
main detectors which played an important role in the development of jets and detailed tests
of QCD at PETRA and PEP. In doing this, however, we will be very brief and refer the interested
readers to the review by Gidal {\it et al.}~\cite{Gidal:1985cr}, which is a
compendium of the properties and performance characteristics of the major high energy physics 
detectors in that epoch, and the review by Lynch~\cite{Lynch:1987}. In alphabetical orders,
these detectors were CELLO, JADE, MARK-J,
PLUTO and TASSO (all located at the PETRA $e^+e^-$ rings at DESY, Hamburg), DELCO,
the High Resolution Spectrometer HRS, the Magnetic Calorimeter MAC, MARK II, MARK III, and the
Time Projection Chamber TPC (all located at the PEP $e^+e^-$ ring at SLAC, Stanford). As already
mentioned in the introduction, PETRA started data runs in 1978 with the maximum beam energy
of 23.6 GeV and PEP came a little later in 1980 having the maximum beam energy of 15 GeV. These
detectors were involved in measurements for almost a decade, ending their runs as LEP started
taking data at higher energies.

\subsection{Jet-like distributions from the weak decays of heavy quarks}
\label{sec:hqjets}
The process $e^+ e^- \to q \bar{q} g$ leads to $p_T$-broadening of the quark jets, leading 
eventually to three-jet topologies as the centre-of-mass energy increases. There is another source
of $p_T$-broadening in $e^+e^-$ annihilation due to the production of a heavy 
quark-antiquark pair  $e^+ e^- \to Q \bar{Q}$, and the subsequent weak decays of the
 heavy quarks/hadrons. For the
centre-of-mass energies available at the $e^+e^-$ colliders PEP and PETRA, the
heavy quarks whose production and decays had to be correctly taken into account were
the charm-anticharm ($c\bar{c}$) and bottom-antibottom ($b\bar{b}$) pairs. 
Sampling the theoretical predictions of the
top quark mass in the PEP and PETRA era, most guesses put
 it around 10 - 15 GeV~\cite{Georgi:1978fu,Fritzsch:1979zq}.
Hence, the production of a top-antitop pair was widely anticipated at these
 colliders~\cite{Brandelik:1979bv,Berger:1979wn},
and their characteristic jet topologies were worked out in the context of the
Cabibbo-Kobayashi-Maskawa (CKM)
6-quark model~\cite{Cabibbo:1963yz,Kobayashi:1973fv}. However, as subsequent developments showed,
there were no top quarks to be seen at PETRA and PEP (or at LEP).
 Thanks to the Fermilab-Tevatron~\cite{:2009ec}, the
top quark has a measured mass of about 173 GeV.

The event topology in $e^+e^-$ annihilation is
sensitive to the onset of $Q\bar{Q}$ threshold. The data in the center-of-mass energy in the
range $9.4 ~{\rm GeV} \leq \sqrt{s} \leq 17.0 ~{\rm GeV}$ was analysed~\cite{Ali:1979upa}
in terms of the measures of jettiness,
$\langle S \rangle $ and $\langle 1-T\rangle$, which showed a clear step as the $B\bar{B}$ threshold is crossed.
The data taken by the PLUTO collaboration~\cite{Berger79} at 9.4 GeV at DORIS, and at 13.0
 and 17.0
GeV at PETRA by the PLUTO~\cite{Berger:1979bp} and TASSO~\cite{Brandelik:1979cj}
 collaborations were well described by a theoretical
 Monte Carlo~\cite{Ali:1978uy} taking into account the production processes $e^+ e^- \to c \bar{c}$ and
 $e^+ e^- \to b \bar{b}$, with the subsequent non-leptonic decays $c \to s u \bar{d}$ and
$b \to c \bar{u} d$, following the CKM theory of weak decays. 
 
The effects of heavy quark production and decays above their respective thresholds on the
jet distributions are taken into account by a three-step modifications of the light
quark pair production and subsequent fragmentation~\cite{Ali:1978uy}. The heavy quark
 mass enters the
 Lorentz-invariant density matrix for $e^+e^- \to Q \bar{Q}$ (here $Q^2=s$):
\begin{equation}
\vert M\vert^2= \frac{\alpha^2}{Q^4} [ (\ell_+p_1)(\ell_-p_2) + (\ell_+p_2)(\ell_-p_1)
 + m_Q^2 Q^2/2]~,
\end{equation} 
where $\ell_-(\ell_+)$ is the electron (positron) momentum and $p_1(p_2)$ is the momentum
of $Q(\bar{Q})$, and the quark mass is denoted by $m_Q$.  In the second step, the heavy
quark (antiquark) fragments into a heavy hadron and a number of light hadrons, determined by
a function $f_Q^H(z)$, which peaks increasingly near $z \to 1$, as $m_Q$ increases.
In the third step, the heavy hadrons decay, dominantly non-leptonically, modelled on the
quark transitions
$Q(p) \to q_1(q_1) + \bar{q}_2(q_2) + q_3(q_3)$~\cite{Ali:1978kn}.  These effects are
 important quantitatively for jet physics for the lower PETRA energies 
(typically $\leq 30$ GeV). 

%%%%%%%%%%%%%%%%%%%%%%%%%%%%%%%%%%%%%%%%%%%%%%%%%%%%%%%%%%%%%%%%%%%%%
\subsection{3-jet events and cross sections at PETRA}
\label{sec:gluon-jet}
%%%%%
As discussed earlier, the characteristic feature of the process $e^+ e^- \to q\bar{q}$ with the
subsequent fragmentation of the quarks and the antiquarks into a jet of hadrons
is that it leads to a two-jet configuration. In QCD,  the diagrams shown
in Fig.~\ref{fig:qqg} modify this picture. These corrections being proportional to
$\alpha_s(Q^2)$, the QCD coupling constant at the scale $Q^2$, are small.
 However, the process $e^+ e^- \to q \bar{q} g$
may reflect itself in a structure of the final states that topologically
is different from the dominant process $e^+ e^- \to q \bar{q}$. The radiated gluon
provides a new (non-local) mechanism for producing large-$p_T$ hadrons, which,
unlike the $p_T$ of the hadrons generated in the process $e^+ e^- \to q \bar{q}$,
is expected to increase with the $e^+ e^-$ centre-of-mass energy. Thus, broadening
of the transverse momentum of the hadrons with increasing centre-of-mass energy
is a consequence of gluon bremsstrahlung. It was argued 
in~\cite{THgluon}, that a corollary of this phenomenon is that
a third  jet should exist in the direction of the large $p_T$ particle. In particular,
if there is enough phase space available, i.e. for large enough $Q$,
a  three-jet topology in the shape of ``Y'' (Mercedz-Benz symbol) should
emerge, clearly distinguishable from the (dominant) oblate cigar topology
corresponding to two-jet events.

The calculation for the process $e^+ e^- \to q(p_1) + \bar{q}(p_2) + g(p_3)$,
shown in the upper two Feynman diagrams in Fig.~\ref{fig:qqg} leads to the
following (Dalitz) distribution:
\begin{equation}
\frac{1}{\sigma_0} \frac{d^2\sigma}{dx_1 dx_2}=\frac{\alpha_s(Q^2)}{2 \pi} C_F 
\frac{x_1^2 + x_2^2}{(1-x_1)(1-x_2)}~,
\label{eq:egr77}
\end{equation}
where $Q^2=4E^2$, $x_i=E_i/E=2E_i/Q$, and $E_i$ are the energies of the quark,
antiquark, and gluon, with $x_1+x_2+x_3=2$, and $\sigma_0$ is
the lowest order $e^+e^- \to {\rm hadron}$ cross section given in Eq.~(\ref{eq:sigma-0}). 
The differential cross section in Eq.~(\ref{eq:egr77}) diverges near the end-points
$x_{1,2} \to 1$, and indeed has infra-red and collinear divergences. We shall discuss finite
2-jet and 3-jet cross sections in the next subsection, but for the present discussion
these divergences can be removed by a reasonable cut-off procedure, such as a
 cut-off $Q_0^2$ on the invariant masses $s_{13}=Q^2(1-x_2)$ and $s_{23}=Q^2(1-x_1)$,
yielding a finite lowest order three-jet fraction.
\subsection{Experimental evidence of three-jet events at PETRA}
\begin{figure}
\center{
\resizebox{0.95\columnwidth}{!}{
\includegraphics{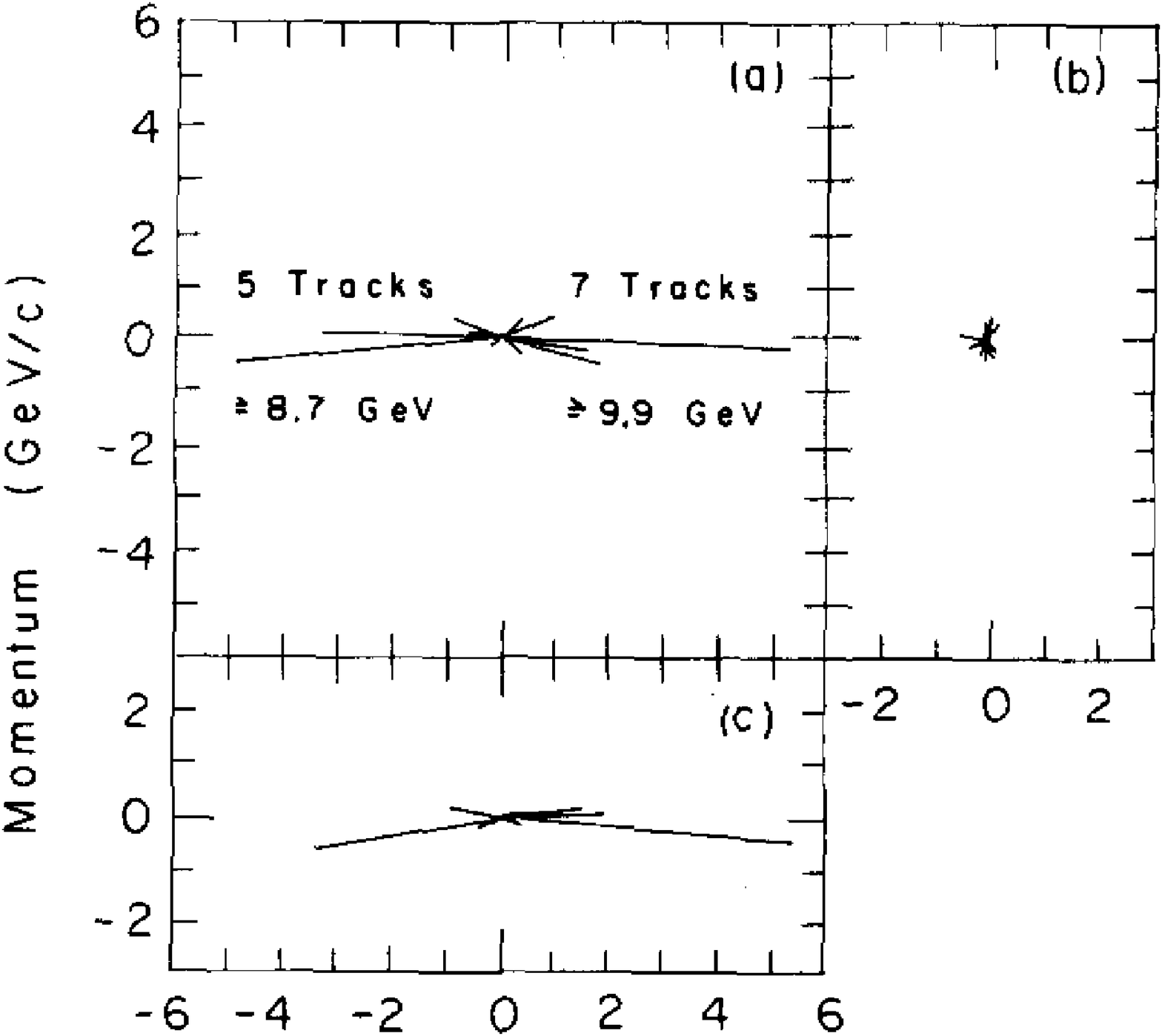}
\includegraphics{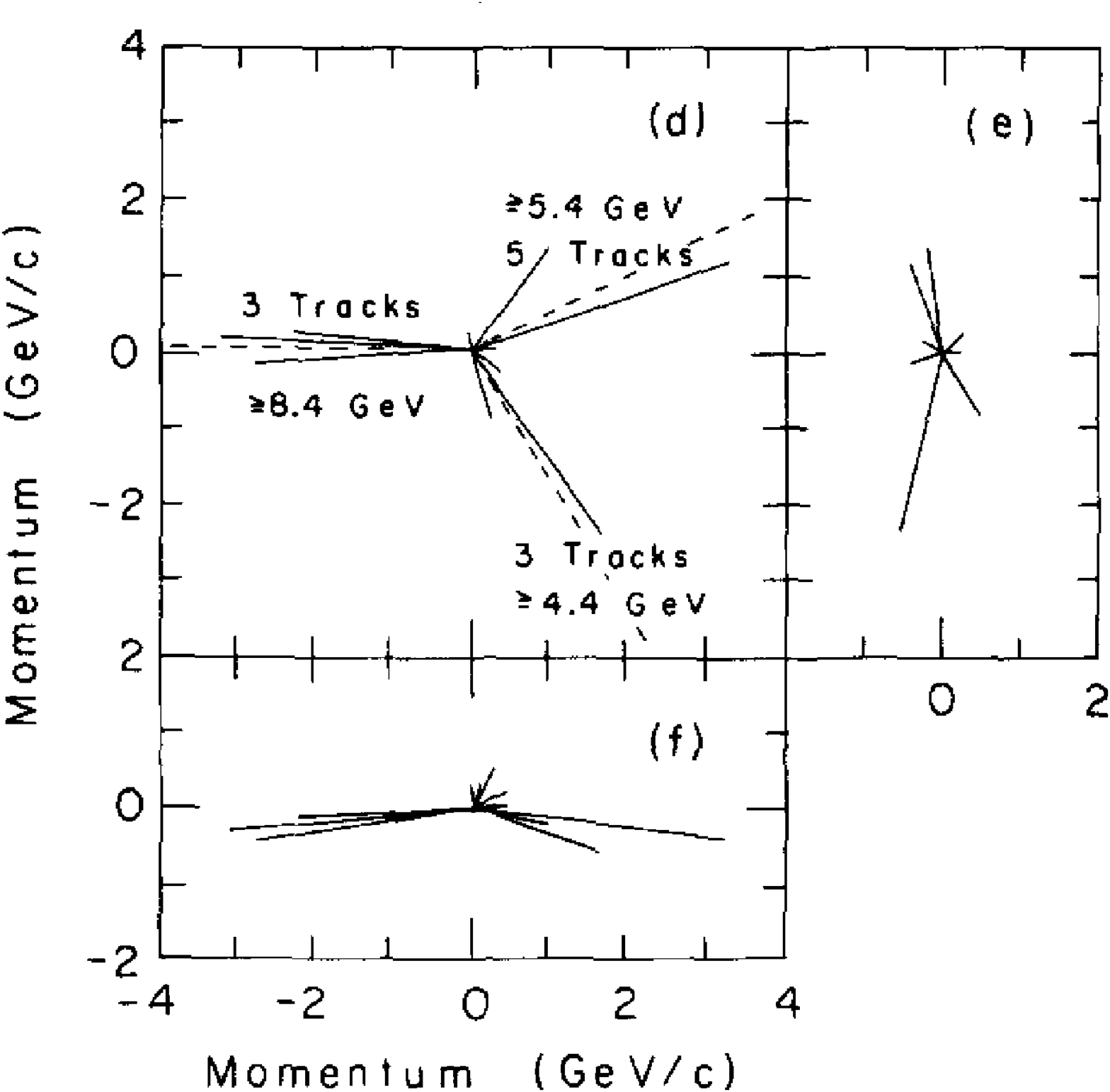}}}
\caption{Momentum space representation of a two-jet event (a) - (c) and a three-jet event
 (d) - (f) 
in each of the three projections, (a),(d) $\hat{n}_2$ - $\hat{n}_3$ plane; (b),(e)
$\hat{n}_1$ - $\hat{n}_2$ plane; (c), (f) $\hat{n}_1$ - $\hat{n}_3$ plane.
Here  $\hat{n}_i$ are the three axes of the sphericity tensor. (From TASSO~\cite{EXP4gluon-TASSO}). 
       }
\label{fig:TASSO-3j}
\end{figure}

While valuable tests of QCD were performed in studies of the $\Upsilon$ decays, based on
the underlying mechanism $\Upsilon \to 3 g$ and the subsequent fragmentation of the
gluons, three-jet events were first observed in
$e^+e^-$ annihilation at PETRA in 1979 by the four experimental
 collaborations: TASSO~\cite{EXP4gluon-TASSO}, MARK-J~\cite{EXP4gluon-MARK-J},
 PLUTO~\cite{EXP4gluon-PLUTO} and JADE~\cite{EXP4gluon-JADE}.  The process
 $e^+e^- \to q \bar{q} g$ leads to planar events, the search of three-jet events in these
experiments was concentrated mainly in demonstrating the excess of planar events
compared to the estimates based on the 2-jet final states around $\sqrt{s}=27$ GeV, where
most of the early experiments at PETRA were carried out.
 Such quantitative analyses were
backed up by topologically well separated 3-jet events. Fig.~\ref{fig:TASSO-3j} shows
momentum-space representation of a representative two-jet and three-jet event measured
by the TASSO collaboration, analysed on the basis of sphericity tensor and
jettiness~\cite{WuZo}.

\begin{figure}
\center{
\resizebox{1.0\columnwidth}{!}{
\includegraphics{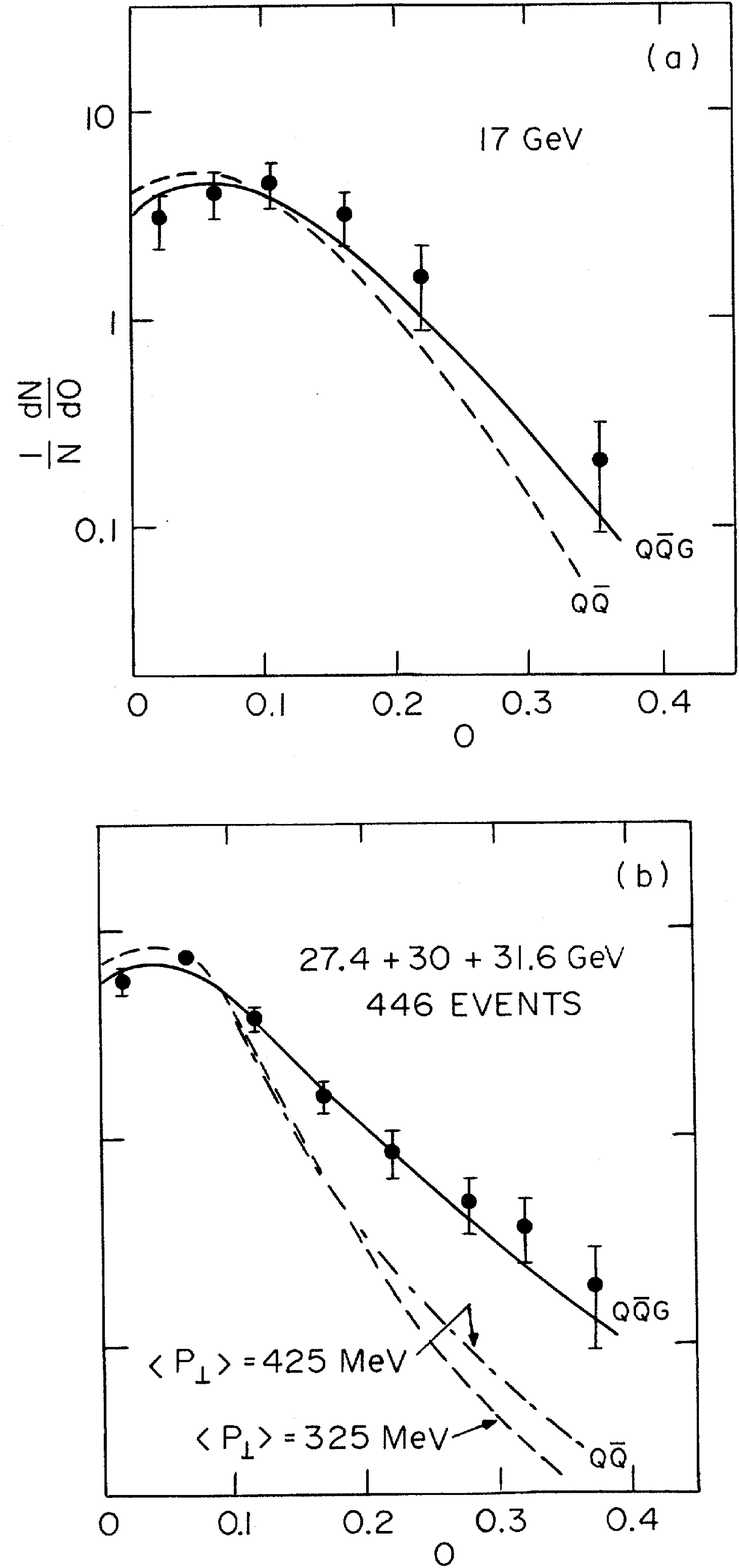} $\;\;\;\;\;\;\;\;\;\;\;$ \includegraphics{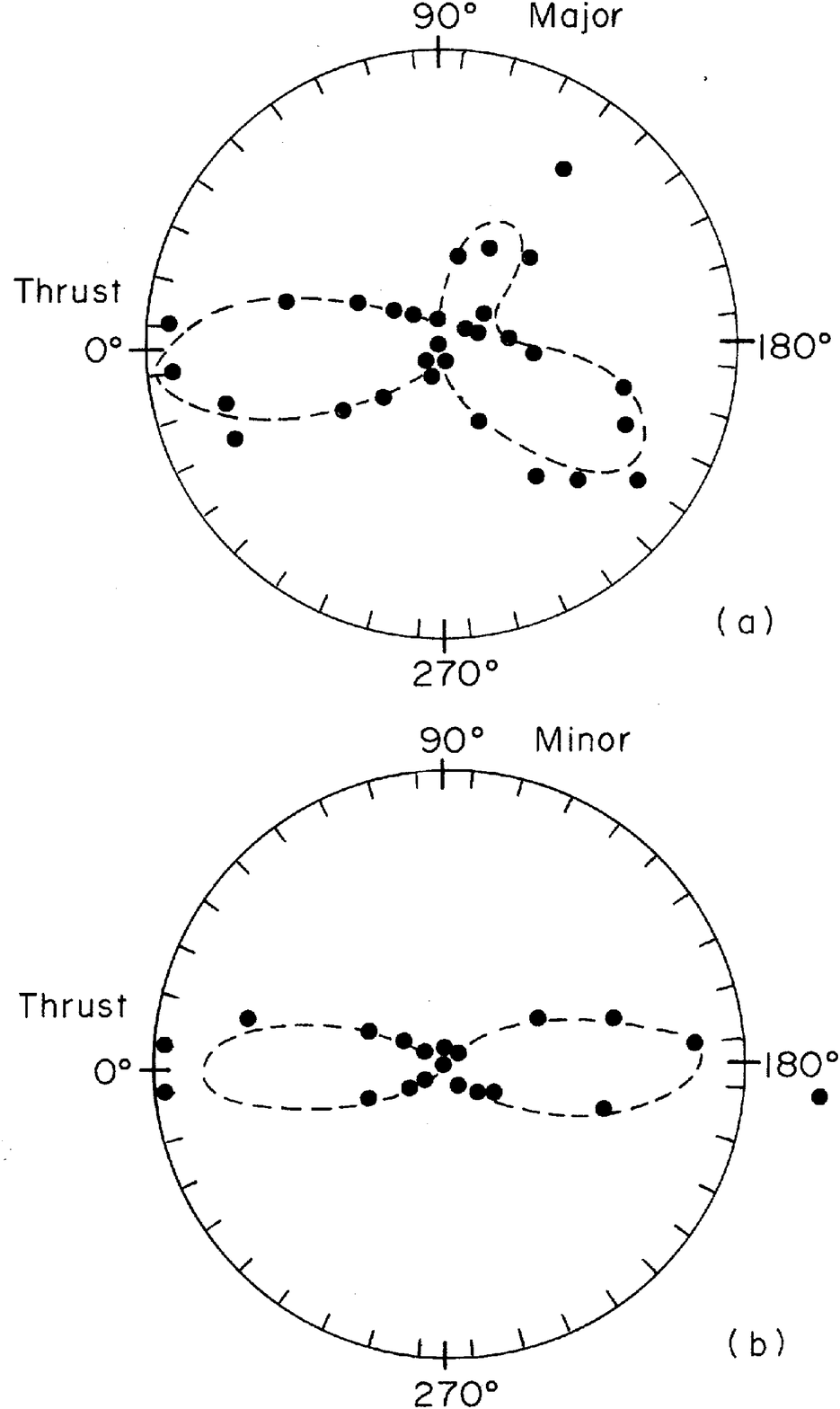}}}
\caption{Left-hand frames: Normalised Oblateness distribution at $\sqrt{s}=17$ GeV (a), and
at $\sqrt{s}=27.4$ - $31.6$ GeV (b). The solid curves are the predictions based on a
$Q\bar{Q}g$ model and the dashed curves are based on the $Q\bar{Q}$ model with
$\langle p_T\rangle=325$ MeV (denoted as $q\bar{q}g$ and $q\bar{q}$, respectively, in this review).
 The dashed-dotted curve in (b) is the $Q\bar{Q}$ model prediction with
 $\langle p_T \rangle = 425$ MeV ($Q=u, d, s, c, b$).
Right-hand frames: Energy flow in the event plane defined by (a) the thrust and the major axes,
and (b) by the thrust and the minor axes with the events satisfying the cuts
thrust $<0.8$ and oblateness $>0.1$ at $\sqrt{s}=27.4$ - $31.6$ GeV.
The energy value is proportional to the radial distances.; dots are the experimental measurements
 (From MARK-J~\cite{EXP4gluon-MARK-J}). 
        }
\label{fig:MARK-J-3j}  
\end{figure}

 The MARK-J measurement of the distributions in oblateness (defined below)
at $\sqrt{s}=17$ GeV and at higher energies (27.4 + 30 + 31) GeV
are shown in Fig.~\ref{fig:MARK-J-3j}(a) and  Fig.~\ref{fig:MARK-J-3j} (b),
respectively. For this measurement, the coordinate system is
defined by the thrust axis $\vec{e_1}$, the major axis $\vec{e_2}$, which is in the plane
 perpendicular to $\vec{e_1}$, and is in the direction along which the projected energy in
that plane is maximised, and the minor axis, $\vec{e_3}$, which is orthogonal to both
 $\vec{e_1}$ and $\vec{e_2}$. Oblateness is then defined as
\begin{equation}
{\cal O}=F_{\rm major} - F_{\rm minor}~,
\end{equation}
where $F_{\rm major}=\sum_{i}\vec{p_i}.\vec{e_2}/\sum_i|p_i|$ and 
$F_{\rm minor}=\sum_{i}\vec{p_i}.\vec{e_3}/\sum_i|p_i|$.
The two frames on
the r.h.s. of this figures show the energy flow in the event plane defined by the thrust and
major axes (upper frame) and by the thrust and the minor axes (lower frame).
 These measurements were compared with the
$q\bar{q}$ (two-jet) and $q\bar{q}g$ (three-jet) Monte Carlo models~\cite{PHENgluon1,PHENgluon2},
 and clearly favoured
the  $q\bar{q}g$ description, in a statistically significant way. 

\begin{figure}
\center{
\resizebox{0.75\columnwidth}{!}{
\includegraphics{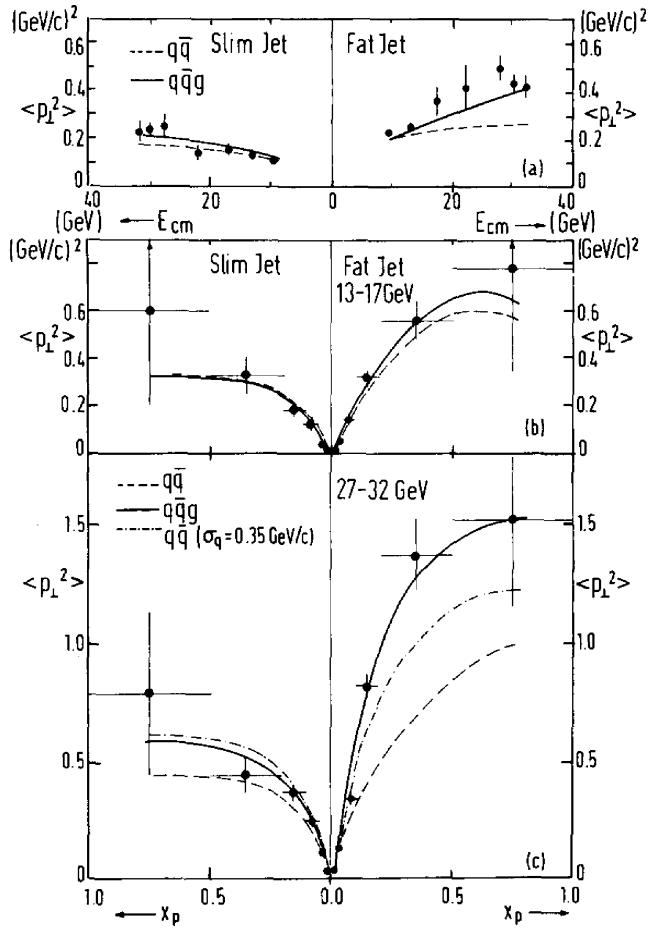}}}
\caption{Average observed $\langle p_T^2\rangle$ of charged particles in the slim and
fat jets as
 a function of the c.m. energy (a). Sea-gull plots ($\langle p_T^2\rangle$ as a
function of $x_p=p/p_{\rm beam}$, where $p=\sqrt{p_\parallel^2 +p_T^2}$)
 for slim and fat jets in two separate energy range (b), (c). 
The solid and dashed curves are
 $q\bar{q}g$ and $q\bar{q}$ predictions, respectively. In (c), the dashed curve
corresponds to  $\sigma_q=0.247$ GeV (default value) and the dash-dotted curve to
 $\sigma_q=0.35$ GeV.
 (From PLUTO~\cite{EXP4gluon-PLUTO}). 
       }
\label{fig:PLUTO-3j}
\end{figure}

PLUTO studied the averages  of the momenta of the charged particles
  $\langle p_\parallel\rangle$, where $p_\parallel$ is the longitudinal momentum,
$\langle p_\perp\rangle$ and $\langle p_\perp^2\rangle$, measured relative to the thrust
axis of the event as a function of the c.m. energy. Their analysis showed that the
quantities $\langle p_\parallel\rangle$ and $\langle p_\perp\rangle$ are not very
discriminative between the $q\bar{q}$ and $q\bar{q}g$, but the energy dependence of
$\langle p_\perp^2\rangle$ is better described if gluon bremsstrahlung is included.
To study this effect in more detail, they distinguished for every event the two jets which
are separated by a plane perpendicular to the thrust axis. The jet with the lower (higher)
 average $\langle p_\perp\rangle$is called the slim (fat) jet. Fig.~\ref{fig:PLUTO-3j}(a)
shows $\langle p_\perp^2\rangle$ of the charged particles as a function of the c.m. energy, 
where the average is taken over the charged hadrons in all slim (fat) jets. For the slim jet
 the $q\bar{q}$ and $q\bar{q}g$ predictions from the Monte Carlo~\cite{PHENgluon1} are very similar
 and the data are in agreement
 with both. For the fat jet, however, the data clearly favour $q\bar{q}g$, and $q\bar{q}$ is 
ruled out. Fig.~\ref{fig:PLUTO-3j}(b) and  \ref{fig:PLUTO-3j}(c) show the so-called
``sea-gull plot'', obtained by plotting the variable $x_p=p/p_{\rm beam}$ and   
$\langle p_\perp^2\rangle$, at lower c.m. energies 13 and 17 GeV and at higher
energies 27.6, 30 and 31.6 GeV, respectively. At the lower energy
 (Fig.~\ref{fig:PLUTO-3j}(b)), there is very little difference between $q\bar{q}$ and
 $q\bar{q}g$ predictions. For the higher energies(Fig.~\ref{fig:PLUTO-3j}(c)), $q\bar{q}g$
predicts a genuine one-sided jet broadening caused by the gluon jet; the effect is quite
dramatic, especially at high $x_p$. TASSO collaboration~\cite{EXP4gluon-TASSO} has done a very
similar analysis. 

\begin{figure}
\center{
\resizebox{0.75\columnwidth}{!}{
\includegraphics{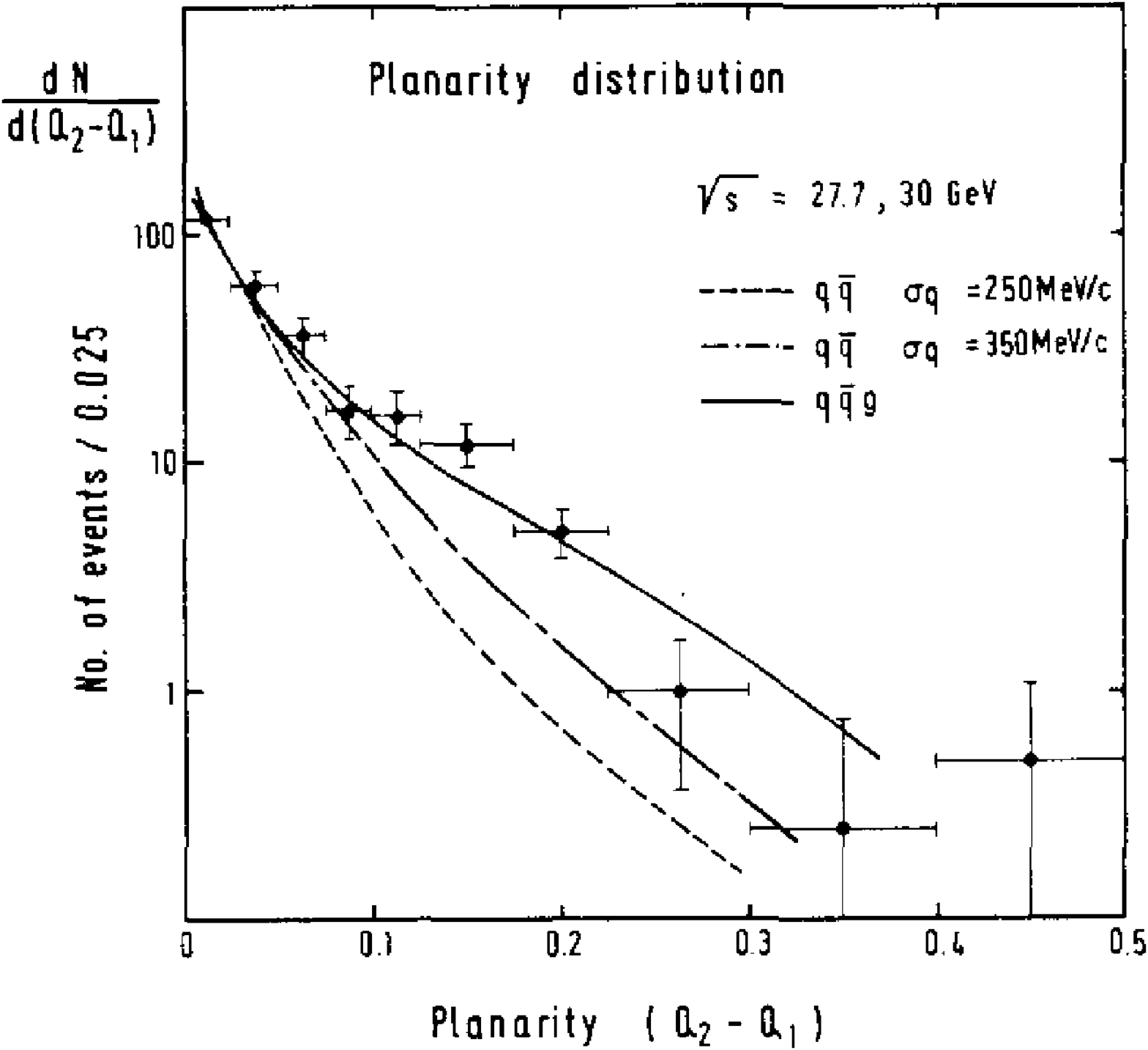}}}
\caption{The planarity distribution compared with the model predictions 
for $e^+e^- \to q\bar{q}$ and  $e^+e^- \to q\bar{q}g$ at $\sqrt{s}=27.7, 30$ GeV.
 (From
JADE~\cite{EXP4gluon-JADE}). 
       }
\label{fig:JADE-3j}
\end{figure}

The JADE analysis is based on the normalised sphericity tensor $S_{\alpha\beta}$ (defined in
Eq.~(\ref{eq:Salfabeta}))
and the resulting eigenvalues $Q_1, Q_2, Q_3$ obtained by diagonalising this tensor on an
event by event basis. The variables which play a central role 
in this analysis are the sphericity $=3/2(Q_1+Q_2)$ 
and planarity $=(Q_2-Q_1)$. Fig.~\ref{fig:JADE-3j} shows the planarity 
distribution $dN/d(Q_2-Q_1)$ measured by JADE at $\sqrt{s}=27.7$ and 30 GeV. Their data
are compared with a $q\bar{q}$ model, with $\sigma_q=250$ MeV and 350 MeV, both of 
which fail to describe the data. The $q\bar{q}g$ model describes the data well.

The results reviewed in this section were the first measurements
through which the effect of a third (gluon) jet was convincingly established in
$e^+e^-$ annihilation. This is an important milestone in the confirmation of QCD
in which jet physics played a central role. From a theoretical point of view,
observation of the gluon jet was inevitable. Like many other discoveries in particle
physics, this discovery needed high energy $e^+ e^-$ beams, particle detectors well equipped to
measure the characteristics of the hadrons, and data analysis techniques.
This was the work of dedicated teams of machine builders and experimental physicists who should be
credited with the discovery.
For the interested readers we refer to individual accounts leading to the discovery
of the gluon 
jets~\cite{Stella:2010ne,Schopper:1980jd,Wu:1992vc,Branson:1994eu,Soding:1996zk,Ellis:2009zz,Soding:2010zz}, but
stress that this list of references is by no means exhaustive.

%%%%%%%%%%%%%%%%%%%%%%%%%%%%%%%%%%%%%%%%%%%%%%%%%%%%%%%%%%%%%%%%%
\subsection{Quantitative studies of QCD at PETRA and  PEP}
\label{sec:PEP-PETRAJets}
Subsequent to the discovery of the gluon jet, the four PETRA collaborations, JADE,
 Mark-J, PLUTO
(later replaced by CELLO) and TASSO collaborations made many more measurements
in $e^+e^-$ annihilation to hadrons,
in which further evidence for the gluon jet was presented,
 These prompted quantitative
studies of QCD for inclusive jet-observables, like thrust and the Fox-Wolfram shape 
variable~\cite{Fox:1978vu}
 etc., and for jet topology, like the 2-jet and 3-jet rates etc.
Also the gluon spin was determined, following a suggestion in~\cite{Ellis:1978wp}.
One important issue was
the universality of the quark-gluon coupling $\alpha_s(Q^2)$, i.e. to check
whether the same value is obtained independent of the observables and the
measurements of  $\alpha_s(Q^2)$ at various values of $s=Q^2$ were consistent
with the evolution anticipated by the renormalisation group. These attempts
to obtain $\alpha_s(Q^2)$ required the calculation of next-to-leading order corrections
to the topological jet-rates and inclusive jet-observables,
 and also required a better understanding of
the non-perturbative models used to interpret the experimental data. These theoretical
and  phenomenological studies often took the form of detailed Monte Carlo programs
without which no realistic comparison of theory and experiment was possible. In fact,
since the days of experimentation at PETRA and PEP, Monte Carlo based theoretical
frameworks have become indispensable for the quantitative analysis of data, as
witnessed later at LEP, HERA and the Tevatron, and now at the LHC.

 In $O(\alpha_s(Q^2)$, 2-jet cross sections defined by a jet-resolution criterion,
such as the Sterman-Weinberg jet-cones or the jet invariant mass, receive contributions
from the virtual corrections to the process $e^+e^- \to q \bar{q}$, and soft or collinear
configurations from the processes $e^+ e^- \to q \bar{q} g$.
 In $O(\alpha_s^2(Q^2))$, the 3-jet cross sections receive
contribution from the virtual corrections to $e^+ e^- \to q \bar{q} g$ and soft and
collinear configurations from the 4-parton processes $e^+ e^- \to q \bar{q} gg$
and $e^+ e^- \to q \bar{q}  q \bar{q}$. The hard and non-collinear
configurations in the 4-parton processes
   give rise to 4-jet cross sections, with the leading
contribution arising  in $O(\alpha_s^2(Q^2))$, whose rates were calculated
 in~\cite{Ali:1979rz,Ali:1979wj} including the quark mass effects. They were important to
check the non-abelian character of QCD, as discussed later.
The first complete next-to-leading order correction to event shapes up to order
$\alpha_s^2$ were undertaken by Ellis {\it et al.}~\cite{Ellis:1980nc,Ellis:1980wv}. They presented
their results in terms of the tensor
\begin{equation}
\theta^{ij}= \sum_a (p_a^i p_a^j)/|p_a|) (\sum_a |p_a|)^{-1}~,
\label{eq:ERT-C}
\end{equation}
where $p_a^i$ are the components of the centre-of-mass three-momentum of hadrons $a$, and the
sum runs over all hadrons. The eigenvalues of $\theta$ are determined by the characteristic
equation
\begin{equation}
\lambda^3 - \lambda^2 +\frac{1}{3} C\lambda -D/27=0,~~~0 \leq C, D \leq 1~.
\end{equation}
The quantities $C$ (also called the Fox-Wolfram shape variable~\cite{Fox:1978vu})
and $D$ are symmetric functions of the eigenvalues, defined as
 \begin{equation}
C\equiv 3(\lambda_1\lambda_2 + \lambda_2\lambda_3 + \lambda_3 \lambda_1),
 ~~D=27\lambda_1\lambda_2\lambda_3~,
\end{equation}
Integrating $\frac{1}{\sigma}\frac{d\sigma}{dC}$ in the range
$\frac{1}{2} < C < 1$ yields
\begin{equation}
\frac{1}{\sigma_0} \int_{0.5}^{1.0} dC \frac{d \sigma}{dC}=C_1 \frac{\alpha_s(Q^2)}{\pi}
(1 + C_2 \frac{\alpha_s(Q^2)}{\pi})~.
\end{equation}
Numerically,  $C_1=2.8$ and $C_2=18.2 \pm 0.7$ for five quark flavours~\cite{Ellis:1980nc,Ellis:1980wv}. Thus, 
large corrections are obtained for the Fox-Wolfram shape variable, $C$. 

\begin{figure}
\center{
\resizebox{0.75\columnwidth}{!}{
\includegraphics{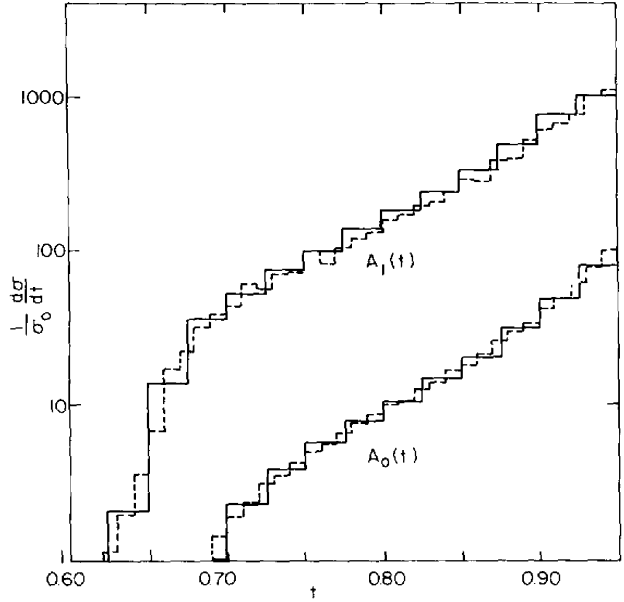}}}
\caption{The scalar functions $A_0(t)$ and $A_1(t)$ for the Thrust distribution as
 defined in Eq.~\ref{eq:thrust2}. Solid histograms (based on Ellis, Ross and
Terrano~\cite{Ellis:1980nc}), dashed histograms (Vermaseren {\it et al.}~\cite{Vermaseren:1980qz}).
(From~\cite{Ellis:1981re}). 
       }
\label{fig:thrust-2}
\end{figure}

Another, and experimentally widely studied, example of an inclusive distribution
is thrust. In $O(\alpha_s^2)$, this was first calculated by Vermaseren {\it et al.}~\cite{Vermaseren:1980qz}, 
and verified subsequently by Ellis and
Ross~\cite{Ellis:1981re} and by others~\cite{Kunszt:1981rp,Clavelli:1981yh,Ali:1981tm}, using the 
earlier work reported in~\cite{Ellis:1980nc,Ellis:1980wv},
In  next-to-leading order NLO in $\alpha_s$, thrust-distribution  in $e^+e^- \to {\rm hadrons}$ is
given by the following expression
\begin{equation}
\frac{d\sigma}{dT}= A_0(T) \frac{\alpha_s(Q^2)}{\pi} + A_1(T)(\frac{\alpha_s(Q^2)}
{\pi})^2~,
\label{eq:thrust2}
\end{equation} 
where the functions $A_0(T)$ and $A_1(T)$ are shown in Fig.~\ref{fig:thrust-2}
(note that the variable $t$ used in these plots is the same as $T$ used in the text).
The shapes of $A_0(T)$ and $A_1(T)$ are rather similar, but the $O(\alpha_s^2(Q^2)$
corrections to the thrust-distributions are also numerically large. Integrating the
distribution in Eq.~(\ref{eq:thrust2}) up to $T=0.85$ yields
\begin{equation}
\frac{1}{\sigma_0} \int_{0.5}^{0.85} dT \frac{d\sigma}{dT}=K_1 \frac{\alpha_s(Q^2)}
{\pi}(1+ K_2 \frac{\alpha_s(Q^2)}{\pi})~.
\end{equation}
Numerically, $K_1=1.156$, $K_2=17.6 \pm 0.3$ for five quark flavours,
which for $\alpha_s(Q^2)=0.13$ at $\sqrt{s}=35$ GeV yields a correction of
about 70\%.
Theoretical calculations from~\cite{Ellis:1980nc} were later implemented in the independent jet
Monte Carlo~\cite{PHENgluon2} and used to determine $\alpha_s(Q^2)$ from the inclusive
measurements. The first such determination using the 
thrust and oblateness distributions, measured by the TASSO~\cite{Brandelik:1980zw} and
MARK-J~\cite{Newman:1980ck} collaborations, respectively, 
yielded~\cite{Ali:1981tm} $\alpha_s(Q=35 GeV)= 0.128 \pm 0.013$ from the TASSO data and
$\alpha_s(Q=35 GeV)= 0.120 \pm 0.010$ from the MARK-J data.

 Subsequently, an enormous effort
 has gone into estimating  the effects from the jet resolutions, choice of jet variables, 
and fragmentation models. Also, the statistical significance of the
data from the experiments at PETRA and PEP increased enormously over the time.
 An observable studied
intensively in theory and experiments at PETRA is the energy-energy correlation (EEC)
and its asymmetry (AEEC).  EEC is a measure of the energy flow involving two
calorimeters subtending solid angles $\Omega$ and $\Omega^\prime$ with respect to the 
incoming $e^+e^-$ axis. Keeping the orientation between the two calorimeter cells fixed
$(=\chi )$, the differential distribution in $\cos \chi$ can be expressed as
\begin{equation}
\frac{1}{\sigma} \frac{d\Sigma^{EEC}}{d \cos \chi} =\frac{1}{\sigma} \sum \int\frac{d\sigma}
{dx_i dx_j d \cos \chi} x_i x_j dx_i dx_j~,
\end{equation}  
where $x_i$ are the scaled energies in terms of the c.m. energy $\sqrt{s}$. The experimental
configurations with fixed angle between the calorimeters $\chi$ and the polar angle of one 
of the calorimeters $\theta$ are calculable in perturbative QCD~\cite{Basham:1978bw,Basham:1978zq}.
However, most experimental measurements were carried out for the averaged EEC, obtained by
integrating over $\cos \theta$, for which perturbative QCD yields the following 
expression (for $m_q=0$)
\begin{equation}
\frac{1}{\sigma_0} \frac{d\Sigma^{EEC}}{d \cos \chi}= \frac{\alpha_s(Q^2)}{\pi} F(\xi)~,
\end{equation} 
where $\xi=\frac{1-\cos \chi}{2}$ and $F(\xi)$ is given by~\cite{Basham:1978bw,Basham:1978zq}
\begin{equation}
F(\xi)= \frac{(3-2\xi)}{6\xi^2(1-\xi)} \left[2(3-6\xi + 2\xi^2) +\ln(1-\xi) + 3\xi(2-3\xi) 
                                                 \right]~.
\end{equation}
The averaged (obtained by integrating over $\cos \theta$) AEEC cross section has an obvious definition
\begin{eqnarray}
\frac{1}{\sigma_0} \frac{d\Sigma^{AEEC}}{d \cos \chi} &\equiv& 
\frac{1}{\sigma_0} \frac{d\Sigma^{EEC}(\pi -\chi)}{d \cos \chi}
-\frac{1}{\sigma_0} \frac{d\Sigma^{EEC}(\chi)}{d \cos \chi}\nonumber\\
&=& \frac{\alpha_s(Q^2)}{\pi} \left[(F(1-\xi) -F(\xi)\right]\equiv \frac{\alpha_s(Q^2)}{\pi}A(\xi)~.
\end{eqnarray}
Effects of quark masses in the EEC and AEEC cross sections were calculated 
in~\cite{Ali:1982ub,Ali:1983au,Csikor:1983dt,Cho:1984rq}. The $O(\alpha^2_s(Q^2)$ corrections to these
distributions were calculated numerically~\cite{Ali:1982ub,Ali:1983au,Richards:1982te,Richards:1983sr}.
Restricting the angular range to $-0.95 < \cos \chi <0.95$, where the non-perturbative effects
are relatively small, the NLO corrections 
to the EEC cross-section were found to be moderate,
 typically $O(35\%)$, but the corresponding corrections to the AEEC were  small,
typically $O(10\%)$, giving reasons
to be optimistic about the convergence of perturbative QCD in these variables, particularly the
AEEC. 

\begin{figure}
\center{
\includegraphics[width=0.45\textwidth, height=8.0cm]{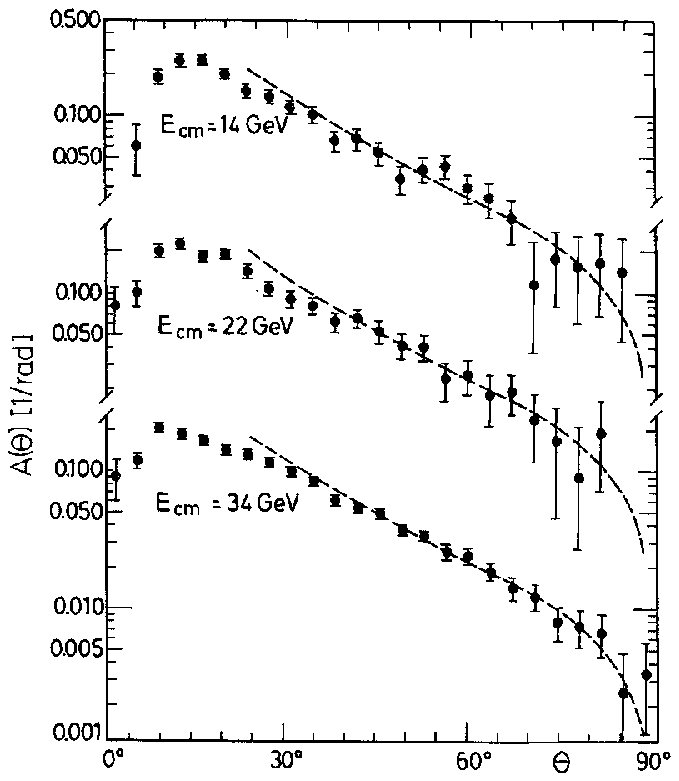} 
$\;\;\;$
\includegraphics[width=0.50\textwidth,height=8.5cm]{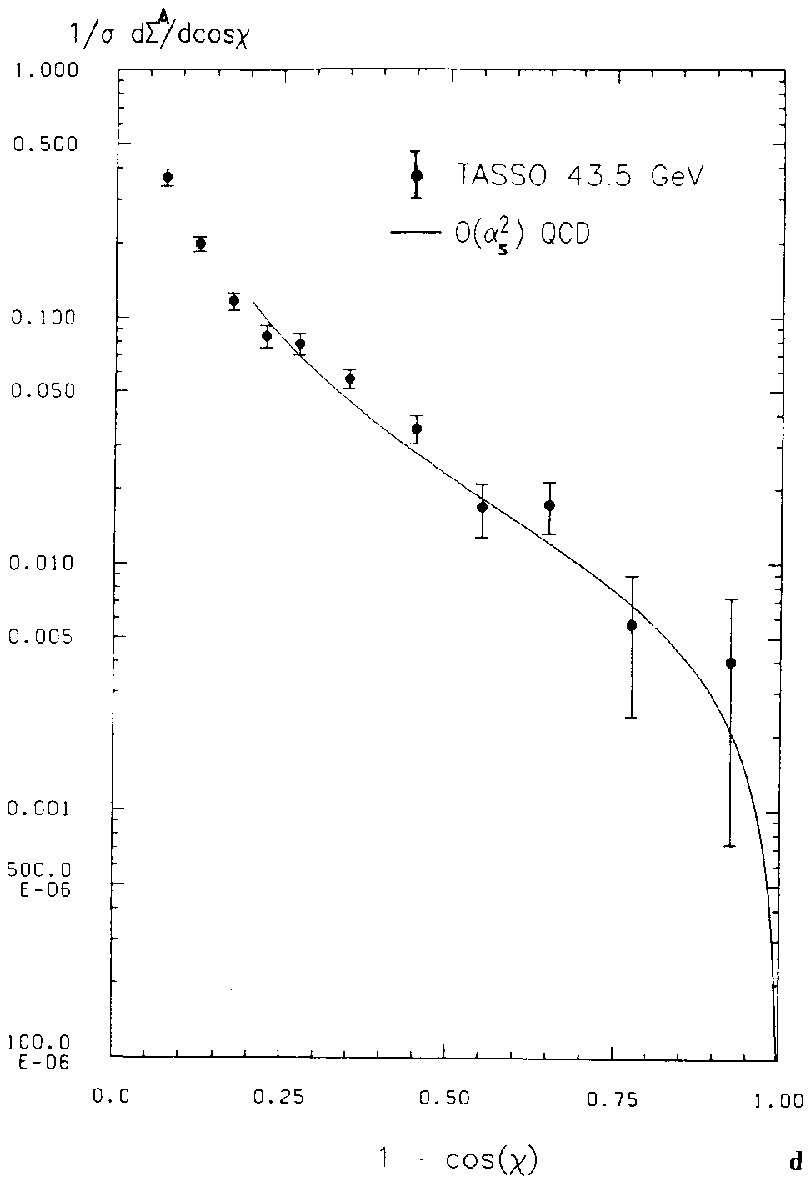}
%$\;\;\;$
\caption{The asymmetric part of the energy energy correlation cross section
 measured by the
 JADE~\cite{Bartel:1984uc} and ~[TASSO]~\cite{Braunschweig:1987ig}
  collaborations at PETRA and comparison with the perturbative
QCD calculations including $O(\alpha_s^2(Q^2))$ corrections from ~\cite{Ali:1982ub}.
The distribution in the upper left-hand frame
from the JADE collaboration shows the corrected asymmetry $A(\theta)$ vs. $\theta$, measured at
$\sqrt{s}=14, 22 $ and 34 GeV. The upper right-hand frame from  TASSO shows the measurements at
$\sqrt{s}=43.5$ GeV and comparison with the perturbative QCD (solid curve).
\label{fig:AEEC-PETRA}
        }}
\end{figure}

Measurements of the EEC and AEEC were undertaken by all four experiments at PETRA:
JADE, MARK-J, TASSO, and PLUTO. The AEEC measurements 
have been used to determine  $\alpha_s(Q^2)$ by comparing them with the NLO
 expression\cite{Ali:1982ub}.  The extracted values of $\alpha_s(Q^2)$ are found to be:
$\alpha_s(Q^2= (34~{\rm GeV})^2)=0.115 \pm 0.005$~[JADE]~\cite{Bartel:1984uc},
$\alpha_s(Q^2= (34~{\rm GeV})^2=0.13$~[MARK-J]~\cite{Adeva:1983ur},
$\alpha_s(Q^2= (34.8~{\rm GeV})^2)$\\
$=0.125 \pm 0.005$~[TASSO]~\cite{Braunschweig:1987ig},
and $\alpha_s(Q^2= (34.6~{\rm GeV})^2)=0.125 \pm 0.005$~[PLUTO]~\cite{Berger:1985xq}. Within errors, these
values of $\alpha_s(Q^2)$ are consistent with each other, and with the ones from oblateness and thrust
distributions, given earlier.
Representative distributions from the  JADE~\cite{Bartel:1984uc} and  [TASSO]~\cite{Braunschweig:1987ig}
 collaborations are shown in
 Fig.~\ref{fig:AEEC-PETRA}, in which
$A(\theta)$ vs.~$\theta$ and $1/\sigma d\Sigma^{\rm A}/d\cos \chi$ vs. $1-\cos \chi$ are plotted, respectively. These measurements are
compared with the perturbative QCD expression, calculated to $O(\alpha_s^2(Q^2))$ and the
agreement is impressive for $(\theta, \chi) > 30^\circ$. For $(\theta, \chi) < 30^\circ$,
 one needs to implement non-perturbative effects as well as the resummation of the large
logs to all orders in perturbation theory.  

\begin{figure}
\center{
\includegraphics[width=0.48\textwidth, height=9.5cm]{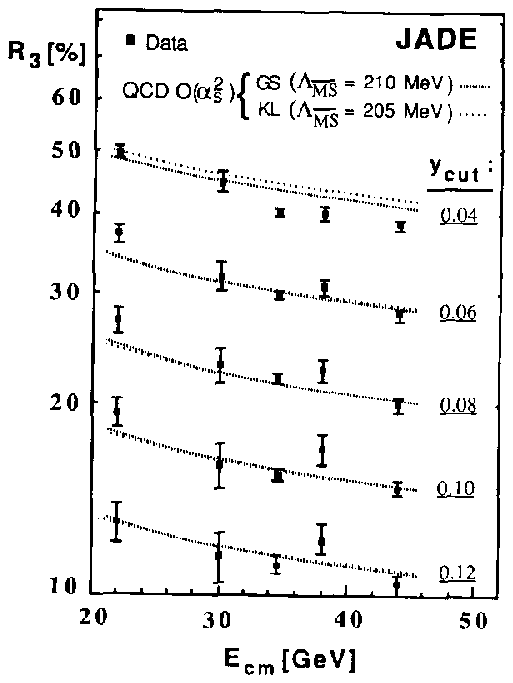} 
$\;\;\;$
\includegraphics[width=0.48\textwidth,height=9cm]{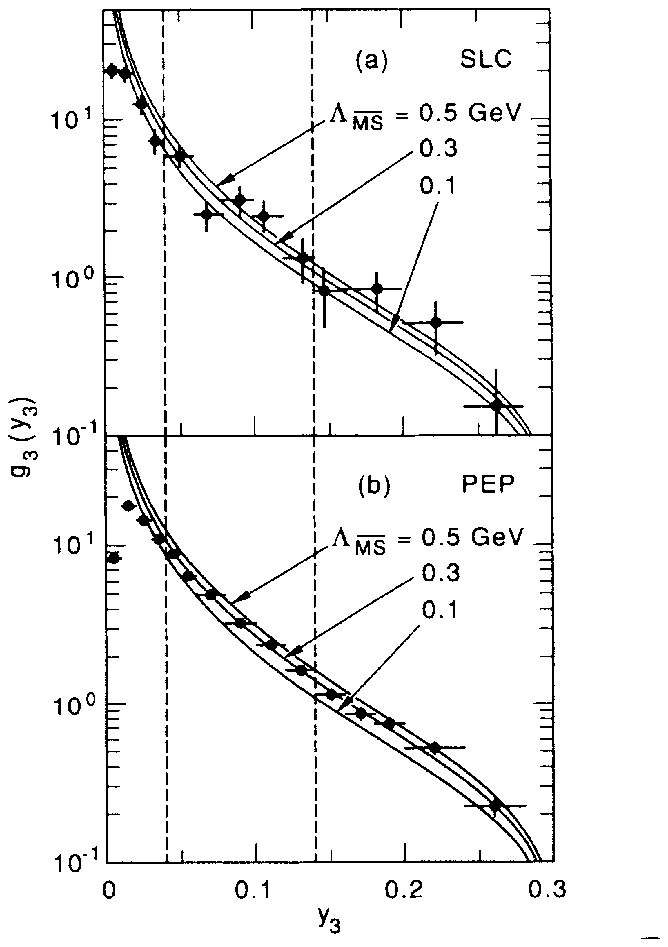}}
\caption{Left-hand frame: Three-jet event rates measured by the JADE collaboration~\cite{Bethke:1988zc}
as a function of the c.m. energy $E_{\rm cm}$ [GeV]
for the indicated values of the jet resolution parameter $y_{\rm cut}$, together with the
predictions of the order $\alpha_s^2$ perturbatve calculations by Gottschalk and Shatz (GS)
and Kramer and Lampe (KL). Right-hand frame: Experimental distribution $g_3(y_3)$ as a function of $y_3$ at (a) $\sqrt{s}=91$ GeV and
  $\sqrt{s}= 29$ GeV measured by the MARKII collaboration~\cite{Komamiya:1989hw}. The $y_3$ range used in
 the fit for the determination of $\alpha_s$ is defined by the two dashed lines. The curves are
second order perturbative calculations  with the indicated values of $\Lambda_{\overline{MS}}$.
        }
\label{fig:JADE-MARKII-3jet}
\end{figure}

A lot of  experimental effort went also in studying the topological
cross sections (jet multiplicity) in $e^+e^-$ annihilation experiments at PETRA, PEP,
 TRISTAN and later at
LEP. Theoretical distributions to these topologies were calculated in a series of
papers~\cite{Kramer:1986sg,Fabricius:1980fg,Fabricius:1981sx,Gutbrod:1983qa,Kramer:1986mc,Gutbrod:1987jt}.
Making use of this theoretical work, the JADE collaboration measured $\alpha_s(Q^2)$ in a
limited range of $\sqrt{s}$ using the three-jet rate and established the running of
$\alpha_s(Q^2)$. Defining the fractional three-jet rate 
 $R_3=\sigma_{\rm 3-jet}/\sigma_{\rm tot}$ as a function of $y_{\rm min}$, which is a
cut-off parameter such that $y_{ij} \geq y_{\rm min}$ for any pair of partons $i$ and $j$
and $y_{ij}= M_{ij}^2/s$, the measured jet-rate was fitted to the expression
\begin{equation}
R_3 (y_{\rm min})= C_1 \alpha_s(Q^2) + C_2 \alpha_s^2(Q^2)~,
\end{equation} 
where $C_1$ and $C_2$ are $y_{\rm min}$-dependent constants calculated by Kramer and
Lampe (called KL below)
 in~\cite{Kramer:1986mc}. The JADE measurements for $R_3(y_{\rm min})$ as a
function of $\sqrt{s}$ in the range $20 < \sqrt{s} < 44$ GeV are shown in 
Fig.~\ref{fig:JADE-MARKII-3jet} (left-hand frame)~\cite{Bethke:1988zc}. They follow nicely the RG-prescribed running of
$\alpha_s(Q^2)$ with $\Lambda_{\overline{MS}}=205$ MeV for $0.04 < y_{\rm min} < 0.12$ using KL,
 with almost the
same value $\Lambda_{\overline{MS}}=210$ MeV using a calculation by Gottschalk and Shatz
(called GS)~\cite{Gottschalk:1984vy}.
An even more convincing measurement of the running of $\alpha_s(Q^2)$ was presented
 by the MARK II
collaboration~\cite{Komamiya:1989hw} at PEP and SLC. They determined $\alpha_s(Q^2)$
 from the differential three-jet
rate $g_3(y_3)|_{y_3=y_{\rm cut}}= \frac{\partial}{\partial y_{\rm cut}}f_2(y_{\rm cut})$,
 where $f_2(y_{\rm cut})$ is the fraction of two-jet events defined by the jet resolution
 $y_{\rm cut}$. Their result for $g_3(y_3)$ is shown as a function of $y_3$ in
Fig.~\ref{fig:JADE-MARKII-3jet} (right-hand frame) for two values $\sqrt{s}=91$ GeV (SLC)
 and $\sqrt{s}=29$ GeV (PEP). The three
curves shown are the predictions of KL~\cite{Kramer:1986mc} for three different
values $\Lambda_{\overline{MS}}=0.1, 0.3$ and 0.5 GeV. Here $\Lambda_{\overline{MS}}$
refers to the QCD scale parameter in a specific renormalisation scheme,
the so-called modified minimal subtraction scheme $\overline{MS}$~\cite{Bardeen:1978yd}.
For further reading of the technical issues of renormalisation and schem-dependencies
at a nonspecialist level, we refer to a review on perturbative QCD~\cite{Brock:1993sz}.
 These measurements yielded 
$\alpha_s(Q^2)=0.123 \pm 0.009 \pm 0.005$ at $Q=\sqrt{s}=91$ GeV and 
$\alpha_s(Q^2)=0.149 \pm 0.002 \pm 0.007$ at $Q=\sqrt{s}=29$ GeV, The running of
 $\alpha_s(Q^2)$ is
clearly established. A comparison with the values of $\alpha_s(Q^2)$ determined from the
measurements of the AEEC cross section at PETRA energies, discussed earlier, also shows
 that non-perturbative
effects at these energies are observable dependent and not negligible.

These investigations were extended to jet rates of higher multiplicity, i.e. four-jet
 and five-jet.
An earlier paper along these lines is due to the JADE collaboration, in which $n$-jet
 rates $(n=2,3,4,5)$
were presented~\cite{Bartel:1986ua}. At this time, NLO corrections to the 4-jet rates
and even LO predictions for the
5-jet final states did not exist. The data were compared with the
 leading-logarithmic- approximation
(LLA) model . Similar studies based on the MARK II data at PEP  ($\sqrt{s}=29$ GeV)
 are found in~\cite{Bethke:1989ma} and~\cite{Bethke:1989jr} using the
 so-called ``optimised perturbation theory'', i.e., by fitting the scale.

An earlier attempt to establish the non-abelian nature of QCD from a study of multijet
 events was made by the AMY collaboration at the TRISTAN $e^+e^-$ storage ring at the
 KEK laboratory~\cite{Park:1989fq}. Their data showed a clear preference for QCD in
 contrast to an abelian model. In addition, they showed the running of $\alpha_s(Q^2)$
 by measuring the 3-jet rate $R_3$ at $\sqrt{s}=50$ to 57 GeV by comparing their
 measurements with those of the JADE collaboration~\cite{Bethke:1988zc}  and the
TASSO collaboration~\cite{Braunschweig:1988ug} at PETRA taken at lower c.m. energies. 
Other publications towards a determination of $\alpha_s$ from PEP and PETRA are
for example by MARK II~\cite{Lohr:1982wh} and CELLO~\cite{Behrend:1989jh}.
%
%%%%%%%%%%%%%%%%%%%%%%%%%%%%%%%%%%%%%%%%%%%%%%%%%%%%%%%%%%%%%%%%%%%
\subsection{String- and String-like effects in Jets}
\begin{figure}
\center{
\includegraphics[width=0.48\textwidth, height=9.5cm]{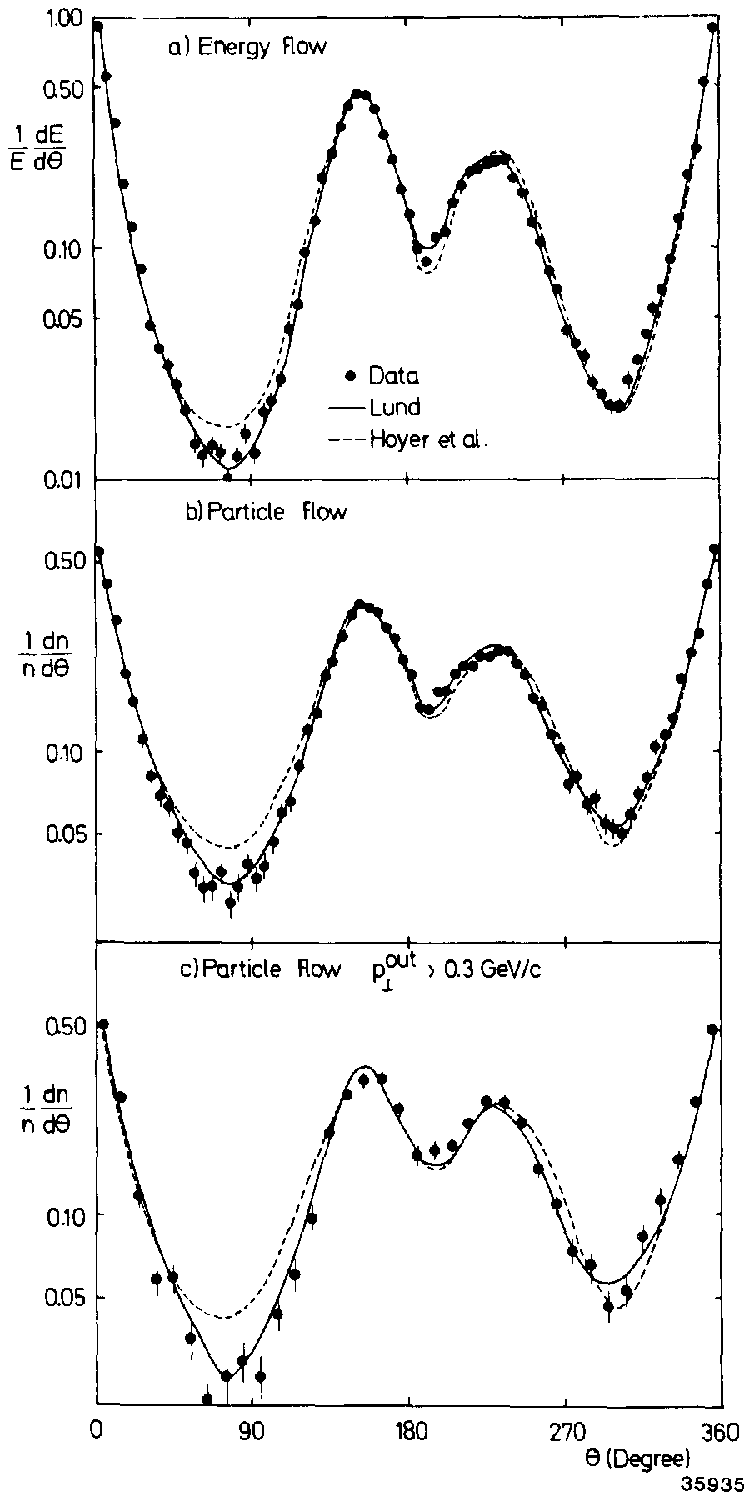}
\includegraphics[width=0.48\textwidth,height=9.5cm]{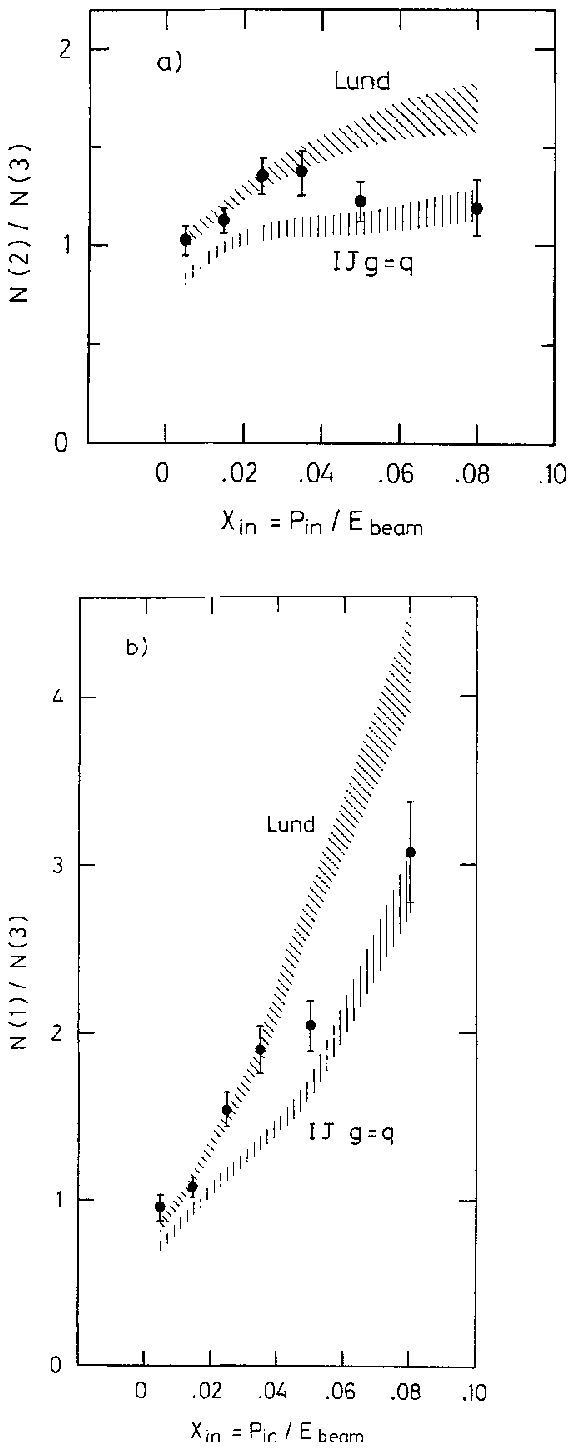}}
\caption{Left-hand frames: (a) The normalised energy flow $1/E dE/d\theta$ in the 
three-jet events compared with two
model predictions. (b) The normalised charged particle flow $1/n dn/d\theta$.
 (c) $1/n dn/d\theta$
with $p_T^{\rm out} > 0.3$ GeV. Here $n$ is the total number of particles used in each plot 
(JADE collaboration~\cite{Bartel:1983ij}).
Right-hand frames: Ratios of particle densities in the angular gaps between the jet axes,
 defined by
 $0.25 < \psi_j^\prime < 0.75$ as a function of $x_{\rm in}$. The calculation of the IJ ($g=q$)
and of the LUND models are shown as shaded bands. a) $N(2)/N(3)$ and b)$N(1)/N(3)$ for
the three-jet event sample (TASSO collaboration~\cite{Althoff:1985wt}).
  }
\label{fig:JADE-string}
\end{figure}

%
%%%%%%%%%%%%%%%%%%%%%%%%%%%%%%%%%%%%%%%%%%%%%%%%%%%%%%%%%%%%%%%%%%%
%
The data taken by the experimental collaborations at PEP and PETRA have been used also
to investigate non-perturbative effects in the jet profiles with the view of testing
various phenomenological models available in the 1980's. This was important, since
depending on the observables considered, non-perturbative effects influenced also
the measurement of $\alpha_s$. Several groups~\cite{Bartel:1981kh,Bartel:1983ij,Aihara:1984du,Althoff:1985wt}
 have used three-jet ($q \bar{q} g $) events to study the impact
 of hard gluon bremsstrahlung on the hadronisation process. In these studies they
observed the so-called string effect~\cite{Andersson:1983ia}, predicting a depletion of
 particles in the angular region between the quark and antiquark jet
relative to the particle flow in the regions between the quark and gluon jets and the
antiquark and gluon jets. In  Fig.~\ref{fig:JADE-string} (left-hand frames), we
show the measurements of the normalised energy flow $(1/E) dE/d\theta$ in planar three-jet events and
the normalised charged  particle flow  in these events undertaken 
by the JADE collaboration~\cite{Bartel:1983ij} between $\sqrt{s}=30$ GeV and 36 GeV at PETRA. 
 These distributions allowed one 
to distinguish between a hadronisation model~\cite{PHENgluon1}
in which the fragmentation proceeds along the parton momenta (the
independent jet IJ model) and the model in which the fragmentation takes place along
the colour-anticolour axes (the LUND string model~\cite{Lund}), discussed earlier. Only the leading order
 ($O(\alpha_s) $)
matrix elements were taken into account for the gluon bremsstrahlung process ($e^+e^- \to q \bar{q}g$),
which were encoded  in these fragmentation models.
As seen in this figure,  JADE data on the energy and charged particle flow are better reproduced by 
fragmentation along the colour axes~\cite{Lund}.

A similar analysis was undertaken somewhat later in 1985 by the TASSO
 collaboration~\cite{Althoff:1985wt}. In this case, the three-jet events produced in
 $e^+e^-$ annihilation
 into hadrons at 34.4 GeV were compared with the $O(\alpha_s^2)$ perturbative QCD
calculations convoluted with two different models of fragmentation (IJ and Lund).
 The analysis was undertaken in terms of the ``reduced azimuthal angles''
$\psi_j^\prime$ and $x_{\rm in}=p_{\rm in}/E_{\rm beam}$, where $p_{\rm in}$ is the particle
 momentum
projected into the event plane. The $\psi_j^\prime$ are defined as
\begin{equation}
\psi_j^\prime =\frac{\psi -\Phi_i}{\Phi_k-\Phi_i}~~, i,j,k=1,2,3~{\rm and~cyclic}~, 
\label{eq:reduced-psi}
\end{equation}
where the particle under consideration is located between jets $i$ and
 $k$ $(\Phi_i < \psi < \Phi_k)$.
The reduced angles $\psi_j^\prime$ run from 0 and 1. The subscript $j$ denotes the
 angular region opposite
 to the jet $j$. The analysis was restricted to $x_{\rm in} < 0.1$ and the data
were divided in two samples $x_{\rm in}< 0.04$ and $0.04 <x_{\rm in} <0.1$. The result of the
TASSO analysis is  displayed in  Fig.~\ref{fig:JADE-string} (right-hand frames) showing
 that the distribution of
low energy (soft) hadrons in the 3-jet plane is better described by the LUND colour
fragmentation model than by the independent jet model. The opposite is true for more energetic
particles flowing between the 3 jets.

\begin{figure}
\center{
{\hspace*{2.0cm}\includegraphics[width=0.47\textwidth, height=4.8cm]{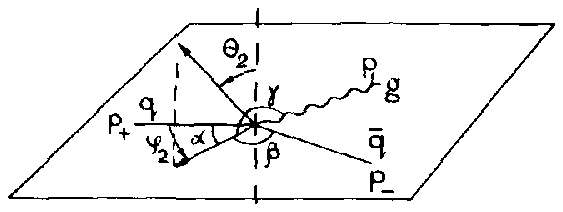}}
%$\;\;\;$
{\hspace*{-1.0cm}\includegraphics[width=0.44\textwidth, height=4.5cm]{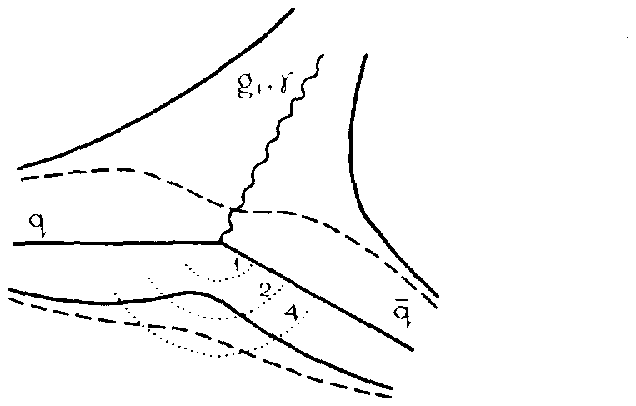}}
%$\;\;\;$
\includegraphics[width=0.95\textwidth, height=9.5cm]{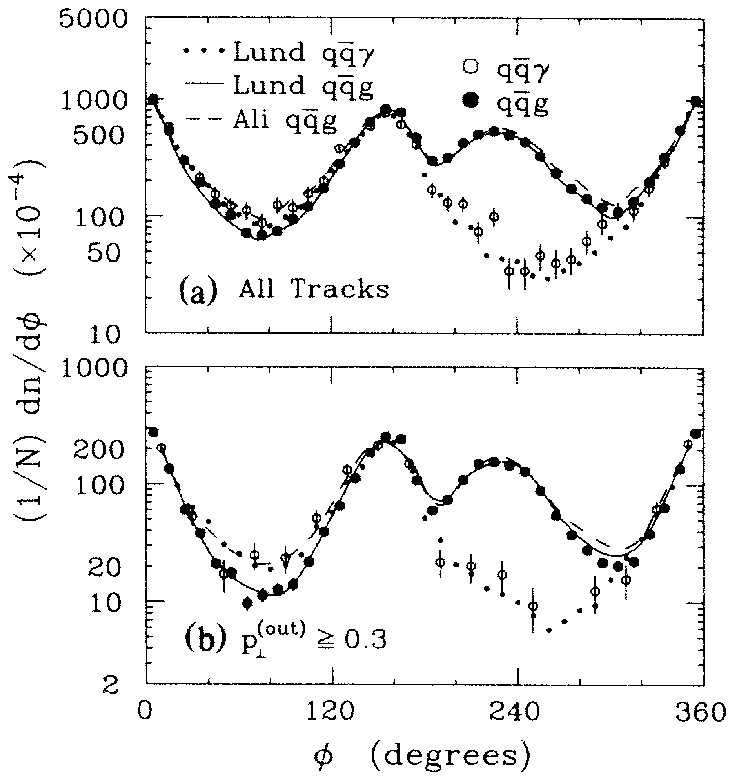}}
\caption{Upper left-hand frame: Kinematics of non-jet radiation in three-jet events;
Upper right-hand frame: 
Directivity diagram of the soft particle flows, projected on to
the $q\bar{q}\gamma$ (dashed lines) and $q\bar{q}g$ (solid line) event planes. Dotted
circles show
the constant levels of density flow [$W(\phi)=1,2,4 $] (from Azimov et
al.~\cite{Azimov:1986sf}).
Lower frame: The charged-track density as a function of the event-plane angle $\phi$.
The angular region between $\phi=0^\circ$ and $\phi=150^\circ$ separates the $q$
and $\bar{q}$ for
the $q\bar{q}\gamma$ events and for 65\% of the $q\bar{q}g$ events (from the MARK II
collaboration~\cite{Sheldon:1986gy}). }
\label{fig:MARKII-string}
\end{figure}

The ``string effect'' was subsequently attributed to the coherence of soft gluon emission
 from the
$q\bar{q}g$ system -- a characteristic feature of the non-abelian nature of
 QCD~\cite{Azimov:1986sf}.
This is illustrated by contrasting the case of a soft gluon emission (assumed here as
 $g(p_2)$)
in $e^+ e^- \to q (p_+) + \bar{q}(p_-) + g(p_1) +g(p_2)$ from the process in which 
the gluon $g(p_1)$
is replaced by a photon, i.e., $e^+ e^- \to q (p_+) + \bar{q}(p_-) + \gamma(p_1) +g(p_2)$.
 The angular
distribution of the soft gluon (antenna pattern)in the case of
 $e^+ e^- \to q \bar{q} \gamma$ is given by
\begin{equation}
W_{+-}(\phi_2) \equiv 2C_F a_{+-}  V(\alpha, \beta) =\frac{4 C_F a_{+-}}
{\cos \alpha - \cos\beta}
\left(\frac{\pi-\alpha}{\sin \alpha} -\frac{\pi-\beta}{\sin \beta} \right)~,
\label{eq;qedflow}
\end{equation} 
where $\alpha=\phi_2$ and $\beta=\theta_{+-} - \phi_2$ (see the kinematics shown in the
upper left-hand frame in Fig.~\ref{fig:MARKII-string});
$a_{ik}=1-(\vec{n}_i.\vec{n}_k)$, with $\vec{n}_i$ being the unit vector in the direction of $\vec{p}_i$,
and $\theta_{+-}$ is the angle between the $q$ and $\bar{q}$ directions. Replacing $\gamma(p_1)$ with
a gluon $g(p_1)$ changes the angular distribution essentially due to the antenna element $g(p_1)$
 participating in the emission as well.
 One now obtains ($\gamma=\theta_{+1}+ \phi_2 $):
\begin{equation}
W_{\pm 1}(\phi_2)=N_c[a_{+1}V(\alpha, \gamma) +a_{1-}V(\alpha,\gamma)]
+ (2C_F-N_c)a_{+-} V(\alpha,\beta)~.
\label{eq;qcdflow}
\end{equation}
The (soft) particle flow according to these two configurations   is illustrated
in Fig.~\ref{fig:MARKII-string} (upper right-side frame)
 showing that the flow opposite to the direction of $\vec{n}_1$ is appreciably
 lower for the case of a gluon than for a photon due to the destructive interference
 in the case of QCD ($q\bar{q} g g $).

 This phenomenon can be qualitatively understood . Omitting the small contribution from 
the second terms in Eq.~(\ref{eq;qcdflow}), one reduces this equation to the sum of two
 independent quark
 antennas ($+,1$) and ($-,1$). Therefore, in this approximation, the total particle flow
 can be obtained
by the simple incoherent composition of two ``annihilations'' $e^+e^- \to q \bar{q}$,
 boosted from their
 respective rest frames to the overall $q\bar{q}g$ c.m. frame. It is clear that the
 angular region between
the $q$ and $\bar{q}$ will be depopulated as it is opposite to the boost direction of
 both two-jet configurations.
This perturbation theory based scenario ($3=2+2 $ + Lorentz boost) then coincides with
 the fragmentation of the
gluon in the process $e^+ e^- \to q \bar{q} g$ events in the LUND fragmentation model.
 The independent jet
model misses this, as the gluon fragments independently on its own. Consequently,  the
 Lorentz boost
effect is absent. 

 The colour coherence study of $e^+e^-$ jets by Azimov et al.~\cite{Azimov:1986sf} suggested
 an
interesting experimental test in the form of particle flows in three-jet ($q\bar{q}g $)
 and radiative
 two-jet ($q\bar{q}\gamma)$ events by observing the negative contribution of the third
antenna. This test was carried out by the MARK II collaboration at PEP at
 $\sqrt{s}=29$ GeV~\cite{Sheldon:1986gy} with the result that in the angular region between
the quark and antiquark jets
fewer charged tracks were observed in the two-jet events than in the radiative
 three-jet events. Their result is shown in Fig.~\ref{fig:MARKII-string} (lower two frames).

To end this review of the studies of jets at PETRA and PEP, we briefly discuss the 
angle ordered perturbation theory, as this approach has been used to develop a parton shower
Monte Carlo~\cite{Herwig1}. In this approach, the phase space of soft gluon emission is
 restricted using an angle ordering criterion, which allows one to take into account the
interference (colour coherence) approximately, and hence it reproduces the string and
string-like effects discussed above. Both the LUND fragmentation model (PYTHIA in its
modern incarnation) and the parton shower
Monte Carlo models (such as HERWIG) describe the $e^+e^-$ data adequately. However, the
main drawback of these models is that they do not (easily) match with the fixed order
perturbation theory in next-to-leading and higher orders. The main obstacle is that fixed
order perturbation theory has soft and collinear singularities that give rise to logarithmic
enhancement of higher order contributions. These enhanced terms should be summed to all orders. 
However, there is no unique way of doing this. For example, the $p_T$-ordered and the angular ordered
showers can both be arranged to resum these logarithms. Matching with a fixed order
perturbation theory is more easily achieved in $p_T$ ordered showers which, however, do not have
the colour coherence needed by the low-energy $e^+e^-$ data. It is the other way around with
the angle-ordered showers. We will discuss these aspects further in the next section.

%%%%%%%%%%%%%%%%%%%%%%%%%%%%%%%%%%%%%%%%%%%%%%%%%%%%%%%%%%%%%%%%%%%
\section{Jets in QCD and at LEP}
\label{sec:LEPJets}
%%%%%%%%%%%%%%%%%%%%%%%%%%%%%%%%%%%%%%%%%%%%%%%%%%%%%%%%%%%%%%%%%%
In this section we review the salient features of jets at LEP which were helpful
in testing some of the basic elements of QCD more precisely. Just as in the preceding section,
we recall the main detectors at LEP, ALEPH, DELPHI, OPAL, and L3, which collected large data
samples (typically, 4 million hadronic events around the $Z$ resonance for each of the
four LEP experiments. In the second stage, LEP2, the beam energy was increased to about
103 GeV. These detectors proved to be very powerful tools in carrying out precision electroweak
and QCD physics.

\subsection{Quark/Gluon cascades}
\begin{figure}
\center{
\resizebox{0.75\columnwidth}{!}{
\includegraphics{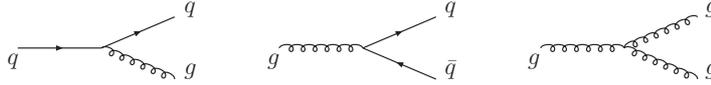}}}
\caption{The three basic splitting processes of quarks and gluons
         into pairs of quarks and gluons.
        }
\label{fig:F_split}
\end{figure}

Electric charges which are accelerated, reduce their energy by 
radiating photons preferentially collinear with the flight 
direction of the charge. This is a general feature of gauge 
theories and, specifically, collinear radiation is predicted 
in QCD processes in which quarks and gluons are produced with
high energies. If the observed partons carry away a fraction
$z$ of the parent partons, the splitting functions \cite{split}, 
{\it cf.} Fig.~{\ref{fig:F_split}},  
\begin{eqnarray}
dP \, [q \to q+g(z)] &=& 
   \frac{\alpha_s}{2\pi}  \,
   C_F \, 
   \frac{1+(1-z)^2}{z} \,
   dz \, \frac{dQ^2}{Q^2} 
   \equiv\frac{\alpha_s}{2\pi}  \,
     P_{gq}(z) \,
      dz \, \frac{dQ^2}{Q^2}~,                 \nonumber \\
 dP \, [q \to q(z) +g] &=& 
   \frac{\alpha_s}{2\pi}  \,
   \left[ C_F \, \frac{1+z^2}{(1-z)_+} + 2 \delta(1-z)\right]\,
      dz \, \frac{dQ^2}{Q^2} 
   \equiv\frac{\alpha_s}{2\pi}  \,
     P_{qq}(z) \,
      dz \, \frac{dQ^2}{Q^2}~,                 \nonumber \\
dP \, [g \to q(z)+\bar{q}] &=&
   \frac{\alpha_s}{2\pi}  \,
      T_R \; [z^2+(1-z)^2]        \,
      dz \, \frac{dQ^2}{Q^2}
      \equiv\frac{\alpha_s}{2\pi}  \,
     P_{qg}(z) \,
      dz \, \frac{dQ^2}{Q^2}~, \nonumber \\                                
dP \, [g \to g+g(z)] &=&
   \frac{\alpha_s}{2\pi}  \,
  \left(  2 C_A \, [\frac{1-z}{z} + z(1-z) + \frac{z}{(1-z)_+}]
        +[\frac{11}{2} -\frac{n_f}{3}]\delta(1-z)\right) 
   dz \, \frac{dQ^2}{Q^2} \nonumber\\
   &\equiv&\frac{\alpha_s}{2\pi}  \,
     P_{gg}(z) \,
      dz \, \frac{dQ^2}{Q^2}~,
\label{eq:dglap}                       
\end{eqnarray}
universally predict collinear splittings, with $Q^2 \simeq z(1-z) E^2 
\Theta^2$ denoting the invariant mass of the final parton pair. 
The notation $[F(z)]_+$ defines a distribution such that for any sufficiently
regular function $f(z)$,
\begin{equation}
\int_0^1dz f(z)[F(z)]_+ = \int_0^1 dz (f(z) - f(1))F(z)~.
\end{equation}
 The bremsstrahlung
 splittings $q \to qg$ and $g \to gg$ preferentially  
generate soft radiation spectra in the limit $z \to 0$. The group
characteristics are $C_F = 4/3$, $C_A =3$ and $T_R = 1/2$ for
SU(3)$_C$ of QCD.  

Repeated splittings generate cascades of collimated quarks and
gluons. Since the lifetime of the final-state pair in the 
splitting processes is long, $\tau^\ast \sim E/Q^2$, the cascade 
is expected to be described by a sequence of probabilities and not 
by interfering quantum-mechanical amplitudes. If the branching 
occurs at a value $Q^2$ without any radiation between the initial
maximum value $Q^2_{max}$ and $Q^2$, the probability is given by
\begin{equation}
d {\mathcal{P}}_{a \to bc} = \frac{\alpha_s}{2 \pi} \,
  P_{a \to bc} (z) \, dz \, \frac{dQ^2}{Q^2}        \;
  \exp \left[ - \sum_{b',c'} \int^{Q^2_{max}}_{Q^2}
       \frac{dQ'^2}{Q'^2} 
       \int dz^\prime \,
       \frac{\alpha_s}{2 \pi} \, \hat{P}_{a \to b'c'} (z^\prime) \right]~,
\label{eq:sud} 
\end{equation}
where the exponential Sudakov factor~\cite{Sudakov:1954sw} accounts for the non-radiation
probability. Here $\hat{P}_{a \to b'c'} (z^\prime)$ are the same functions as 
$P_{a \to b'c'} (z^\prime)$, defined in Eq.~(\ref{eq:dglap}) except for the regularization terms
at $z^\prime \to 1$.

However, the branching probability Eq.(\ref{eq:sud}) is refined
by an important coherence effect. If in electrodynamics 
a photon splits into an electron$-$positron pair, the pair can emit 
photons only at angles less than the angle between the charged pair 
as photons propagating at larger angles would see coherent 
electron$+$positron states which, being neutral, cannot radiate. 
As a result, the emission angles are ordered in the sequence $\Theta_1
> \Theta_2 > ...$ This effect is also predicted in QCD \cite{Herwig1}.
The only difference arises from the fact that the coherent
superposition adds up the colour charges of the daughter partons 
to the non-zero value of the parent colour charge so that 
wide-angle splitting is generated at a non-zero rate. In addition 
to the angular ordering, non-resolved infrared radiation restricts 
the energy fractions of the partons in the cascades. These 
restrictions on energies and angles can be mapped into the boundary 
values of the Sudakov integral after re-expressing the invariant mass
by the angle between the momenta of the daughter partons.
 
The cascading of the primordial quarks and gluons affects 
the observed hadron distributions within the jets. In particular,
energy spectra are softened through the cascading mechanism, multiplicities increase strongly with
energy, and quark and gluon jets will develop different profiles.
Formulated for simplicity by neglecting the change in $k_T$, and restricting to 
one parton species, the energy 
dependence of the fragmentation function is described by the DGLAP
equation \cite{split,DGL,DGL-2,DGL-3}:
\begin{equation}
\frac{\partial D(z,Q^2)}{\partial \log Q^2 / \Lambda^2}
     = \frac{\alpha_s(Q^2)}{2 \pi} \int^1_z \frac{d\zeta}{\zeta}
       P(\zeta) D \left(\frac{z}{\zeta}, Q^2 \right) \,. 
\label{eq:fragE}
\end{equation}
The splitting function $P(\zeta)$ consists of two parts (see, Eq.~(\ref{eq:dglap}). The first
part $P$ describes the standard component and accounts for the 
accumulation of particles at $z$ generated by the splitting of partons at 
$\zeta \ge z$, the second part accounts for the loss of particles at $z$ 
due to splitting to smaller energy values. In the parton model, the function $D$ depends
only on the variable $\frac{z}{\zeta}$ but not on $Q^2$. This would  yield an
scale-invariant fragmentation function $D(\frac{z}{\zeta})$.  In QCD,
this is obviously modified.  The solution of the above equation 
leads to striking effects which modify the predictions of the 
scale-invariant parton model. In particular:

{\it (i)} For large $z$ values beyond 0.2 the spectrum decreases
with increasing energy while the particles accumulate at small $z$. The loss
of particles by splitting at large $z$ is bigger than the gain 
by splitting from yet higher $\zeta$ values. This is naturally 
opposite at small $z$ values.

\begin{figure}
\centering
\includegraphics[width=0.48\textwidth, height=6.5cm]{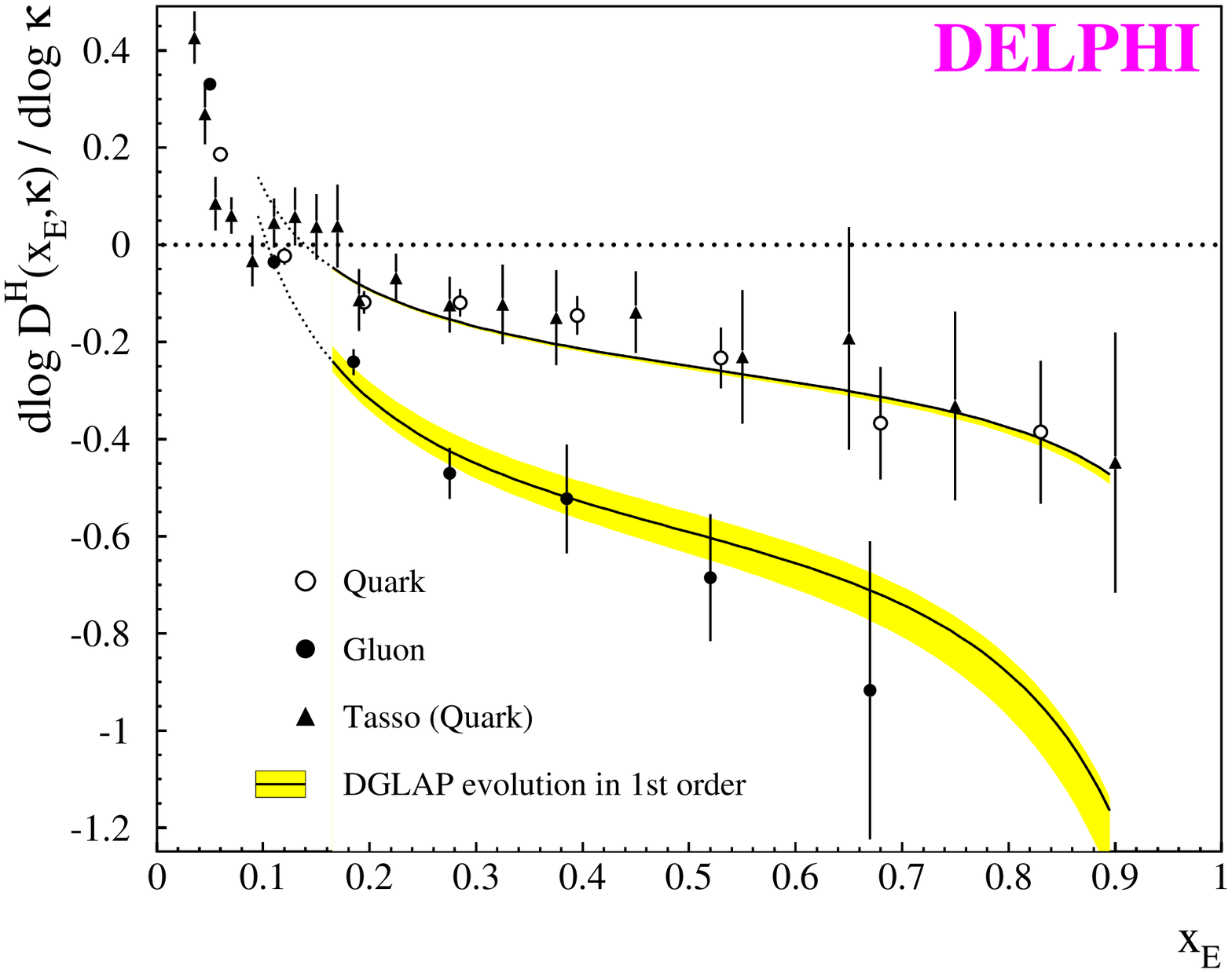} 
$\;\;\;$
\includegraphics[width=0.48\textwidth,height=6.5cm]{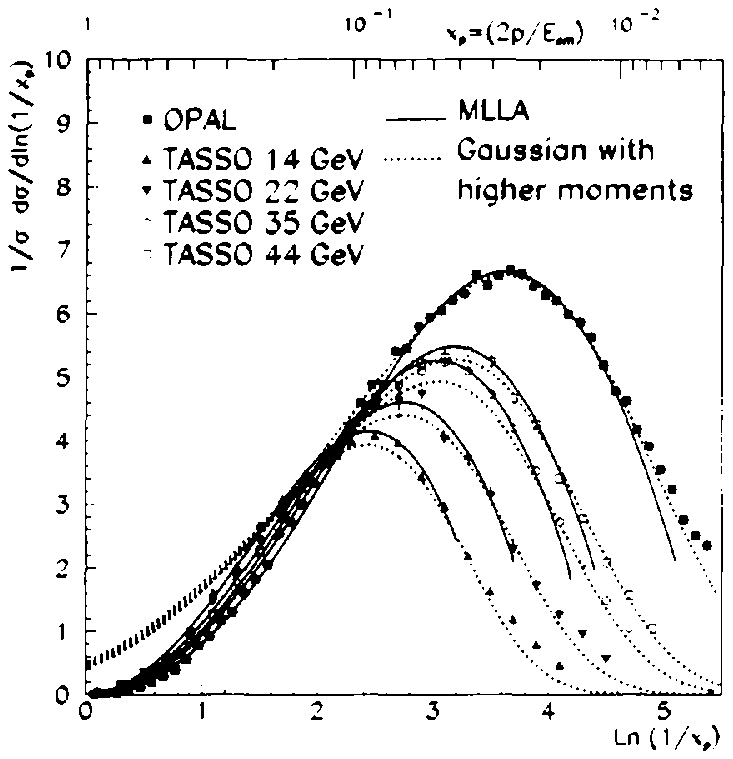}
\caption{(a) Scaling violations in quark and gluon jets (using the DURHAM algorithm).
The solid curves show the DGLAP evolution and the dotted parts of the curves are
extrapolation outside the fit range [DELPHI~\cite{Abreu:1999af}].   
(b) Measurements of the $\ln(1/x_p)$ distributions  for the center-of-mass energies
between 14 and 91 GeV and comparison with the MLLA~\cite{Azimov:1985by} and
a Gaussian~\cite{Fong:1989qy}. This figure is referred as the
humpback plateau of the gluon fragmentation function for small $x$ values in the
text [OPAL~\cite{hback}]. 
        }
\label{fig:2.2ab}
\end{figure}

{\it (ii)} The constant plateau (characterized by $D(z)$ as in Eq.~(5)) in the
parton model generates a
multiplicity of particles which increases logarithmically with
the length of the plateau $\sim \log (\sqrt{s}/m_h)$. Multiple
splittings raise the multiplicity much more strongly. Solving
Eq.~(\ref{eq:fragE}) for the multiplicity, given by the 
integrated fragmentation function, predicts a rise with energy
stronger than exponential:
\begin{equation}
n(s) \sim \exp  ({\alpha}^{-1/2}_s (s)) 
     \sim \exp (\log^{1/2} \! s/\Lambda^2)        \,.
\end{equation}
In addition, the flat plateau in $\log z^{-1}$ is deformed 
to a humpback of Gaussian character~\cite{Azimov:1985by}, with centre 
$[\log z^{-1}]_{max} \sim \log s/\Lambda^2$
and width $\sigma \sim \log^{3/4} \! s/\Lambda^2$.
Experimental proof for the energy dependence of the fragmentation
function $D(z,Q^2)$ and the formation of the humpback at small $z$ is
presented in Fig.{\ref{fig:2.2ab}}.  

{\it (iii)} While the first splitting of a quark jet $q \to qg$ is 
determined by $C_F = 4/3$, the first splitting of a gluon jet 
$g \to gg$ is determined by the bigger Casimir invariant $C_A =3$.
Thus, the large average colour 
of gluons compared with quarks should increase the multiplicity 
of gluon jets with regard to quark jets asymptotically in the ratio 
$C_A / C_F = 9/4$. Similarly, since $dN_{g/q} \sim C_{A/F} \, d \log \Theta$, 
the angular widths of quark and gluon jets are different, 
$\Theta_g = \Theta^{C_F/C_A}_q$, {\it viz.} gluon jets
are wider than quark jets. Even though the asymptotic limit 
has not been reached yet, the particle multiplicities in gluon jets
have been shown significantly larger than in quark jets \cite{mult}: 
$n_g/n_q > 3/2$. For a recent discussion of particle multiplicities, we
refer to~\cite{Dremin:2000ep}.

{\it (iv)} Small-angle gluon radiation off heavy quarks $Q = c,b$ 
is suppressed compared to light quarks \cite{DokK}, and the logarithmic
 enhancement of the particle yield is restricted
to infrared gluon configurations.
%
%%%%%%%%%%%%%%%%%%%%%%%%%%%%%%%%%%%%%%%%%%%%%%%%%%%%%%%%%%%%%%%%%%
%
\subsection{Multijets at LEP}
\label{sec:LEP}
Increasing the energy from the PETRA regime of about $\sqrt{s} =$
46 GeV to the LEP regime by factors of two and five in the two phases 
of LEP, $Z$-boson runs with $\sqrt{s} =$ 91 GeV and
beyond with $\sqrt{s}$ up to 206 GeV, provided two opportunities: the
experimental analysis of multijet final states \cite{LEP} and 
the study of the jet properties over a large range
in energy \cite{PETRA.LEP}. This allows a more precise measurement of two fundamental
characteristics of QCD~\cite{WZ}, the running of the QCD 
coupling with energy as predicted by asymptotic freedom, and the
three-gluon coupling, a fundamental ingredient of asymptotic freedom. We discuss
these measurements below in turn.
\subsection{Inclusive jet observables and determination of $\alpha_s(M_Z)$ at LEP}
\label{sec:LEP-Observables}
All four experiments at LEP, DELPHI~\cite{Abreu:1999rc},
 OPAL~\cite{Pfeifenschneider:1999rz}, L3~\cite{Achard:2002kv}, and
 ALEPH~\cite{Heister:2003aj}, undertook measurements of
the inclusive jet (or event shape) variables and their moments. In these analyses, 
the next-to-leading order ($O(\alpha_s^2)$) perturbative QCD calculations for the
 inclusive observables and event shape distributions, discussed in the context of three-jet
 events at PETRA, were augmented by theoretical estimates in the next-to-leading-log
 approximation (NLLA)~\cite{Catani:1991kz,Catani:1991bd} and others in which the
 $O(\alpha_s^2)$ and NLLA schemes were combined~\cite{Catani:1992ua}. It is helpful
to explain the NLLA in more detail. For a generic shape variable, $y$,
well-defined in pertrubation theory (i.e., infrared and collinear safe), the typical
leading behaviour at small $y$ is
\begin{equation}
\frac{1}{\sigma}\frac{d\sigma_n}{dy} \sim \alpha_s^n \frac{1}{y} \ln^{2n-1} \frac{1}{y}~.
\end{equation}
The normalized event shape cross section $R(y)$ defined as
\begin{equation}
R(y)\equiv\int_0^ydy \frac{1}{\sigma} \frac{d\sigma}{dy}~,
\end{equation}
then has the expansion
\begin{equation}
R_n(y) \sim \alpha_s^n \ln^{2n} (1/y)\equiv \alpha_s L^{2n}~.
\end{equation} 
Whenever $L$ is large, one can improve the range and accuracy of perturbative
predictions by identifying these logarithmically-enhanced terms and resum them to
all orders. For a class of variables for which the leading logarithmic
contributions exponentiate, i.e., variables which in the small-$y$ range yield the
logarithm of the shape cross-section in the form $\ln R(y) \sim Lg_1(\alpha_s L)$,
where $g_1(\alpha_s L)$ has a power series expansion in $\alpha_sL$, one has~\cite{Catani:1992ua} 
\begin{equation}
R(y) =C(\alpha_s) \Sigma (y, \alpha_s) + D(y, \alpha_s)~,
\end{equation}
with $C(\alpha_s)= 1+ \sum_{n=1}^{\infty} C_n (\alpha_s/2\pi)^n$ and
$\ln \Sigma(y,\alpha_s)= Lg_1(\alpha_sL) + g_2(\alpha_sL) + \alpha_s g_3(\alpha_s L) + ...$.
The function $g_1(\alpha_s L)$ resums all the leading contributions of the form
$\alpha_s L^{n+1}$ (defining the LAA approximation) and $g_2(\alpha_s L)$ contains the
next-to-leading logarithmic corrections $\alpha_s^n L^n$ (defining the NLLA) and so on. 

\begin{figure}
\center{
\resizebox{0.75\columnwidth}{!}{
\includegraphics{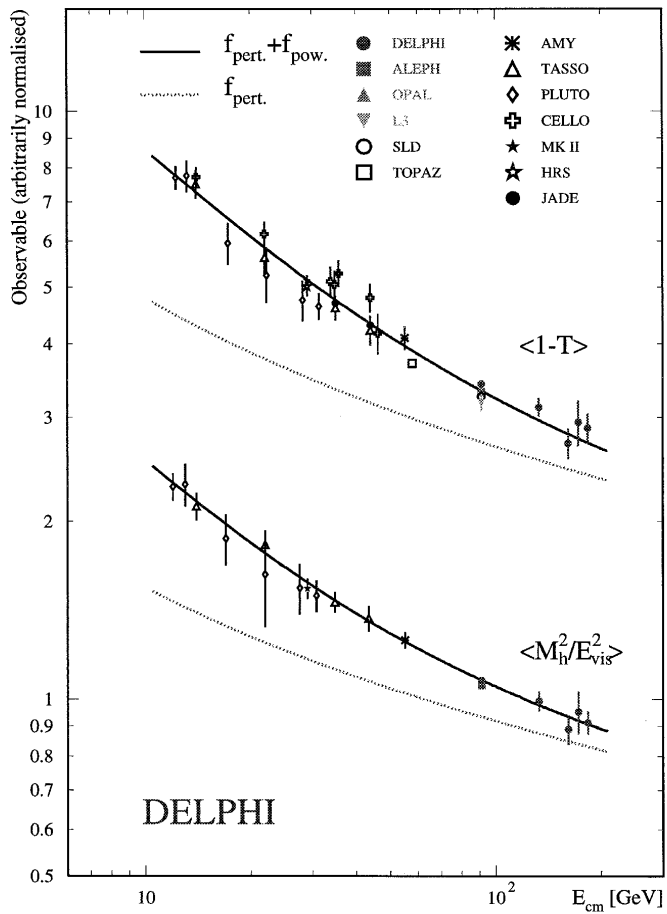}}}
\caption{Measured mean values of the observables
 $\langle 1 -T \rangle$ and $\langle M_h^2/E_{\rm vis}^2 \rangle$
as a function of the center-of-mass energy. The solid lines present the results of the
fits including power corrections and the dotted lines show the perturbative part only.
 (From DELPHI~\cite{Abreu:1999rc}).        }
\label{fig:PLB456-DELPHI}
\end{figure}

\begin{figure}
\centering
\includegraphics[width=0.48\textwidth, height=7.0cm]{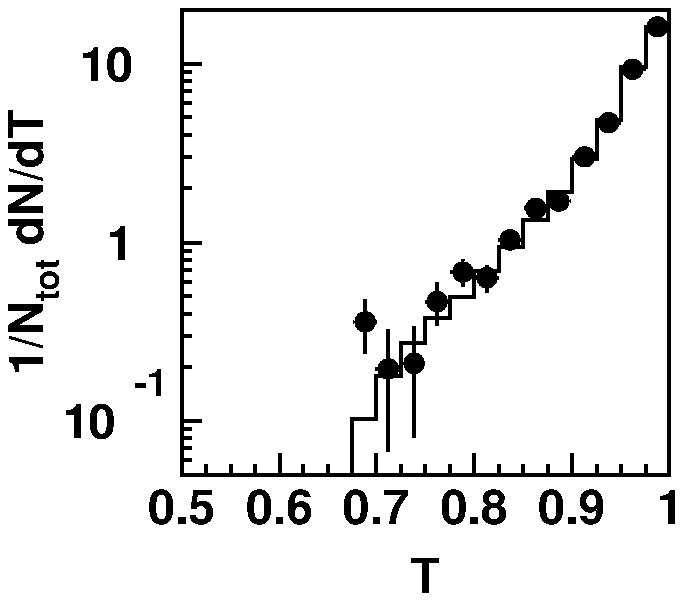} 
$\;\;\;$
\includegraphics[width=0.48\textwidth,height=6.5cm]{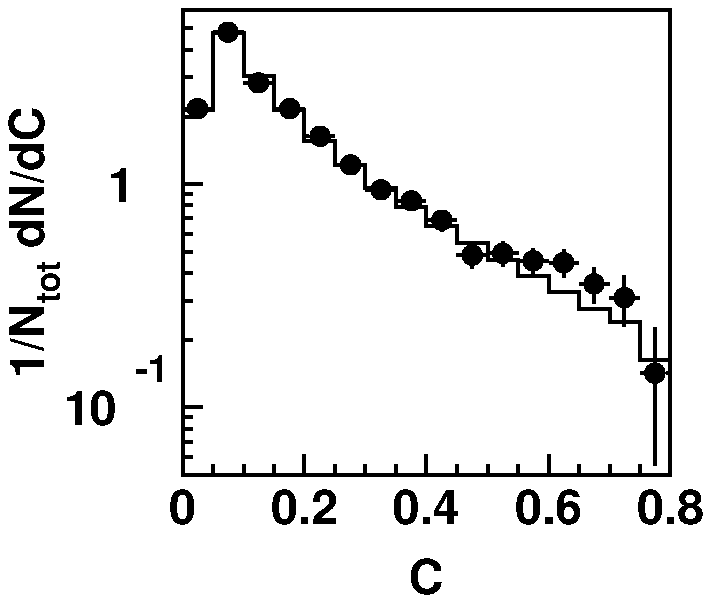}
\caption{Measured distributions of thrust, T, (left-hand frame) and the $C$-parameter in
comparison with QCD predictions at $\sqrt{s}=$206.2 GeV [From L3~\cite{Achard:2002kv}].
        }
\label{fig:L3-TC}
\end{figure}

 For the hadronisation effects, either the Monte Carlo based hadronisation models,
PYTHIA~\cite{Pythia,Sjostrand:2007gs}, HERWIG~\cite{Herwig} and ARIADNE~\cite{Lonnblad:1992tz}
were used, or, alternatively, non-perturbative power-correction formulae derived in
~\cite{Dokshitzer:1995zt,Dokshitzer:1997iz,Dokshitzer:1998pt} were employed. This latter
ansatz provides an additive term to the perturbative QCD estimate in mean event shape
variables. Studying the energy-dependence of these mean variables $\langle f\rangle$,
defined as
\begin{equation}
\langle f \rangle = \frac{1}{\sigma_{\rm tot}} \int f \frac{df}{d \sigma} = 
\langle f_{\rm pert} \rangle + \langle f_{\rm pow}(\alpha_0)\rangle~,
\end{equation}
yielded a measurement of $\alpha_s(\mu)$ and $\alpha_0$, the non-perturbative parameter
characterising the power corrections. We have discussed the  $O(\alpha_s^2)$ calculations of
$f_{\rm pert}$ for several observables (thrust, the Fox-Wolfram shape variable $C$, etc.)
in section 5.4. Explicit formulae for  $f_{\rm pow} $ are given  by
 Dokshitzer {\it et al.}~\cite{Dokshitzer:1995zt,Dokshitzer:1997iz,Dokshitzer:1998pt},
and can also be seen, for example, in the DELPHI analysis~\cite{Abreu:1999rc} of the LEP2
data. For related discussion of event shapes in $e^+e^-$ annihilation and deep inelastic
scattering, we refer to~\cite{Dasgupta:2003iq}.
 A typical measurement along these lines is shown in Fig.~\ref{fig:PLB456-DELPHI},
in which the measured mean values of $\langle 1-T\rangle$, and the scaled heavy jet mass
$\langle M_h^2/E_{\rm vis}^2 \rangle$ are shown as a function of the center-of-mass energy together
with the results of the fits. The dotted curves in these figures show the perturbative
QCD part only. It is obvious from this figure that even at the highest LEP2 energy,
non-perturbative power corrections are not small.  The fits yield $\alpha_s(M_Z)=0.1191
\pm 0.0015 \pm 0.0051$ from $\langle 1-T \rangle$ and a very consistent value from the
other observable $\langle M_h^2/E_{\rm vis}^2 \rangle$. However, the value of $\alpha_0(2~{\rm GeV})$,
the measure of power corrections, differs by about 20\% from the two measurements, 
showing considerable non-universality in the parametrisation of $f_{\rm pow}$
by Dokshitzer {\it et al.}~\cite{Dokshitzer:1995zt,Dokshitzer:1997iz,Dokshitzer:1998pt}.
Along the same lines, the L3 collaboration measured the mean values of several global shape
parameters. For all these variables, the same theoretical
 frameworks~\cite{Catani:1991kz,Catani:1991bd,Catani:1992ua} as discussed 
above in the context of the DELPHI measurements were used. To compare these calculations
at parton level with the experimental measurements, the effects of hadronisation and
decays were corrected for using the JETSET PS (parton shower) Monte Carlo program. We display 
in Fig.~\ref{fig:L3-TC}
the measured distributions in thrust and the variable $C$ at $\sqrt{s}=206.2$ GeV
and comparison with the QCD fits, showing excellent agreement. To determine $\alpha_s$
at each energy, the formalism in ~\cite{Catani:1992ua} is used, which yielded
$\alpha_s(M_Z)=0.1227 \pm 0.0012 \pm 0.0058$~\cite{Achard:2002kv}.
%
%%%%%%%%%%%%%%%%%%%%%%%%%%%%%%%%%%%%%%%%%%%%%%%%%%%%%%%%%%%%%%%%%%%
\subsubsection{Jet rates}
\begin{figure}
\center{
\includegraphics[width=0.75\textwidth, height=6.5cm]{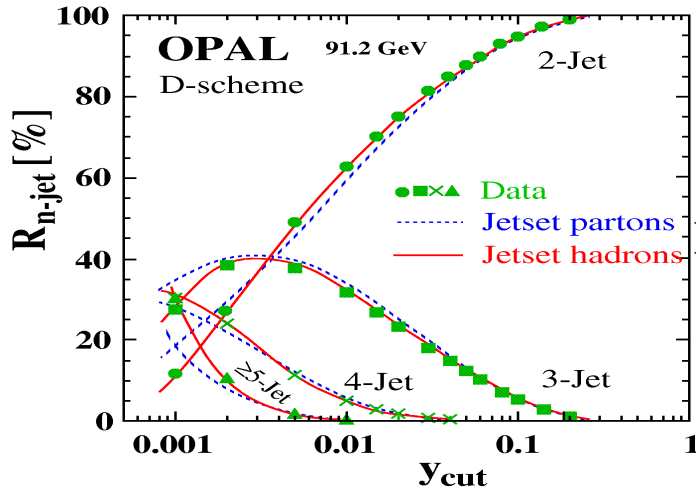} 
\caption{Relative production rates of $n$-jet events defined in the Durham jet
algorithm scheme~\cite{Catani:1991hj} as a function of the jet resolution
 parameter $y_{cut}$. The data are compared to model calculations before
and after the hadronisation process as indicated on the figure~[OPAL\cite{EXPnjets}].}
 \label{fig:OPAL-5jets} 
        }
\end{figure}

\begin{figure}
\centering
\includegraphics[width=0.95\textwidth, height=7.0cm]{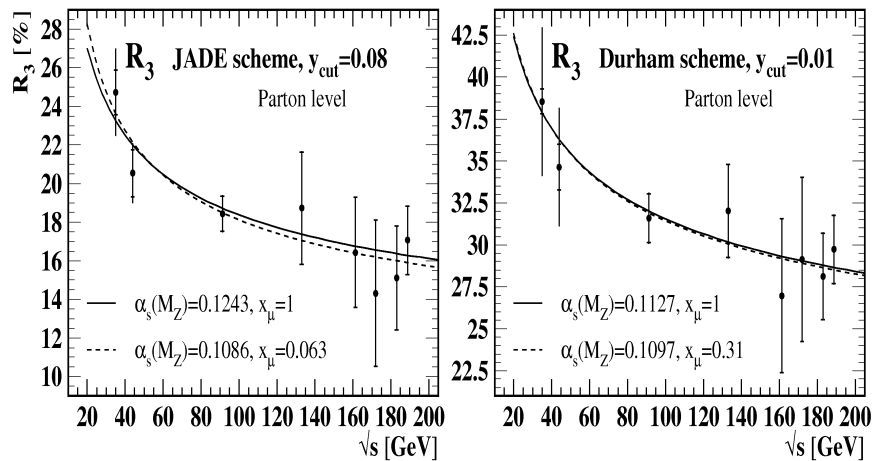} 
\caption{Measured 3-jet fraction as obtained with the JADE (left-hand frame) and
 Durham schemes (right-hand frame) at
parton level versus the c.m.s. energy $\sqrt{s}$. The data shown are from the JADE and OPAL
 collaboration, and the curves are the $O(\alpha_s^2)$ predictions at a fixed scale
(solid lines) and for the fitted values of the scale (dashed lines).
[From OPAL~\cite{Pfeifenschneider:1999rz}.]
        }
\label{fig:OPAL-JADE}
\end{figure}

\begin{figure}
\center{
\resizebox{0.95\columnwidth}{!}{\includegraphics{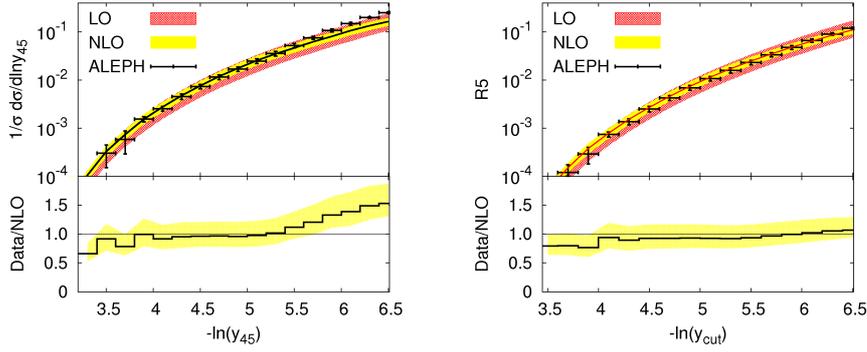}}}
\caption{ALEPH LEP1 data~\cite{Heister:2003aj} compared to leading and next-to-leading order
predictions for $1/\sigma d\sigma/d\ln y_{45}$ plotted against $\ln y_{45}$ (left-hand frame)
 and $R-5$ 
plotted against $\ln (y_{\rm cut})$ (right-hand frame) without hadronisation corrections.
 The uncertainty bands are obtained by varying
the scale in the interval $0.15 M_Z < \mu < 0.6 M_Z $, and the solid lines refer to NLO QCD
 evaluated at $\alpha_s(M_Z)=0.118$ and $\mu=0.3 M_Z$.
 [From Ref.~\cite{Frederix:2010ne}]. 
\label{fig:5jet-FFMZ}}
\end{figure}

\begin{figure}
\center{
\resizebox{0.95\columnwidth}{!}{\includegraphics{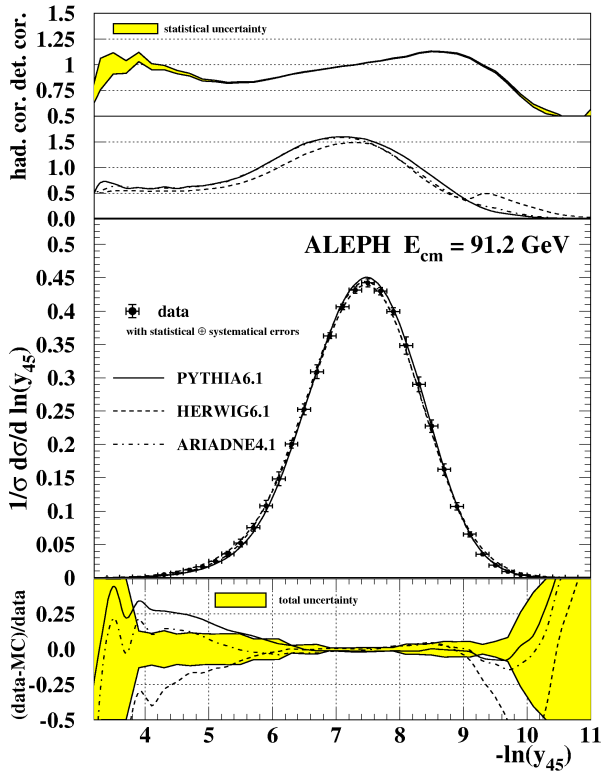}}}
\caption{ALEPH data~\cite{Heister:2003aj} for the $y_{45}$ distribution at LEP1
compared to PYTHIA,HERWIG and ARIADNE Monte Carlo results. The upper frames show 
detector and hadronisation corrections, respectively. The lowest frame shows the relative
 difference between data and event generator predictions. [Figure attributed to H. Stenzel
in Ref.~\cite{Frederix:2010ne}]. 
\label{fig:Aleph-MC}}
\end{figure}

Due to the high energy at LEP1, up to four jets could be resolved 
experimentally. The number of resolved jets depends strongly 
on the criterion by which the jets are defined. Early definitions 
had used the JADE recombination scheme which combined particle pairs 
on the experimental side, and equally quark/gluon parton pairs 
on the theoretical side \cite{Kramer:1986mc}, for scaled invariant masses $M^2_{ij}/s 
= y$ below a cut-off value $\leq y_{cut}$. In the DURHAM scheme~\cite{Catani:1991hj} 
the invariant mass was replaced by $M^2_{ij} = 2 \, \min(E^2_i,E^2_j) \, 
[1 - \cos\theta_{ij}]$, essentially the transverse momentum between 
the particles or partons for small angles. The cut-off value $y_{cut}$ was 
chosen typically from $10^{-1}$ down to $10^{-3}$. Small values of $y_{cut}$
naturally lead to large numbers of jets while the number of jets
is reduced if $y_{cut}$ is increased. The cross section for 3-jet events,
\begin{equation}
\sigma_3[y] = \left( \frac{\alpha_s}{\pi} \right) \sigma_{31} +
              \left( \frac{\alpha_s}{\pi} \right) ^2 \sigma_{32} + 
              \left( \frac{\alpha_s}{\pi} \right) ^3 \sigma_{33}    \,,
\end{equation}
has been calculated up to third order in the QCD coupling \cite{Dissertori:2009qa}. 
NLO corrections to the four-jet rates in the process
$e^+ e^- \to \gamma^*,Z \to 4$ jets were done around 1996 by
Dixon and Signer~\cite{Signer:1996bf,Dixon:1997th} and subsequently by
 Nagy and Trocsanyi~\cite{Nagy:1998bb}.
The experimental number of jets in $Z$ decays is displayed in Fig.~\ref{fig:OPAL-5jets}
and compared with a parton shower Monte Carlo prediction (Jetset partons), and including
hadronisation effects (Jetset hadrons).
Evidently, for $y_{cut}$ below $10^{-2}$ up to four jets can clearly be identified
at LEP1 \cite{EXPnjets}.
A dedicated effort to test QCD and determine $\alpha_s$ was undertaken by the 
combined JADE (at PETRA) and OPAL (at LEP) collaborations~\cite{Pfeifenschneider:1999rz},
giving a considerably larger lever arm in energy from 35 GeV to 189 GeV. The observables used
in this (JADE + OPAL) analysis are exclusively based on the multiplicities of hadronic jets. 
The $n$-jet fractions, $R_n$ were defined using the JADE~\cite{JADE},
 Durham~\cite{Catani:1991hj} and 
Aachen/Cambridge\cite{Wobisch:1998wt} algorithms. We show in Fig.~\ref{fig:OPAL-JADE} 
the three-jet
fraction $R_3$ obtained with the JADE and Durham jet algorithms versus the
 center-of-mass energy.
(The result from the Aachen/Cambridge algorithm  can be seen
 in~\cite{Pfeifenschneider:1999rz}.)
Here, the data from PETRA and LEP are compared with the $O(\alpha_s^2)$ prediction.
 The renormalisation scale dependence is shown by the scale parameter $x_\mu= \mu/\sqrt{s}$,
with the solid lines corresponding to a fixed value $x_\mu=1$, and the dashed lines are the
results obtained with a fitted scale, indicated on the figure. This and related analyses
reported in~\cite{Pfeifenschneider:1999rz} yield a rather precise value for the QCD
coupling constant $\alpha_s(M_Z)=0.1187 ^{0.0034}_{-0.0019}$.  
 At LEP2 (up to $\sqrt{s}=206$ GeV), the highest jet multiplicity 
measured is five, obtained using the variable $y_{\rm cut}$, and inclusive measurements
are available for up to six jets.
To match this data, NLO QCD corrections to five-jet production at LEP have  been carried
out by Frederix {\t et al.}~\cite{Frederix:2010ne}, and the fixed-order perturbative
 results have been compared with the LEP1 data from ALEPH~\cite{Heister:2003aj}.
 Two observables have been used for this comparison:\\
 (i) Differential distribution with respect to the five-jet
 resolution parameter
$y_{45}$, the maximum value of $y_{\rm cut}$ such that a given event is classified as
a five-jet event by the DURHAM jet algorithm~\cite{Catani:1991hj}:  
\begin{equation}
\int_{y_{\rm cut}}^1 dy_{45} \frac{d\sigma}{dy_{45}}=
\sigma_{\rm incl}^{5-{\rm jet}} (y_{\rm cut})~,
\label{eq:sigma-5-jet-incl}
\end{equation}
where $\sigma_{\rm incl}^{5-{\rm jet}}$ is the {\rm inclusive} five-jet production cross
section in $e^+e^-$ annihilation. (ii) Five-jet rate $R_5(y_{\rm cut})$, defined as follows:
\begin{equation}
R_5(y_{\rm cut}) = \frac{\sigma_{\rm excl}^{\rm 5-jet}(y_{\rm cut})}{\sigma_{\rm tot}}~,
\label{eq:sigma-5-jet-excl}
\end{equation}
where $\sigma_{\rm excl}^{\rm 5-jet}(y_{\rm cut}) $ is the exclusive five-jet production 
cross section. This is also calculated using the Durham jet algorithm by requiring that
exactly five jets are reconstructed. Both observables,
 $\sigma_{\rm tot}^{-1} d\sigma/d\ln y_{45}^{-1}$ and $R_5(y_{\rm cut})$,  can be written as
a series in $\alpha_s(\mu)$, with the leading contributions starting in
$O(\alpha_s^3)$. A comparison of the leading order and next-to leading order predictions 
 for $(1/\sigma) d\sigma/d\ln y_{45}$ vs. $\ln (y_{45})$ and the exclusive 5-jet fraction
$R_5$ vs. $\ln (y_{\rm cut})$ is shown in Fig.~\ref{fig:5jet-FFMZ}  with the
ALEPH data in a limited range of these variables (perturbative regime). Hadronisation
 effects have been estimated
using the SHERPA MC~\cite{Sherpa}. As is typical of NLO calculations, scale dependence is
significantly reduced compared to the LO calculations.
 Agreement between data and NLO theory is impressive and has been used to
extract  a value of $\alpha_s$, obtaining $\alpha_s(M_Z)=0.1156^{+0.0041}_{-0.0034}$, which is
in excellent agreement with the world-average discussed below.
 The limitation of using fixed-order perturbative QCD in describing the $e^+e^-$ data can
be seen in the ALEPH data shown in Fig.~\ref{fig:Aleph-MC}, which show a characteristic 
turnover shape around $-\ln y_{45} \simeq 7.5$. In this region, perturbation theory fails
and a resummation (equivalently showers) have to be included to describe the data. This
underscores the importance of having MC generators which include showers.

 In Fig.~\ref{fig:Aleph-MC}, the ALEPH LEP1 data for
 $\sigma_{\rm tot}^{-1} d\sigma/d\ln y_{45}^{-1}$ are compared with the hadron level
 predictions of three event generators (the numbers denote the various versions of
these Monte Carlo programmes),
 PYTHIA6.1~\cite{Sjostrand:2006za}, HERWIG6.1~\cite{Corcella:2000bw}
and ARIADNE4.1~\cite{Lonnblad:1992tz}. Agreement between data~\cite{Heister:2003aj} and
these MCs is generally good.  However, as
shown in the upper frame of fig.~\ref{fig:Aleph-MC}, hadronic corrections are large, varying
from 0.5 to 1.5 in this range. In addition, differences between hadronisation corrections
among the MCs are as large as 25\%.  This deficiency can be overcome to some extent by matching the parton
 shower and  high multiplicity matrix elements, as, for example, proposed
 in~\cite{Catani:2001cc}. This matching procedure has been implemented in the
SHERPA event generator~\cite{Sherpa} and results in improved agreement between the
MC and fixed-order perturbative description.

\subsubsection{The gluon self-coupling} 
The study of the three-gluon coupling in gluon splitting to two gluons 
requires four (or more) jets in $e^+ e^-$ annihilation. A variety of angular correlations 
and energy distributions, see~\cite{THchi4}-\cite{jet42}, can be exploited 
to signal the three-gluon coupling of QCD.

The sensitivity to angular distributions may be illustrated in a transparent 
example~\cite{THchi4}. A virtual gluon, radiated off the quarks 
in the process $e^+ e^- \to q \bar{q} g^\ast$, is polarised preferentially
in the production plane. The subsequent splitting of the virtual gluon 
into two real gluons or a quark-antiquark pair is sensitive to the azimuthal
angle $\phi$ between the $g^\ast$ polarisation vector and the decay planes:
\begin{eqnarray}
n_{gg}       &=& \frac{[1-z+z^2]^2}{z(1-z)}
               + z(1-z) \, \cos 2 \phi~,                                 \nonumber \\
n_{q\bar{q}} &=& \frac{1}{2} \, \left[ z^2 + (1-z)^2 \right] 
               - z(1-z) \, \cos 2 \phi                                 \,.
\end{eqnarray}
As a result, the polarisation vector and the decay plane tend to be aligned
in gluon splitting to two gluons. In contrast, if the gluon splits 
to a quark-antiquark pair, the decay plane tends to orient itself perpendicular 
to the polarisation vector.

\begin{figure}
\center{
\resizebox{0.75\columnwidth}{!}{
\includegraphics[width=0.75\textwidth]{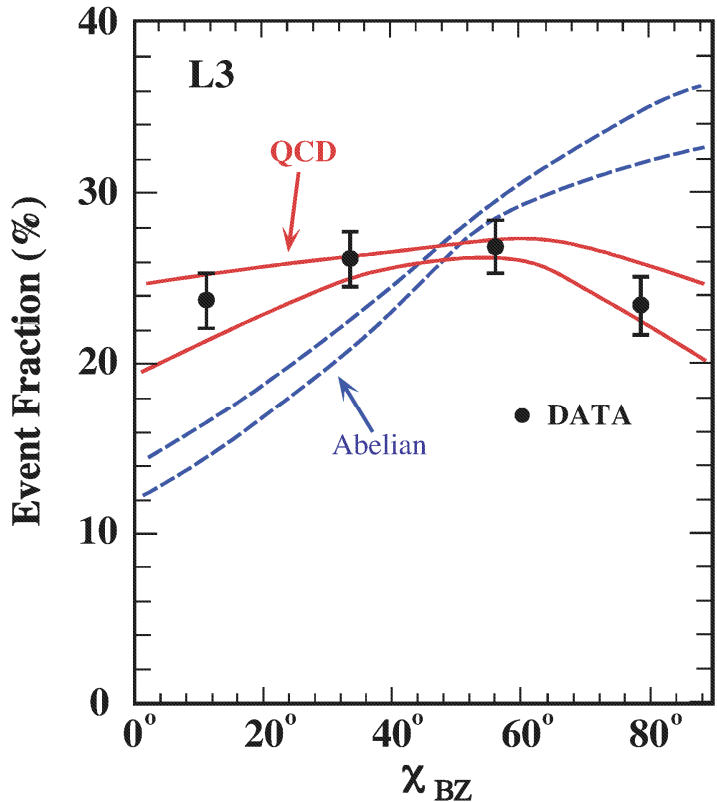}}}
\caption{The distribution  of the azimuthal angle $\chi_{\rm BZ}$ (called $\phi$ in Eq.~(59))
         between the planes formed by the two most 
         energetic jets and the two least energetic jets~\cite{EXPchi4}; 
         the predictions of QCD including the gluon self-coupling are 
         compared with an abelian theory without self-coupling of the 
         gauge fields.
\label{fig:BZ-L3} 
}
\end{figure}

The azimuthal distribution can be studied experimentally by measuring the
angle between the planes spanned by the two hardest and the two softest jets.
In an abelian theory the $\phi$ asymmetry is large and the two planes
would orient themselves perpendicular to each other. By contrast,
since the $\phi$-independent term in gluon splitting to two gluons in QCD 
is large, the azimuthal asymmetry in this process is predicted to be small, 
but the two planes should have a tendency to orient themselves parallel 
rather than perpendicular. This is borne out by experimental analyses~\cite{EXPchi4} 
indeed, as demonstrated in Fig.~\ref{fig:BZ-L3}.   

Quite generally, four jets are produced in $e^+ e^-$ annihilation by
three mechanisms: double gluon bremsstrahlung, gluon splitting to two
gluons, and gluon splitting to a quark-antiquark pair. The cross
section \cite{Ali:1979rz,Ali:1979wj} can be decomposed accordingly:
\begin{equation}
\sigma_4 = \left( \frac{\alpha_s}{\pi} \right) ^2 \, C_F \,
           [C_F \sigma_{bb} + C_A \sigma_{gg} + n_f T_R \sigma_{qq}] \,.
\end{equation}
The first term accounts for double gluon bremsstrahlung $q \to qg$
and $\bar{q} \to \bar{q} g$, the second
for gluon splitting to two gluons $g \to gg$, the third for gluon splitting to 
$n_f$ quark pairs $g \to q \bar{q}$. The Casimir group characteristics 
of the splitting vertices are $[C_F,C_A,T_R] = [4/3,3,1/2]$ in QCD,
while the corresponding characteristics are [1,0,3] in an abelian theory. 
Measurements of their ratios yield \cite{EXPgroup}
\begin{eqnarray}
C_A / C_F &=& 2.29 \pm 0.06 [stat.] \pm 0.14 [syst.]~, \nonumber \\       
T_R / C_F &=& 0.38 \pm 0.03 [stat.] \pm 0.06 [syst.]  \,,
\end{eqnarray}
compared with the theoretical predictions $C_A / C_F = 9/4$ and $T_R / C_F = 3/8$
in QCD. Again we observe a strong signal of the three-gluon coupling in $C_A$, 
far away from zero in the abelian theory.

\subsubsection{QCD coupling and asymptotic freedom}
A large range of energies can be covered in the measurement of the
QCD coupling $\alpha_s(Q^2)$, extending from 29 GeV (at PEP) to 46 GeV (at PETRA) up to
about 206 GeV at LEP; the lever arm can be elongated down to
1.8 GeV by including $\tau$ decays.    

The QCD prediction for the running coupling $\alpha_s(Q^2)$ has been
determined up to the 5th power \cite{Czakon,Ritbergen-97}. Keeping terms up to 2nd order leads to
the following expression 
\begin{equation}
\alpha_s(Q^2) = \frac{1}{\beta_0 \log \left( Q^2/\Lambda^2 \right) }
                 - \frac{\beta_1 \log \log \left( Q^2/\Lambda^2 \right)}
                   {\beta^3_0 \log^2 \left( Q^2/\Lambda^2 \right) } + ...,
\end{equation}
with $\beta_0 = (33-2 n_f)/12\pi$, $\beta_1 = (153-19 n_f)/24\pi^2$, ... and
$\Lambda \approx 200$ MeV denoting the QCD scale at which the coupling grows 
indefinitely.  
An ensemble of observables has been calculated in perturbative expansions
in next-to-leading order NLO up to N$^3$LO. Most accurate are the totally 
inclusive observables, like total cross sections, followed by jet cross 
sections and hadronic shape variables, like thrust. The estimates still
depend significantly on the models  used for calculating the shape variables
in the non-perturbative region, see~\cite{Gehrmann2,Dissertori-2009}, for instance.

\begin{figure}
\center{
\includegraphics[width=0.48\textwidth, height=7.5cm]{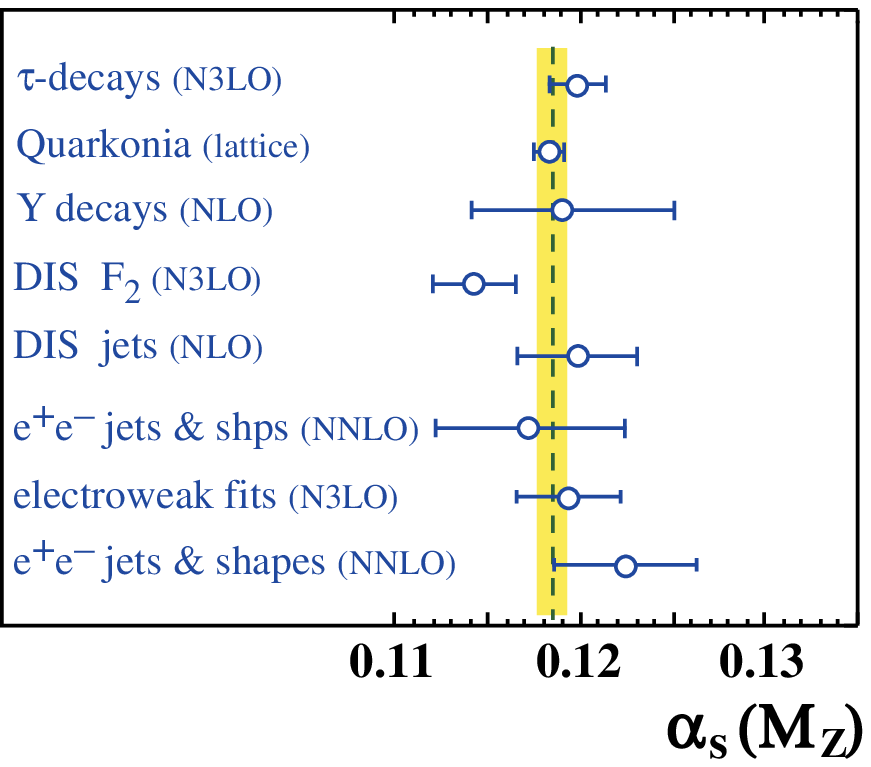} 
$\;\;\;$
\includegraphics[width=0.48\textwidth,height=7.5cm]{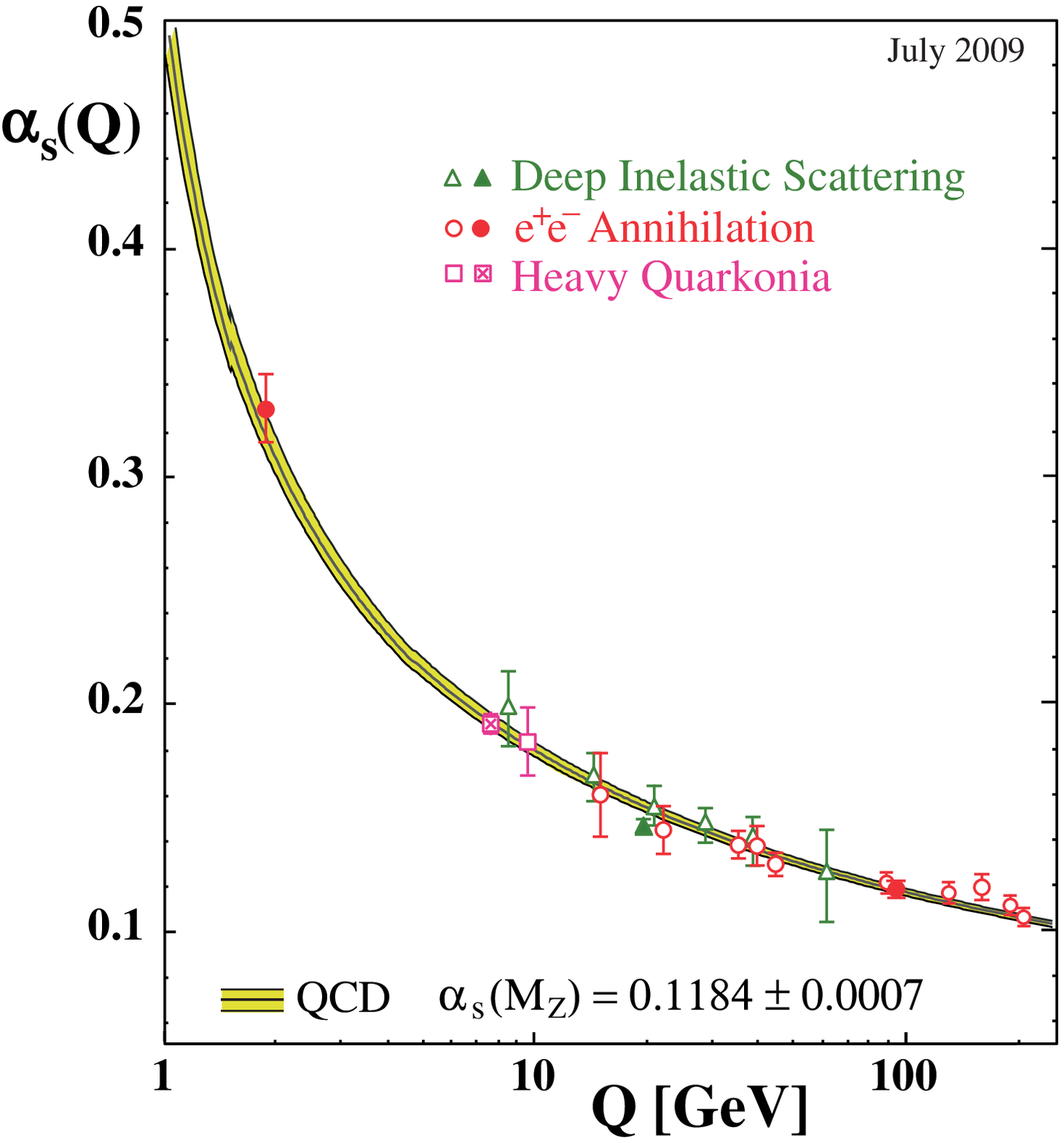}
\caption{Left-hand frame: Summary of measurements of $\alpha_s(M_Z)$, used as input
for the world average value.
Right-hand frame: Evolution of the QCD coupling with energy.
         Among other methods, analyses based on hadron production in
         $e^+ e^-$ annihilation play a leading role up to the highest 
         LEP energies. Both frames  are from Bethke~\cite{Basy}.
        }
\label{fig:3.2}
}
\end{figure}

Combining the experimental measurements with the theoretical apparatus,
the knowledge of the QCD coupling and its evolution with the energy is
summarised in Fig.~{\ref{fig:3.2}}. The lever arm extends from the 
hadronic decays of the $\tau$ lepton throughout the PETRA range up to the highest 
energy values in the second phase of LEP. Including deep-inelastic 
lepton-nucleon scattering and jet production in hadron collisions, 
all the analyses are in remarkable agreement with the theoretical 
expectation from asymptotic freedom \cite{asyfree}. It has become customary to quote
the value of $\alpha_s(\mu)$ measured in experiments by scaling the result to the scale
 $\mu=M_Z$ using the RG equation. This yields the current world
average~\cite{Basy} 
\begin{equation}
\alpha_s(M_Z)=0.1184 \pm 0.0007~.
\end{equation}  

This ends our discussion of jets in $e^+e^-$ annihilation experiments and in QCD.
In summary,  essential parts of QCD jets can now be controlled at the level 
of typically ten percent ($\alpha_s(M_Z)$ is known better than 1\%). {\it Notabene} the basic interactions and the strength 
of the quark-gluon coupling are proven to be asymptotically free. The high level of accuracy
achieved in measuring the gauge couplings - weak, electromagnetic and QCD - is now a diagnostic
tool to probe physics at  scales as high as the the grand unification 
scale. 
\section{Jets as tools}
In the preceding sections, we discussed  the impact which jet physics in
$e^+e^-$ annihilation experiments
had in establishing QCD quantitatively. This progress owes itself to some extent to the fact
that in $e^+e^-$ annihilation the initial state is precisely known.
 This is not the case 
in other high energy collisions, such as the electro- and photoproduction
processes $ep$ and $\gamma p$, as well as the gamma-gamma and hadron hadron
collisions, $\gamma \gamma$, $p\bar{p}$ and $pp$. Here, jets
 could be used as  powerful tools for studying other aspects of high energy
collisions. Examples are the partonic  composition of the proton, i.e., quark
and gluon densities of the proton (and antiproton), the parton distribution functions
(PDFs) of the photon and the QCD coupling at HERA,  Tevatron and the LHC. Yet other
applications of jet physics 
include analyses of the electroweak sector and searches for new heavy
particles in many extensions of the Standard Model (SM) -- the QCD and
electroweak theory of particle physics based on the groups
$SU(3)_c\otimes SU(2)_I\otimes U(1)_Y$ Thus 
the prominent role of jets in studying QCD phenomena extends to 
quite different areas in particle physics.

Before we embark upon  illustrating the use of jets as tools in $ep$,
 $\gamma \gamma$, $p\bar{p}$ and $pp$ collisions, it is worth pointing out that
in these processes, QCD is at work in both the initial and the final states as opposed
to the $e^+e^-$ annihilation processes, where it influences only the final state
distributions and rates. As seen in Fig.~\ref{fig:feynborn} for the DIS process, 
the cross section depends on three components: (i) the probability of finding a
parton in the proton having a fractional longitudinal momentum $x$ (or $ x_{\rm Bj}$),
(ii) the interaction between these partons and the virtual photon, and (iii) the transition
of partons to jets, which theoretically involved the recombination of two partons into
one jet. While perturbative QCD provides a framework
to evolve the PDFs and the fragmentation functions FFs from a low scale $\mu^2=Q_0^2$ to a
high scale $\mu^2=Q^2$, non-perturbative inputs for the PDFs and FFs are required at the
lower scale. This is obtained by parametrising an ansatz at lower scale. The theoretical
tool which enables this is called factorization, a key concept in the application
of QCD to high energy processes~\cite{Collins:1989gx}. Simply stated, factorization of a process (such as
inclusive- or jet-cross sections in deep inelastic scatterings) allows it to be
expressed as the product of a short-distance part, calculable in perturbative QCD, and 
a long distance part, comprising of non-perturbative matrix elements or PDFs.
The universality of the PDFs and the evolution from a lower scale to a higher scale are
process-independent. The
division into the short- and long-distance parts is governed by the factorization scale
$\mu_F$. We illustrate the applications of factorization on the example of deep inelastic
scattering processes, discussed below.
\subsection{$ep$ Collisions}
In DIS, described here by the process $e p \to e +X$, an electron $e$ with momentum
$k$ emits an off-shell photon with momentum $q$ which interacts with a proton of momentum 
$P$. For virtualities of the photon ($Q^2=-q^2>0$) far above the squared proton mass (but
far below the $Z$-boson mass), the differential cross section in terms of the kinematic
variables $Q^2$, $x=Q^2/(2P.q)$ and $y=(q.P)/(k.P)$ is
\begin{equation}
\frac{d^2\sigma}{dx dQ^2} = \frac{4\pi \alpha}{2xQ^2}\left[(1+(1-y)^2)F_2(x,Q^2) -
y^2F_L(x,Q^2)\right]~,
\end{equation}
where $F_2(x,Q^2)$ and $F_L(x,Q^2)$ are proton structure functions, which encode the
interaction between the photon and the proton. The structure functions are not calculable
in perturbative QCD. In the lowest order, i.e., keeping only the Born contribution
as shown in Fig.~\ref{fig:feynborn} (a) where $x=x_{\rm Bj}$,  the structure functions
 are given by
\begin{equation}
F_2(x,Q^2)=x\sum_q e_q^2 f_{q/p}(x)~: ~~F_L(x,Q^2)=0~,
\end{equation}
where $f_{q/p}(x)$ is the PDF for quarks of type $q$
 inside the proton. This result in which $f_{q/p}(x)$ are independent of the
scale $Q$ is the ``quark-parton model'' picture. Hence, in this picture,
also the structure functions $F_2$ and $F_L$ are independent of $Q^2$. Including 
higher order perturbative QCD corrections, the structure function $F_2(x,Q^2)$ has
the form~\cite{Dissertori:2010}
\begin{equation}
F_2(x,Q^2)=x\sum_{n=0}^\infty \frac{\alpha_s^n(\mu_R^2)}{(2\pi)^n}\int_x^1 \frac{dz}{z}
C_{2,i}^{(n)}(z,Q^2,\mu_F^2,\mu_R^2)f_{i/p}(\frac{x}{z},\mu_F^2)~.
\end{equation}
Here we have a series in powers of $\alpha_s(\mu_R^2)$, each term involving a coefficient 
$C_{2,i}^{(n)}$, which can be calculated using Feynman graphs. The scale $\mu_R$ is
called the renormalization scale at which the QCD coupling constant $\alpha_s(\mu_R^2)$
is calculated. The leading order
in $\alpha_s$ QCD Compton scattering and the boson-gluon fusion contributions are shown in
Fig.~\ref{fig:feynborn} (b) and \ref{fig:feynborn} (c), respectively. An important point to note is that
the momentum of the quark when it interacts with the photon and the momentum when
the quark is extracted from the proton may differ. The ratio of these two momenta is
$z$. The $C_{2,i}^{(n)}$ coefficients depend on the ratio $z$, and hence one has to
integrate over $z$ as indicated above. In lowest order, i.e. without including any
perturbative QCD corrections, one has $C_{2,q}^{(0)}=e_q^2\delta(1-z)$ and $C_{2,g}^{(0)}=0$,
and one recovers the ``quark-parton model'' result.

The PDFs $f_{i/p}(\frac{x}{z},\mu_F^2)$ 
depend on  the factorisation scale $\mu_F^2$, and this dependence is governed by the
DGLAP equation. In leading order in $\alpha_s(\mu_F^2)$, this reads as follows
\begin{equation}
\frac{\partial f_{i/p}(x,\mu_F^2)}{\partial \mu_F^2}=\sum_j\frac{\alpha_s(\mu_F^2)}{2\pi}
\int_x^1 \frac{dz}{z} P_{i \to j}^{(1)}(z)f_{j/p}(\frac{x}{z},\mu_F^2)~,
\end{equation}  
where, for example, $P_{q\to g}^{(1)}(z)=T_R(z^2 + (1-z)^2)$, and the others can be
extracted from Eq.~(\ref{eq:dglap}). The coefficient functions are also $\mu_F$-dependent,
and in the leading order in $\alpha_s$, one has 
$C_{2,i}(x,Q^2,\mu_R^2, \mu_F^2)= C_{2,i}(x,Q^2,\mu_R^2, Q^2) -\ln(\frac{\mu_F^2}{Q^2})
\sum_j\int_x^1 \frac{dz}{z} P_{i \to j}^{(1)} C_{2,j}^{(0)}(\frac{x}{z})$. In the
above expressions, the choice of the renormalization and factorization scales is
arbitrary. Varying $\mu_F$ and $\mu_R$ provides an estimate of the scale-dependent
uncertainties. In inclusive DIS processes, the default choice is $\mu_R=\mu_F=Q$. 

The extension of the factorization formalism to the processes with two initial-state
hadrons follows very much along the same lines, and we shall discuss this somewhat
later, as we discuss high energy hadronic collisions.

\subsubsection{Jets in DIS processes}
\begin{figure}
\begin{center}
\resizebox{0.75\columnwidth}{!}{
 \includegraphics{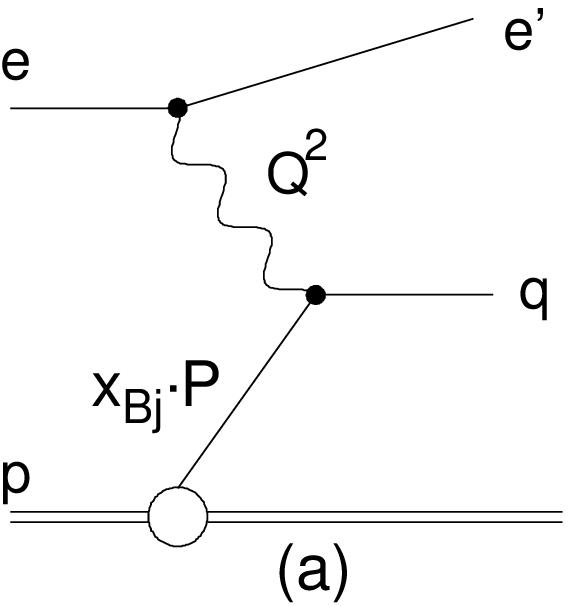}\hskip1.0cm
 \includegraphics{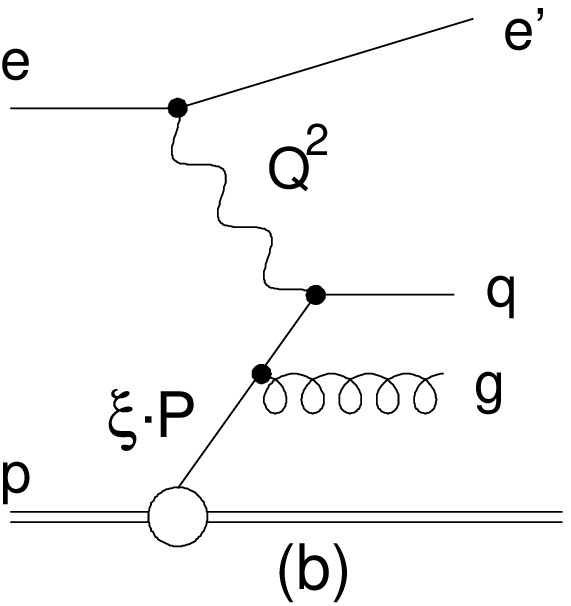}\hskip1.0cm
 \includegraphics{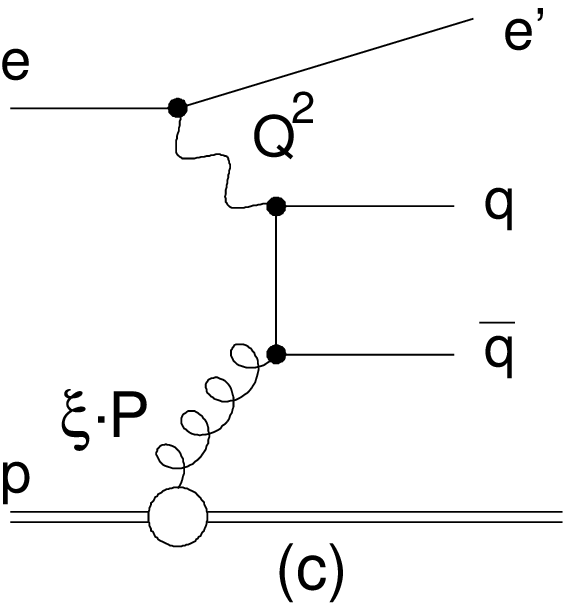}}
\caption{
Deep-inelastic lepton-proton scattering at different orders in
$\alpha_s$: (a) Born contribution $\mathcal{O}(1)$, (b) example of the QCD Compton
    scattering $\mathcal{O}(\alpha_s)$ and (c) boson-gluon fusion $\mathcal{O}(\alpha_s)$.}
\label{fig:feynborn}
\end{center}
\end{figure}

Jet production in neutral
current (NC) deep inelastic scattering at the HERA collider at DESY
provides an important further testing ground for QCD. While inclusive DIS 
 gives  indirect
information on the strong coupling via scaling violations of the proton
structure functions, the production of jets allows one a direct measurement of
the strong coupling constant $\alpha_s$. 
The Born contribution to DIS (Fig.~\ref{fig:feynborn}a) generates no transverse 
momentum in the $\gamma^{*}p$ centre-of-mass frame, where the virtual
boson and the proton collide head on. Significant transverse momentum in
the $\gamma^{*}p$ frame is produced at leading order (LO) in the strong 
coupling $\alpha_s$ by the QCD-Compton (Fig.~\ref{fig:feynborn}b) and the photon-gluon 
fusion (Fig.~\ref{fig:feynborn}c)  processes.
In LO the momentum fraction of the proton carried by the parton 
is given by $\xi = x_{Bj}(1+M_{12}^2/Q^2)$, where $x_{Bj}$ is the 
Bjorken scaling variable $x_{Bj}=Q^2/(Q^2+W^2)$. Here $W$ is the total 
c.m. energy $W^2=(q+P)^2$, $q=$ is the momentum of the virtual photon, $M_{12}$ is
 the invariant mass
of two jets of highest $p_T$ and $Q^2$ is the negative four-momentum
transfer squared of the ingoing and outgoing electron. In the kinematic
region of low $Q^2$ and low $\xi$, the $\gamma^{*}$-gluon
fusion dominates the jet production imparting sensitivity to the gluon
component of the parton density functions (PDFs) of the proton, whereas the
contribution of the QCD-Compton process yields information on the various
quark (antiquark) components  of the proton.

\begin{figure}
\begin{center}
\resizebox{0.75\columnwidth}{!}{\includegraphics{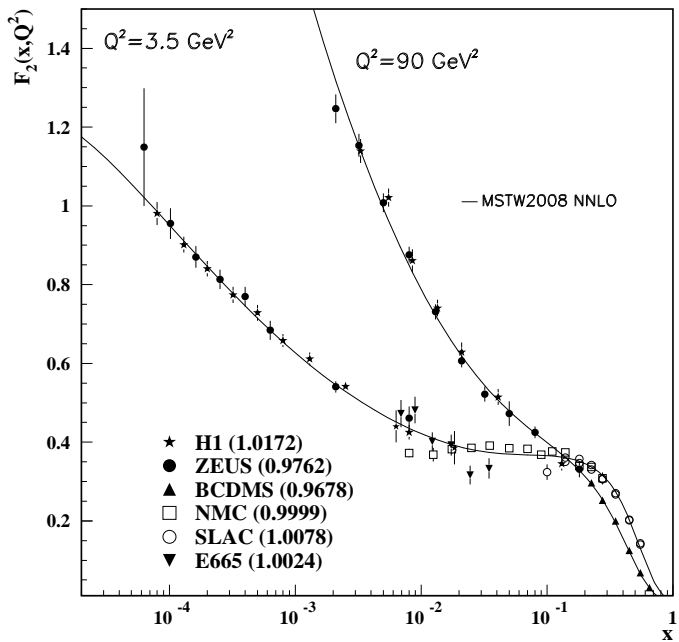}}
\caption{
Proton structure function $F_2^p(Q^2,x)$ given at two $Q^2$ values (3.5 GeV$^2$
and 90 GeV$^2$). The various data sets have been normalised by the factors
shown in the brackets in the key to the plot, which were determined in the
NNLO MSTW2008 global analysis~\cite{MSTW}. (From Amsler {\it et al.} in Ref.~\cite{Amsler}).}
\label{fig:PDG16-2}
\end{center}
\end{figure}

In order to make theoretical predictions on jet productions in
neutral current DIS scattering one needs the PDFs of the proton, provided
mostly by the global analysis collaborations (i.e., analysis in which all
high energy physics measurements relevant for testing QCD and determination of 
various parameters and non-perturbative functions are undertaken), such as 
 CTEQ \cite{CTEQ} and MSTW \cite{MSTW}.  In addition one must have infrared and 
collinear safe parton cross sections, which are known now up to NLO in 
$\alpha_s$~\cite{Catani,Nagy-2001}. An example of the inclusive DIS 
measurements at HERA together with data from lower energy fixed target experiments 
 is shown in  Fig.~\ref{fig:PDG16-2}.  A striking feature of the
 HERA data is the dramatic rise of
the proton structure function $F_2(x,Q^2)$ for increasing $Q^2$ and 
low values of $x$ (typically $x \leq 10^{-2}$).
Almost all of this rise of $F_2(x,Q^2)$is due to the gluon density in the proton.
 This  has profound consequences for high energy $p\bar{p}$ (at the Tevatron)
 and $pp$ collisions (at the LHC).

\begin{figure}
\begin{center}
\resizebox{0.85\columnwidth}{!}{
\includegraphics{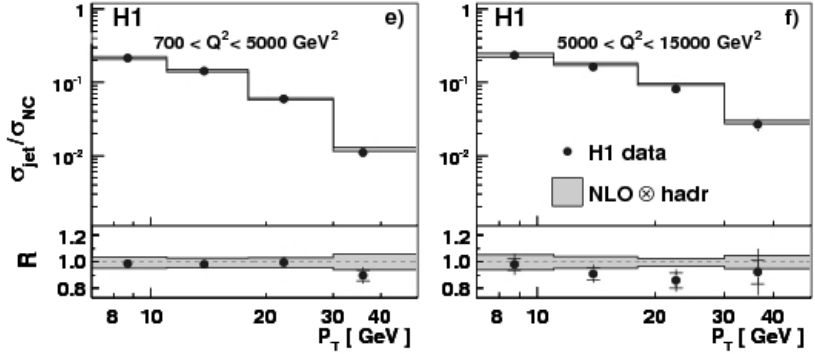}}
\end{center}
\caption{\label{fig:Ijet_Q2ET}The normalised inclusive jet cross
sections measured as a function of the jet transverse momentum in
the Breit frame $P_{T}$ in two regions of $Q^2$ indicated on the frames.
The points are shown at the  average value of $P_T$ within each bin.
(From H1 Collaboration~\cite{H1I}).}
\end{figure}

 DIS jet production 
depends in general on two large scales $Q=\sqrt{Q^2}$ and the $p_T$ of the 
produced jets.
The $ep$ jet data were collected by two detectors at HERA: H1 and ZEUS, resulting from the
collision of electrons or positrons with energy $E_e=27.6$ GeV with protons of energy
$E_p=920$ GeV, providing a center-of-mass energy of $\sqrt{s}\simeq 320$ GeV. 
%In the
%older analysis the produced jets were defined by the cone algorithm.
 In the more
recent analysis the inclusive $k_T$ jet algorithm \cite{Ellis} is used to
combine the particles in the hadronic final state into jets.
 Theoretical predictions (at next-to-leading order) have been corrected
for hadronisation effects, which are
calculated via Monte Carlo Models  with parton showers.
The most recent publication on jet production in DIS comes from the H1
collaboration at HERA, where data up to 2007 are included and $Q^2$ spans the
range $150<Q^2<15000~GeV^2$ \cite{H1I}.
Inclusive jet, 2-jet and 3-jet cross sections, normalised to the neutral current (NC) deep
inelastic scattering cross section, are measured as functions of $Q^2$,
jet transverse momentum and proton momentum fraction. We show
in Fig.~\ref{fig:Ijet_Q2ET} the normalised inclusive jet cross section as a function of
the jet transverse momentum $p_T$ in the Breit frame (defined as the frame in which
$2x \vec{p} + \vec{q}=0$, where $\vec{p}$ and $\vec{q}$ are the 3-momenta of the
proton and virtual photon, respectively) for two ranges of $Q^2$,
$700 < Q^2 < 5000$ GeV$^2$ (shown in the left-hand frame) and   $5000 < Q^2 < 15000$~GeV$^2$
(shown in the right-hand frame).
Agreement between data and theoretical predictions~\cite{Catani,Nagy-2001} is
excellent. The ratio R (of data over theory) lies  near 1 (shown at the bottom of
these frames). Similar
plots for bins with smaller $Q^2$ can be seen in \cite{H1I}. HERA data on inclusive jet
production in DIS~\cite{Chekanov:2002be,Chekanov:2006xr,:2007pb}
have constrained the gluon density in the range $0.01 < x < 0.1 $
The strong coupling $\alpha_s(Q^2)$ has been determined 
and  translated into $\alpha_s(M_Z) =0.1168\pm0.0007({\rm exp.}) 
^{+0.0046}_{-0.0036}({\rm theor.}) \pm 0.0016({\rm PDF})$ using the usual renormalisation 
group equation. This result is competitive with those from $e^+e^-$
data and is in good agreement with the world average \cite{Amsler}.

\begin{figure}
\begin{center}
\resizebox{0.85\columnwidth}{!}{
\includegraphics{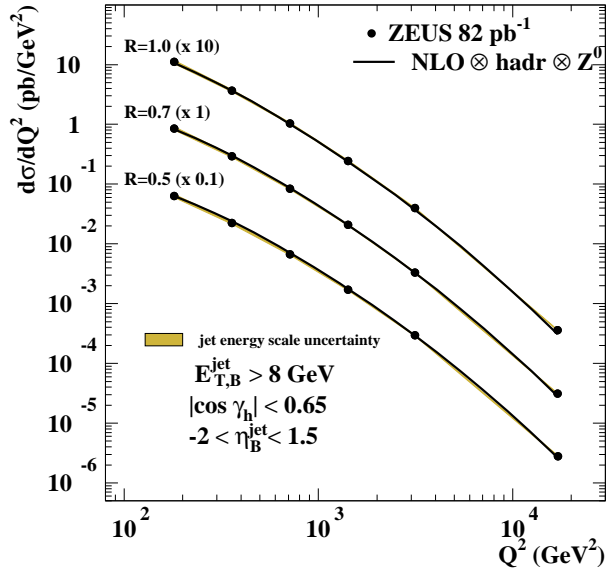}}
\caption{\label{fig:Diff-Q2}
Measured differential cross-section $d\sigma/d Q^2$ for inclusive-jet 
production with $E_{T,B}^{\rm jet} >8$~GeV and $-2<\eta_{\rm B}^{\rm jet}<1.5$ (dots)
for different jet radii, in the kinematic range given by $|\cos \gamma_h|<0.65$.
(From ZEUS Collaboration~\cite{ZEUS1}). 
}
\end{center}
\end{figure}

A similar recent analysis of the ZEUS collaboration has less
integrated luminosity, as it is based only on the data taken from 1998-
2000. However, their data include also results for rather large $Q^2$.
Therefore, $Z$ exchange is included in addition to the $\gamma$-exchange.
The analysis is done in a similar fashion as that of the H1 collaboration
described above. They also studied the inclusive one-jet cross section as a
function of $Q^2$ and $E^{jet}_{T,B}$ (the transverse energy of the jet in the
Breit frame). In addition they also measured this cross section for three
different radii, $R=0.5,0.7$ and $1.0$, used for combining hadrons into
jets with the help of the inclusive $k_T$ cluster algorithm~\cite{Ellis,Catani-algo}. The results are shown in
 Fig.~\ref{fig:Diff-Q2} for 
$d\sigma/dQ^2$ for $E_{T,B}^{jet} > 8$ GeV as a function of $Q^2$.
Further kinematic constraints, namely $|\cos \gamma_h| < 0.65$, where $\gamma_h$ 
corresponds to the angle of the scattered quark in the quark-parton model and is
constructed using the hadronic final state, and the pseudorapidity range
 $-2 < \eta_B^{\rm jet} <1.5$, are indicated on the figure. The pseudorapidity variable,
defined as $\eta=-\ln \tan(\theta/2)$ where $\cos \theta=p_z/|p|$, is approximately
equal to the (longitudinal-boost-invariant) variable called rapidity
 $y=\frac{1}{2}\ln(\frac{E+p_z}{ E-p_z})$ in the limit $|p| \gg m$, and can be measured
when the mass $m$ and the momentum of the particle $|p|$ are unknown.
The NLO QCD predictions with scales $\mu_R=E_{T,B}^{jet}$, $\mu_F=Q$,
corrected to include hadronisation and $Z$ effects,
are compared with the measurements~\cite{ZEUS1}. 
 The calculations reproduce the measured 
differential cross section quite well for all three jet radii considered. 
In this work also $\alpha_s(Q^2)$ has been determined. The result is 
$\alpha_s(M_Z)=0.1207\pm0.0014({\rm stat.}) ^{+0.0035}_{-0.0033}({\rm exp.})
^{+0.0023}_{-0.0023}({\rm theor.})$, which is also consistent with the world average.

We now turn to photoproduction.
At HERA the largest cross section is due to photoproduction, where the beam
(electron or positron) interacts with the proton via the exchange of
a virtual photon with a small virtuality $Q^2 \approx 0$. The spectrum of the
ingoing virtual photon can very well be described by the well-known
Weizs\"acker-Williams formula \cite{Weizsacker,Williams-34,Frixione-93}.

The photoproduction of single jets, dijets and triple jets with high
transverse momenta can be calculated also within perturbative QCD if the
transverse momentum of the jets is large enough to provide the hard scale. 
Besides the larger cross section, as compared to the DIS jet production, the 
photoproduction of jets does not depend on the additional scale $Q$. The 
contributions to the theoretical cross sections which have been calculated up 
to NLO come from two processes: (i) the direct process in which the photon enters 
the hard sub-processes directly by coupling to the quarks, in the same way as 
in deep-inelastic ep scattering (see Fig.~\ref{fig:feynborn}b, c  in LO), and
(ii) the so-called 
resolved process in which the photon fluctuates into partons, quarks or gluon, 
and one of them participates in the hard parton-parton scattering
 process~\cite{Llewellyn Smith:1978dc,Brodsky:1978eq}.
This latter process is equivalent to jet production in hadron-hadron 
collisions, of which the LO hard scattering cross sections for $qq'\to qq'$,
$gg \to gg$ and $gq \to gq$ are written below. The only difference is that the
PDFs of one of the hadrons is replaced by the photon PDF. This process, therefore, is 
sensitive to the parton structure of the proton and the photon. It is one of 
the few processes which can give information on the gluon content of the 
photon.

\begin{figure}
\begin{center}
\resizebox{0.75\columnwidth}{!}{\includegraphics{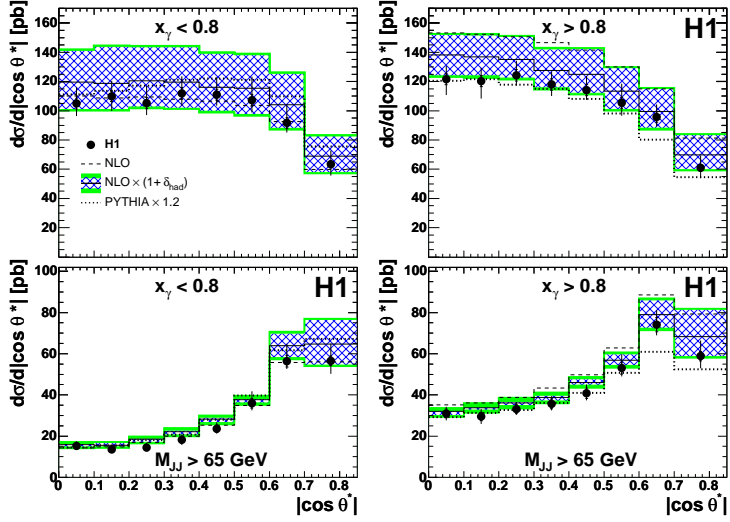}}
\caption{Bin averaged cross sections as a function of $|\cos\theta^{*}|$ for data
(points),  NLO QCD calculations with (solid line) and without (dashed line) hadronisation
corrections $\delta_{\rm{had}}$ and for the PYTHIA Monte Carlo predictions (dotted line)
scaled by a factor of  $1.2$.  The inner (hatched) band of the 
NLO$\times (1+\delta_{\rm{had}})$ result is the scale uncertainty, the outer
(shaded) band is the total uncertainty. The cross sections are shown for two regions
in $x_\gamma$, with and without an additional cut applied on the invariant dijet
mass ($M_{\rm{JJ}}$). (From  H1 Collaboration~\cite{H1photon}). }
\label{fig:costheta} 
\end{center}
\end{figure}

The basic $\gamma$-parton processes which enter the calculation of the
direct process in LO are the following: QCD Compton process: $\gamma q \to gq$, and 
the photon-gluon fusion: $\gamma g \to q\bar{q}$. 
These $\gamma$-parton cross sections have the following simple forms
\begin{eqnarray}
\gamma q \to gq : \frac{d\sigma}{d\cos\theta^{*}} 
\sim e_q^2 \frac{\alpha}{\pi}\frac{\alpha_s}{\pi}\frac{1}{s}\frac{4}{9}
         \left(-\frac{\hat{u}}{\hat{s}}-\frac{\hat{s}}{\hat{u}}\right)~, \\
\gamma g \to q\bar{q}: \frac{d\sigma}{d\cos\theta^{*}} \sim e_q^2
         \frac{\alpha}{\pi}\frac{\alpha_s}{\pi}\frac{1}{2s}
         \left(\frac{\hat{u}}{\hat{t}}+\frac{\hat{t}}{\hat{u}}\right)~,
\end{eqnarray}
where $e_q$ is the charge of the quark with flavour $q$, $\hat{s}=4$,
 $\hat{t}=-2(1 - \cos \theta^*)$, $\hat{u}=-2(1 + \cos \theta^*)$  and $\theta^{*}$ is
 the angle of the dijets in their centre-of mass system. $|\cos\theta^{*}|$ is related to 
the pseudorapidities of the two jets, $\eta_1$ and $\eta_2$ by
\begin{equation}
  |\cos\theta^{*}| = |\tanh(\eta_1-\eta_2)/2)|~.
\end{equation}
There are many observables which have been measured and which can be used
to test the basic parton-parton cross sections for the direct and resolved
process up to NLO \cite{photonNLO,photonNLO-2,photonNLO-3,photonNLO-4}. We shall present only a few taken
from the most recent H1 \cite{H1photon} and ZEUS \cite{ZEUSphoton}
publications. The $d\sigma/d|\cos\theta^{*}|$ distribution has been studied as a function of 
$|\cos\theta^{*}|$ by the H1 collaboration \cite{H1photon} with and without an
additional  cut on the 
invariant mass of the two jets $M_{jj} $  for the direct (resolved) 
enhanced contribution. This analysis is done in terms of a  variable $x_{\gamma}$ defined by,
\begin{equation}
    x_{\gamma} =\frac{1}{2E_{\gamma}} \sum_{i=1}^{2} E_{T,i} e^{-\eta_i}~,
\end{equation}
where $E_{T,1}$ and $E_{T,2}$ are the transverse energies of the two jets with 
the two largest $E_T$'s. In LO the direct contribution is at $x_{\gamma}=1$ 
and the resolved contribution has $x_{\gamma} < 1$. Therefore,
by selecting events with $x_{\gamma} > 0.8$ ($x_{\gamma} < 0.8$) the direct 
(resolved) parts of the cross sections are dominant. The results of the H1 
analysis are shown in Fig.~\ref{fig:costheta} for the two bins of $x_{\gamma}$ 
as a function of $\cos\theta^{*}$, with the upper two frames without a cut on
$M_{jj} $ and the lower two frames with $M_{jj} > 65$ GeV.

\begin{figure}
\begin{center}
\resizebox{0.50\columnwidth}{!}{
\includegraphics{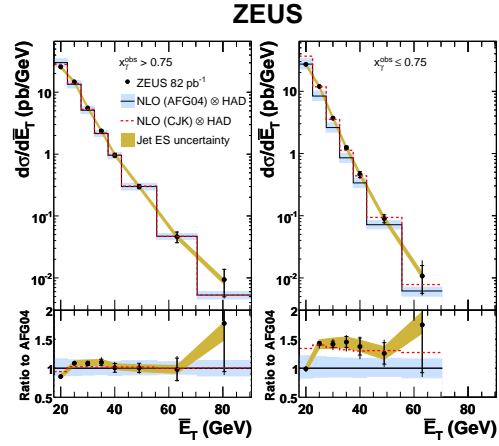}}
\caption{\label{fig:gen_et} Measured cross-section $d\sigma/d\bar{E_T}$ for (a) 
$ x_\gamma^{\rm obs}>$ 0.75 and (b) $ x_\gamma^{\rm obs} \leq$ 0.75 compared
with NLO QCD predictions using the AFG04~\cite{Aurenche:2005da} (solid line) 
and CJK~\cite{Cornet:2004nb} (dashed line) photon PDFs.   The predictions using 
AFG04 are  also shown with their associated uncertainties (shaded histogram). The
ratios to the prediction using the AFG04 photon PDF are shown at the bottom of the figure.
(From ZEUS Collaboration~\cite{ZEUSphoton}).}
\end{center}
\end{figure}

\begin{figure}
\begin{center}
\resizebox{0.50\columnwidth}{!}{
\includegraphics{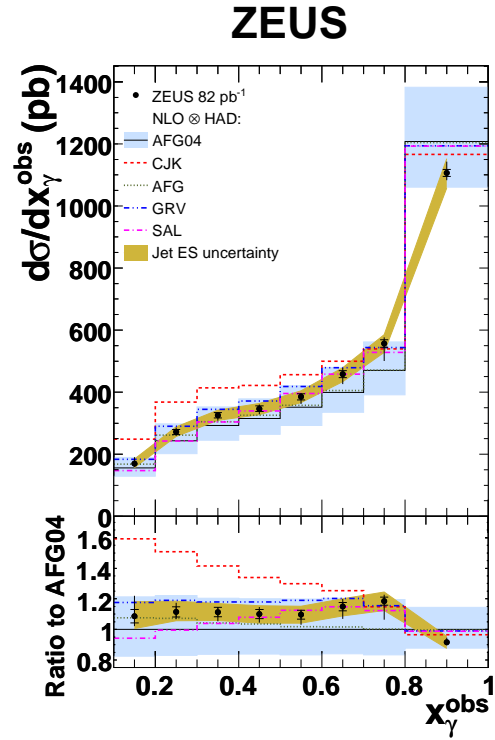}}
\caption{\label{fig:gen_xgamma}
Measured cross section for $d\sigma/dx_\gamma^{\rm obs}$ compared with the NLO QCD predictions
 using the AFG04~\cite{Aurenche:2005da} (solid line), CJK~\cite{Cornet:2004nb} (dashed line),
AFG~\cite{Aurenche:1994in} (dotted line), GRV~\cite{Gluck:1991ee,Gluck:1991jc} (dashed and double-dotted
 line) and SAL~\cite{Slominski:2005bw} (dashed and single-dotted line)
photon PDFs.  The ratios to the prediction using the
 AFG04~\cite{Aurenche:2005da} photon PDF are shown at the bottom of the figure. 
 (From~\cite{ZEUSphoton}).}
\end{center}
\end{figure}

 The cross section $d\sigma/d|\cos\theta^{*}|$
with the $M_{jj}$ cut is sensitive to the dynamics of the underlying 
$\gamma$-parton and parton-parton hard interactions. The cross section in the 
resolved sample $x_{\gamma} < 0.8$ rises more rapidly with $|\cos\theta^{*}|$ 
than that in the direct sample due to the dominance of the virtual gluon 
exchange in the resolved processes (see formulae for parton-parton cross 
sections below). The dependence on $|\cos\theta^{*}|$ and also the
absolute normalisation are well predicted by the NLO calculations.
Similar results have been presented by the ZEUS collaboration 
\cite{ZEUShighmass} by varying the dijet mass $M_{jj}$ and their results have 
been presented in~\cite{ZEUSphotonstr}.
Another example is the cross section $d\sigma/d\overline{E_T}$, where 
$\overline{E_T}$ is the mean transverse energy of the two jets
\begin{eqnarray}
  \overline{E_T} = \frac{1}{2}(E_T^{jet1} + E_T^{jet2})~.
\end{eqnarray}
An example of such a cross section for $x_{\gamma}>0.75$ and $x_{\gamma}<0.75$,
respectively, as presented by the ZEUS collaboration \cite{ZEUSphoton} is shown
in Fig.~\ref{fig:gen_et}.
%%%
%
%
 The cross section is measured up to 
$\overline{E_T} \simeq 80$  GeV, i.e. further out in $E_T$ than in DIS jet 
production. In this figure also results for two  different photon PDFs
(namely, the so-called AFG04~\cite{Aurenche:2005da} and the CJK~\cite{Cornet:2004nb})
 are shown. The most sensitive cross section concerning direct and resolved separation
 is the cross section $d\sigma/dx_{\gamma}^{obs}$, where 
$x_{\gamma}^{obs}$ is the $x_{\gamma}$ defined above with the sum over the two 
jets (therefore "obs" in the notation of $x_{\gamma}$ since not all jets are 
included in the sum). The result from the ZEUS collaboration is shown in 
Fig.~\ref{fig:gen_xgamma}, from which one can see how the data compare with 
different photon PDFs assumed in the NLO prediction. An appreciable dependence 
on these PDFs is seen in the small $x_{\gamma}$ region as one would expect. 
In an earlier analysis the ZEUS collaboration determined also the strong 
coupling $\alpha_s$, just from jet production in $\gamma p$ interactions alone.
The result is $\alpha_s(M_Z)=0.1224\pm0.0001({\rm stat.}) ^{+0.0022}_{-0.0019}({\rm exp.})
^{+0.0054}_{-0.0042}({\rm theor.})$ \cite{ZEUSscaling}, and the variation of 
$\alpha_s$ with the scale $\overline{E_T}$ has been found in good agreement 
with the running of $\alpha_s$ as predicted by QCD. 

Summarising the DIS and photoproduction processes at HERA, we see that QCD and jets
have made very significant impact on the profile of the proton and the photon in terms
of their respective PDFs, which determine the luminosity functions of the parton-parton
scatterings at high energies and hence the hard scattering cross sections of interest.  
 Detailed studies of  $F_2(x, Q^2)$ at HERA have also rekindled 
theoretical interest in the small-$x$ region. The evolution in $\ln(1/x)$
  at fixed value of $Q^2$  is governed by 
 the so-called BFKL equation~\cite{Fadin:1975cb,Balitsky:1978ic}.  Originally developed to study
Regge processes in high energy scatterings and the Pomeranchuk singularity (the QCD Pomeron),
 it can be combined with the DGLAP equation
(for evolution in $Q^2$) to provide a quantitative description of the DIS structure
functions over an enlarged $(x,Q^2)$ domain. Several proposals in carrying out the 
small-$x$ resummation have been considered in the literature, which are comparatively
discussed in a recent working group report~\cite{Dittmar:2009ii}.
 In addition, the evolution in $\ln(1/x)$
leads to soft gluon enhancements, generating 
a dense gluonic system over a limited range of the nucleon
wave function (hot spots). As the gluon occupation number becomes of order $1/\alpha_s$,
non-linear effects present in the QCD Lagrangian become important, leading    
 eventually to the saturation of the gluon density in the nucleons in high energy
 collisions~\cite{McLerran:2010ua}.
This picture of high energy nucleonic wave functions (a high density, nonperturbative
 gluonic system with a weak coupling constant) is called the Color Glass
 Condensate\cite{McLerran:2010uc},
and is of great interest in understanding the QCD aspects of heavy ion collisions, such as
at RHIC and the LHC~\cite{Iancu:2003xm}. 

\subsection{$\gamma \gamma$ collisions}

\begin{figure}
\vspace*{5cm}
\begin{center}
\resizebox{0.50\columnwidth}{!}{\includegraphics{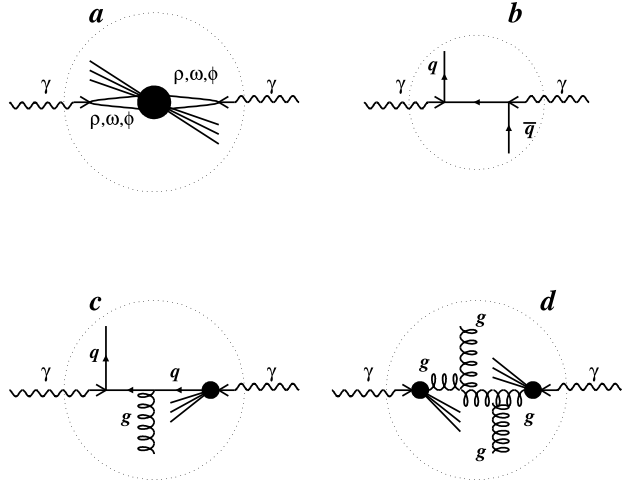}}
\end{center}
\vspace*{-3.5cm}
\caption[]{Main diagrams corresponding to the hadron production in $\gamma \gamma$ collisions
via vector meson interactions (VDM-like, a), point-like interactions (QPM-like, b)
and with one (c) or both (d) photons resolved into partons. (From~\cite{Abdallah:2008zzb}).}
\label{fig:DELPHI-fig1}
\end{figure}

Another area in which jet production has been studied experimentally and
theoretically is photon-photon collisions in the LEP2 energy range.
The two incoming photons are produced in $e^+ e^-$ collisions in the
 anti-tagged mode, i.e. when both the scattered electron and the positron 
escape detection. This is kinematically analogous to the photoproduction
process in high energy $ep$ collisions at HERA.
In $\gamma \gamma \to {\rm hadrons}$, four
classes of events have to be distinguished (see, Fig.~\ref{fig:DELPHI-fig1}).
The variables used in the classification of these events $x_\gamma^+$ and $x_\gamma^-$,
which are analogues of the variable $x_\gamma$ in $\gamma p$ collisions, are defined
 as follows:
\begin{eqnarray}
x_\gamma^+ &=& \frac{\sum_{\rm jets}(E_{\rm jet} + p_{z, {\rm jet}})}
{\sum_{\rm part}(E_{\rm part} + p_{z, {\rm part}})}~,\nonumber\\
x_\gamma^- &=& \frac{\sum_{\rm jets}(E_{\rm jet} - p_{z, {\rm jet}})}
{\sum_{\rm part}(E_{\rm part} - p_{z, {\rm part}})}~,
\label{eq:xgammapm}
\end{eqnarray}
where 'part' corresponds to all detected particles and $E_{\rm jet}$ and $p_{z, {\rm jet}}$
are the two hard-jets energy and the component of jet momentum along the $z$-axis,
 respectively. The four classes are:
\begin{enumerate}
\item Hadron production via vector meson interactions
 (Vector-Meson Dominance Model VDM)~\cite{Brodsky:1972vv,Kwiecinski:1987tb}
 (Fig.~\ref{fig:DELPHI-fig1}a).
\item The  direct domain, where both $x_\gamma^+$ and $x_\gamma^-$ are close to 1. This
domain is mostly populated by the quark-parton model like events
 $\gamma \gamma \to q + \bar{q}$ (Fig.~\ref{fig:DELPHI-fig1}b).
\item The single resolved domain, with the presence of a remnant jet, where only one of the
$x_\gamma^+$ and $x_\gamma^-$ is close to 1 and the other is shifted to some lower value
(Fig.~\ref{fig:DELPHI-fig1}c).
\item The double-resolved domain, where both $x_\gamma^\pm$ are shifted to values below 1 
(Fig.~\ref{fig:DELPHI-fig1}d).
\end{enumerate}
Due to the appearance of the double resolved region, jet production in $\gamma \gamma$
collisions has increased sensitivity to the gluon content of the resolved photon.
This has enormous significance for future high energy $\gamma - \gamma$ collisions,
being entertained in the context of a high energy linear $e^+e^-$ collider.

\begin{figure}
\begin{center}
\resizebox{0.75\columnwidth}{!}{
\includegraphics{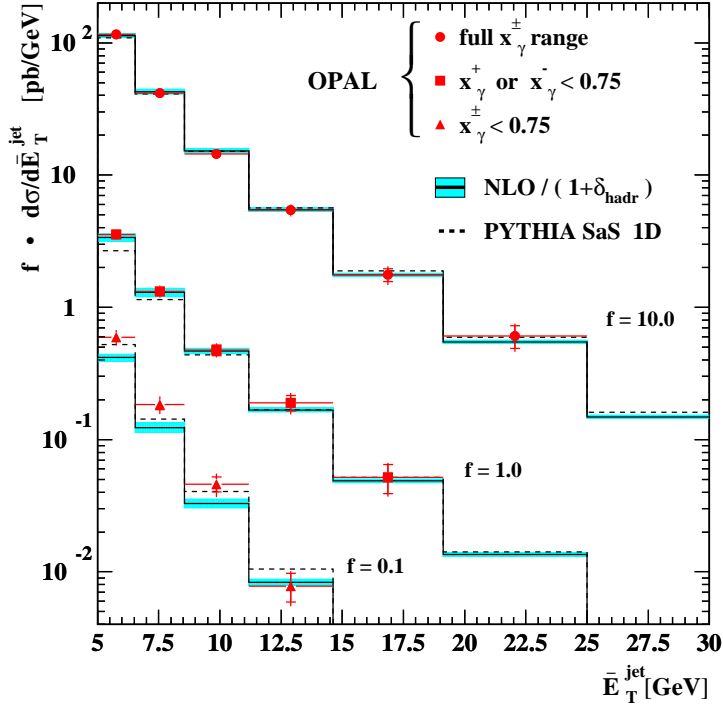}}
\end{center}
\caption{\label{fig:OPAL-etmxs} The dijet cross-section in $\gamma \gamma$ collisions at LEP
as a function of the mean transverse energy $ \bar{E}_T^{\rm jet}$ of the dijet system,
for the three regions in {$x_\gamma^+ - x_\gamma^-$}-space given in the figure.
The factor $f$ is used to separate the three measurements in the figure more clearly.  The
prediction of the LO program PYTHIA  is compared to the data. The NLO calculation is
from~\cite{photonNLO-2}. (From OPAL~\cite{Abbiendi:2003cn}).}
\end{figure}

In the past, dijet production in $\gamma \gamma$ collisions has been studied
experimentally at $\sqrt{s_{ee}}$ from 189 to 209 GeV by the OPAL~\cite{Abbiendi:2003cn} and
DELPHI~\cite{Abdallah:2008zzb} collaborations at  LEP.
To that end a number of observables (differential distributions) have been measured
by introducing $x_\gamma^\pm$-cuts at 0.75 (OPAL) and $x_\gamma^\pm =0.85$ (DELPHI).
 These distributions
include, among others, $d\sigma_{\rm dijet}/d\bar{E}_T^{\rm jet}$, with 
$\bar{E}_T^{\rm jet} = 1/2(E_{T,1}^{\rm jet} + E_{T,2}^{\rm jet})$ and $d\sigma_{\rm dijet}/dx_\gamma$.
The data from both collaborations have been compared with the NLO QCD calculations
based on the work of ~\cite{photonNLO} and are found to be in good agreement.
 This is shown
in Fig.~\ref{fig:OPAL-etmxs} for the differential distribution
 $f \dot d\sigma/d\bar{E}_T^{\rm jet}$, where the factor $f$ is used to visibly separate
the three measurements.
Inclusive jet production in $\gamma - \gamma$ collisions has also been measured by
the L3~\cite{Achard:2004rh} and OPAL~\cite{:2007jx} collaborations at LEP.

\subsection{Proton colliders}

\subsubsection{Fundamental QCD scattering processes}
\begin{figure}
\begin{center}
\resizebox{1.0\columnwidth}{!}{
\includegraphics{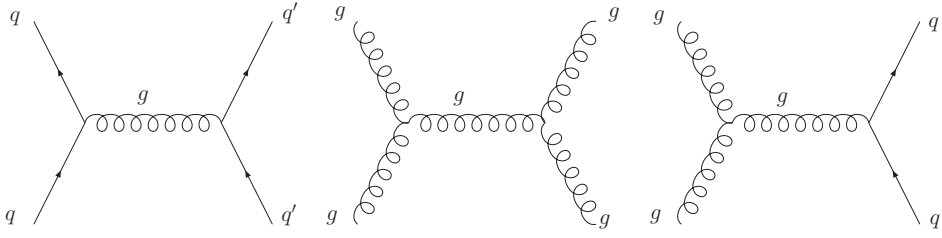}}
\end{center}
\caption{ Representative Feynman diagrams for fundamental QCD processes in hadronic
collisions.
\label{fig:Tevjets}}
\end{figure}

In parallel to electron and photon processes in QED, a large number
of $2\to 2$ scattering processes involving quarks and gluons are predicted 
in QCD, see {\it e.g.} \cite{EllisK}. They give rise to jets at hadron colliders.
 Most interesting are 
the fundamental abelian processes in QED transcribed to the non-abelian extensions in
QCD, like
\begin{eqnarray}
{\rm Rutherford \; quark \; scattering} \;&:&\; qq' \to qq^\prime~,   \nonumber \\     
{\rm Rutherford \; gluon \; scattering} \;&:&\; gg \to gg~,      \nonumber \\
{\rm Super-Compton \; process}           \;&:&\; gq \to gq \,.             \nonumber
\end{eqnarray}
Representative scattering diagrams are depicted in Fig.{\ref{fig:Tevjets}}.
The associated cross sections scale in the energy squared $s$ for
massless initial and final-state quarks, while the angular distributions are given by
\begin{eqnarray}
 qq' \to qq'               \;&:&\; \frac{d\sigma}{d\cos{\theta^\ast}} \sim 
                                   \left(\frac{\alpha_s}{\pi}\right)^2  \frac{1}{s} \,
                                   \frac{4}{9} \, \frac{\hat{s}^2+\hat{u}^2}{\hat{t}^2}~, 
         \nonumber \\
 gg \to gg                 \;&:&\; \frac{d\sigma}{d\cos\theta^\ast} \sim
                                   \left(\frac{\alpha_s}{\pi}\right)^2  \frac{1}{s} \,
                                   \frac{9}{2} \, \left[ 3 - \frac{\hat{s}\hat{u}}{\hat{t}^2}
                                                           - \frac{\hat{s}\hat{t}}{\hat{u}^2} 
                                   -\frac{\hat{t}\hat{u}}{\hat{s}^2} \right]~,   
                  \nonumber \\
 gq \to gq                 \;&:&\; \frac{d\sigma}{d\cos\theta^\ast} \sim
                                   \left(\frac{\alpha_s}{\pi}\right)^2  \frac{1}{s} \,
                                   \left[\frac{\hat{u}^2+\hat{s}^2}{\hat{t}^2} 
                                   - \frac{4}{9} \, \frac{\hat{s}^2+\hat{u}^2}{\hat{s}\hat{u}} 
                                   \right]                                                       \,,   
\end{eqnarray}
and the variables  $\hat{s}, \hat{u}$  and $\hat{u}$ have been defined earlier.
One should notice the three-gluon coupling already in LO.
These amplitudes generate the expected
Rutherford singularities $\sim d\theta^{\ast 2} / \theta^{\ast 4}$ for forward scattering 
$\hat{t} \to 0$, and analogously for backward scattering $\hat{u} \to 0$. 

Calculating the experimentally observed
cross sections at hadron colliders requires three essential steps, which we have
already outlined in the context of calculating the DIS cross sections, namely
(i)  the hard $2 \to 2$ scattering processes, including NLO QCD corrections,
(ii)  flux of the incoming partons, determined in terms of the PDFs of the protons (and
antiprotons), discussed earlier in the context of DIS scattering at HERA, and
(iii) hadronic (non-perturbative) corrections.
% and in the case of heavy hadrons their weak decays also.
 Here also QCD plays an important role in terms of the scale dependence
of the PDFs and FFs. Thus, for example, the cross section of the hadron-hadron
scattering with the four-momenta of the two colliding hadrons $P_1$ and $P_2$
can be written as~\cite{EllisK}
\begin{equation}
\sigma (P_1,P_2)= \sum_{i,j} \int dx_1 \int dx_2 f_{i}(x_1, \mu_F^2) f_j(x_2,\mu_F^2)
\hat{\sigma}_{ij}(p_1,p_2, \alpha_s(\mu_R^2), Q^2/\mu_F^2, Q^2/\mu_R^2)~,
\end{equation}  
where the hard interaction between the partons $i$ and $j$ is given by
$\hat{\sigma}_{ij}$ and $p_1=x_1P_1$ and $p_2=x_2P_2$ are the momenta of the two
partons.

\subsubsection{Jets in hadron colliders and tests of QCD}
\begin{figure}
\begin{center}
\resizebox{0.45\columnwidth}{!}{
\includegraphics{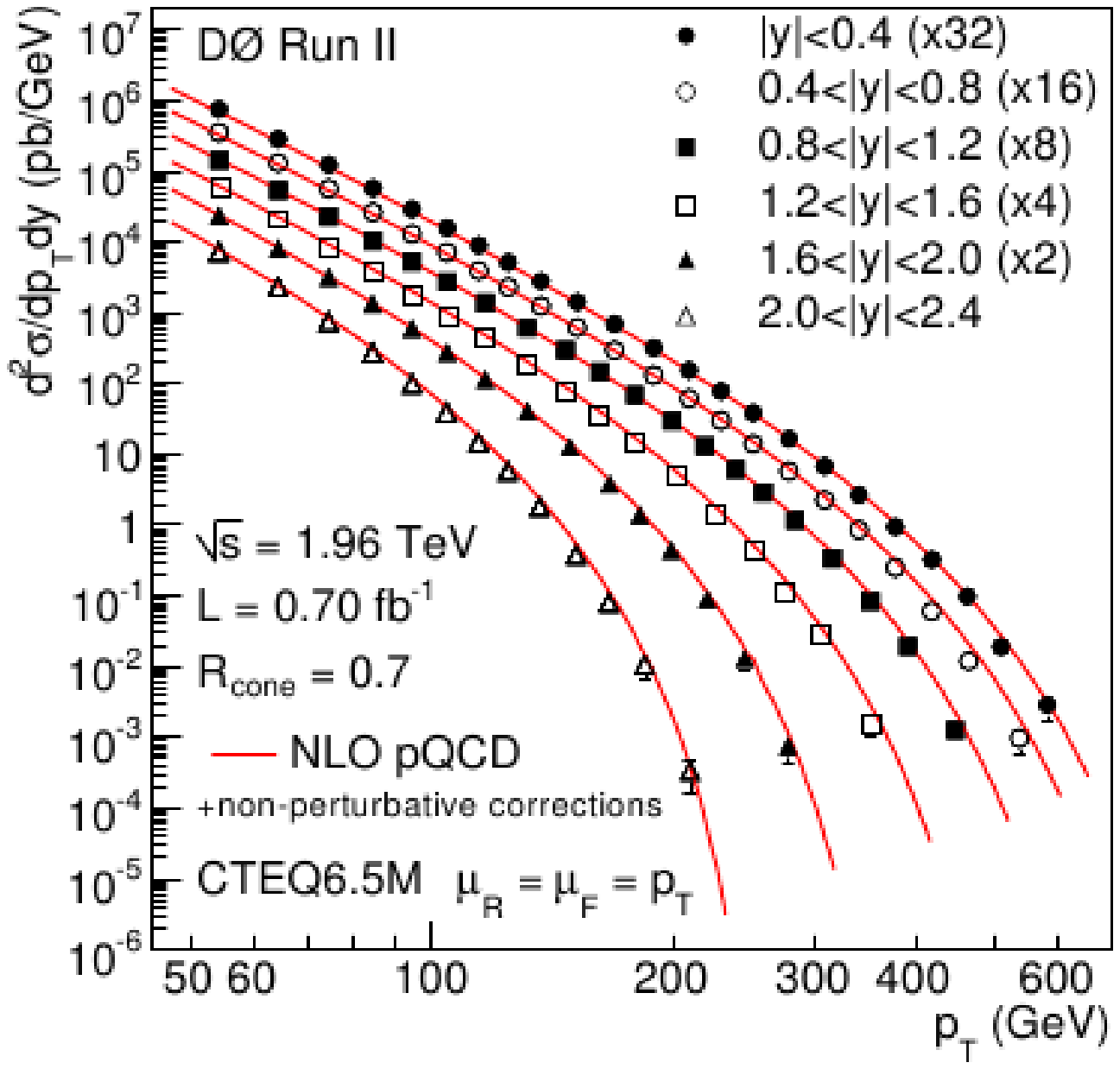}}
\resizebox{0.45\columnwidth}{!}{
\includegraphics{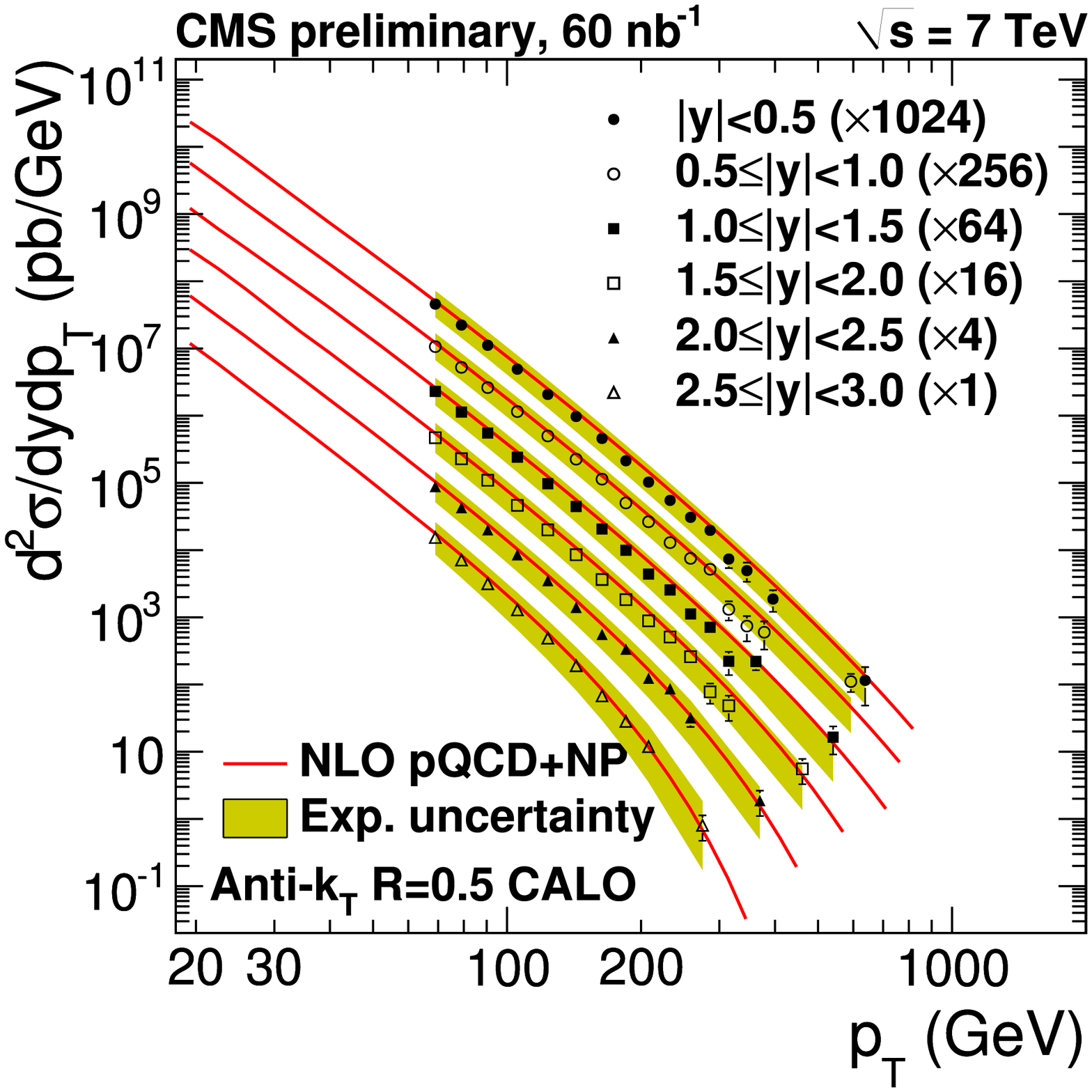}}
\end{center}
\caption{Left-hand frame: Transverse energy distribution of jets at the Tevatron 
             \cite{Tevjet}. 
Right-hand frame: Comparison between the measured $p_T$ spectra by the CMS
 collaboration at the LHC and theory predictions
for calorimeter jets~\cite{CMS-PAS-QCD-10-011}. For better visibility the spectra
in both frames  are multiplied by arbitrary factors
indicated in the legend.
        }
\label{fig:CMSjets}
\end{figure}

An example for inclusive jet production at the Tevatron measured by the D0 collaboration is shown in
 Fig.~\ref{fig:CMSjets} (left-hand frame). Similar measurements have been done by the CDF collaboration
at the Tevatron.
The dominant contribution at small $p_T$ 
 can be traced back to Rutherford gluon scattering $gg \to gg$.
This result is naturally expected since, on average, the gluon colour charges are significantly
larger than the quark colour charges and, as discussed earlier, the gluon flux
for low values of $x$  by far exceeds the quark flux of high-energy protons. 
These jet cross sections can be exploited to determine the gluon distribution 
of the proton and to measure the QCD coupling~\cite{BhattiLinc}. By combination 
with other measurements the two observables are disentangled in the Tevatron
measurements. The gluon flux extracted this way is large, as anticipated, and 
the QCD coupling is compatible with the world average. This~\cite{Tevjet}, and related
 measurements~\cite{Aaltonen:2008eq,Abulencia:2007ez}
impact on the proton PDFs and have been used in updating this information~\cite{CTEQ,MSTW}. 
In particular, they provide constraints on the gluon (and quark) distributions in the
 domain $0.01 < x < 0.5$. A detailed  discussion of jets and comparison
of data and theory at the Tevatron can be seen in~\cite{Ellis:2007ib}.

Very soon, similar but  more sensitive 
analyses will also be undertaken at the LHC and a beginning has already been made.
In Fig.~{\ref{fig:CMSjets}} (right-hand frame), we
show a comparison between the measured $p_T$ spectra by the CMS
collaboration~\cite{CMS-PAS-QCD-10-011} at the LHC with
$\sqrt{s}=7$ TeV and an integrated luminosity of 60 $nb^{-1}$ for the calorimeter jets and
theory predictions at the next-to-leading (NLO) order 
accuracy, using an anti-$k_T$ jet algorithm with $R=0.5$. Data are divided in several
intervals of rapidity $y$ bins. Theory predictions are based on NLOJET++~\cite{Nagy:2001fj}
with CTEQ-6.6~\cite{CTEQ} sets of parton
 distribution functions (PDF). The non-perturbative (NP) corrections are estimated 
using two different hadronisation models, PYTHIA~\cite{Sjostrand:2006za}
and HERWIG++~\cite{Bahr:2008pv}, with the mean of the two predictions
taken as the correction. Despite currently modest LHC luminosity, jets having
transverse momenta up to 800 GeV are measured and the agreement with QCD is excellent. 
An in-depth review discussing the physics basis and use of the general purpose Monte
Carlo event generators for hadronic collisions at the LHC is available~\cite{Buckley:2011ms},
to which we refer for a comprehensive discussion.

Experiments at the LHC have opened a window to sub-energies in the TeV range 
for studying jet phenomena in QCD, enabling searches for
physics beyond-the-SM in a number of such extensions.  Both ATLAS and CMS have searched
 for new heavy particles, which manifest
themselves as narrow resonances in their data collected at the LHC at $\sqrt{s}=7$ TeV. 
Such new states may include an excited composite quark $q^*$, expected in theories with
quark substructure~\cite{Eichten:1983hw,Baur:1987ga,Baur:1989kv};
 an axigluon predicted by chiral colour-models~\cite{Frampton:1987dn,Bagger:1987fz};
a flavour-universal colour-octet coloron~\cite{Chivukula:1996yr,Simmons:1996fz};
or a colour-octet techni-$\rho$ meson predicted by models of extended technicolor
and topcolor-assisted tecnicolor ~\cite{Lane:1991qh,Lane:2002sm,Foadi:2007ue,Belyaev:2009}.

\begin{figure}
\center{
\resizebox{0.75\columnwidth}{!}{
\includegraphics{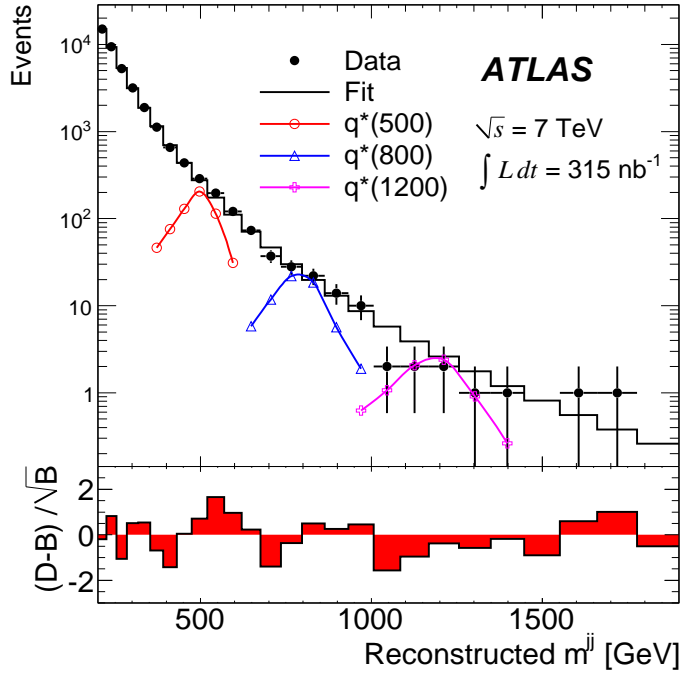}}
}
\caption{\label{fig:LHCjj} The data (D) dijet mass distribution (filled points) fitted using a
binned background (B) distribution described in the text (histogram). The predicted excited
quark $q^*$ signals for excited quark masses of 500, 800 and 1200 GeV are overlaid,
and the significance of the data-background difference is shown (from ATLAS collaboration
~\cite{Collaboration:2010bc}).}
\end{figure}

The dijet invariant mass $(m_{jj})$ is an observable which is particularly sensitive to
such new objects. This was studied already at the Tevatron in $p\bar{p}$ collisions with
negative results, exemplified by the CDF limit on the mass of the excited quarks $q^*$ 
in which a mass range $260 < m_{q^*} < 870$ GeV  was excluded at
95\% C.L.~\cite{Aaltonen:2008dn}. ATLAS has extended this exclusion range to higher $q^*$
masses, with the range $0.40 < m_{q^*} < 1.26$ TeV now excluded using $pp$
collisions~\cite{Collaboration:2010bc}.  Fig.~\ref{fig:LHCjj} shows the
predicted signal for $q^*$ masses of  500, 800, and 1200 GeV satisfying all event selection
cuts. No signal of $q^*$ is found and the data are in excellent agreement with the
background estimates based on the SM. 

Similar measurements of the dijet invariant mass spectrum and search for new particles
decaying to dijets have been performed by the CMS collaboration~\cite{Khachatryan:2010jd}.
The highest observed dijet mass by CMS at $\sqrt{s}=7$ TeV is 2.13 TeV. No deviations are
found  from QCD up to this dijet mass. In particular, string
resonances with a mass less than 1.67 TeV have been excluded by the current CMS
measurements at 95\% C.L. The sensitivities to the narrow resonances in the
dijet mass will increase substantially
with the increase in the LHC luminosity and energy. For example, for
the anticipated luminosity of 1 fb$^{-1}$ at $\sqrt{s}=7$ TeV, the expected limits are all in
the range of 2.5 to 3.5 TeV.

\subsubsection{Physics of the top quark and $(W^\pm$,$Z)$ bosons using jets}
\begin{figure}
\begin{center}
\resizebox{0.45\columnwidth}{!}{
\includegraphics{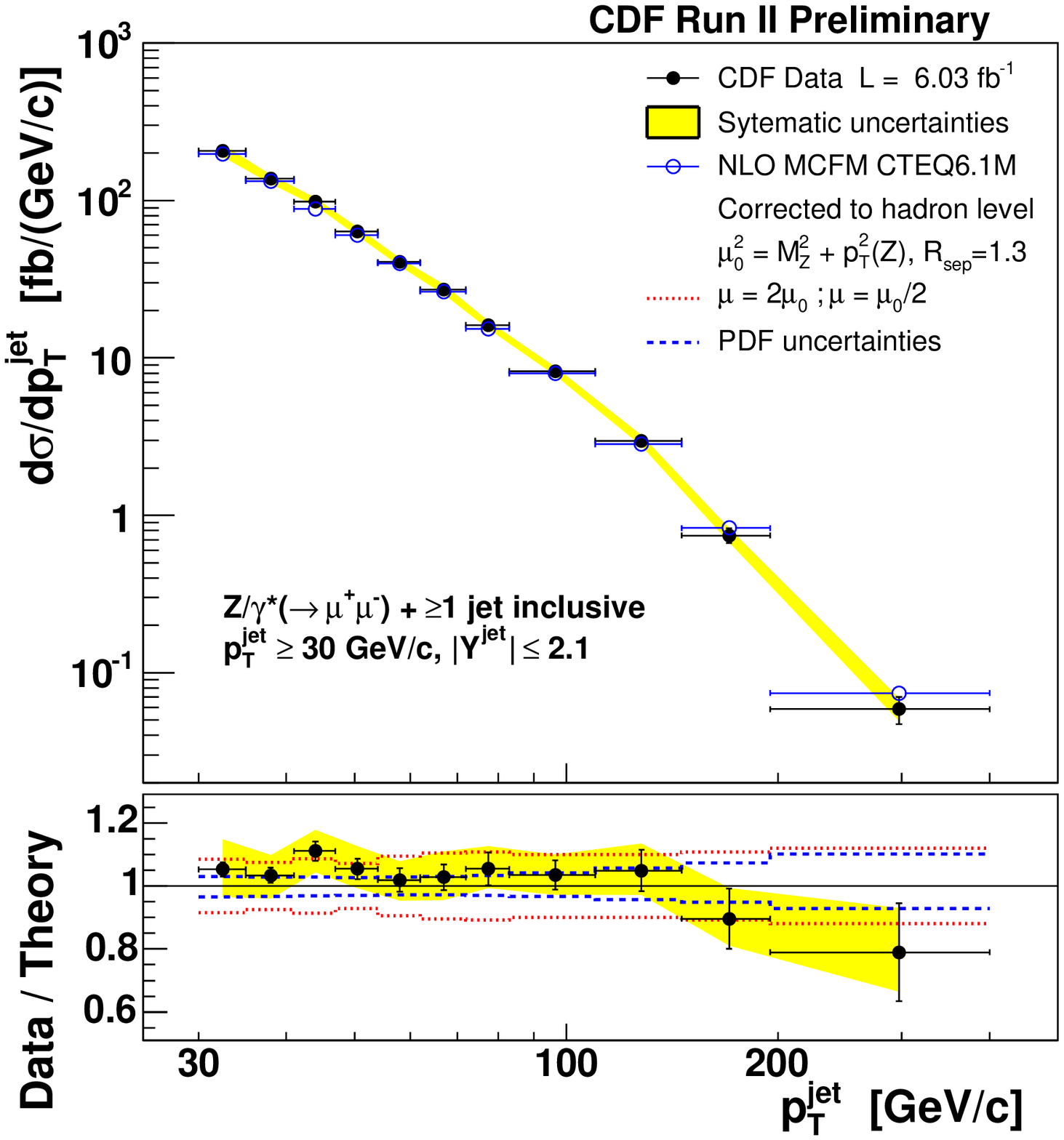}}
\resizebox{0.45\columnwidth}{!}{
\includegraphics{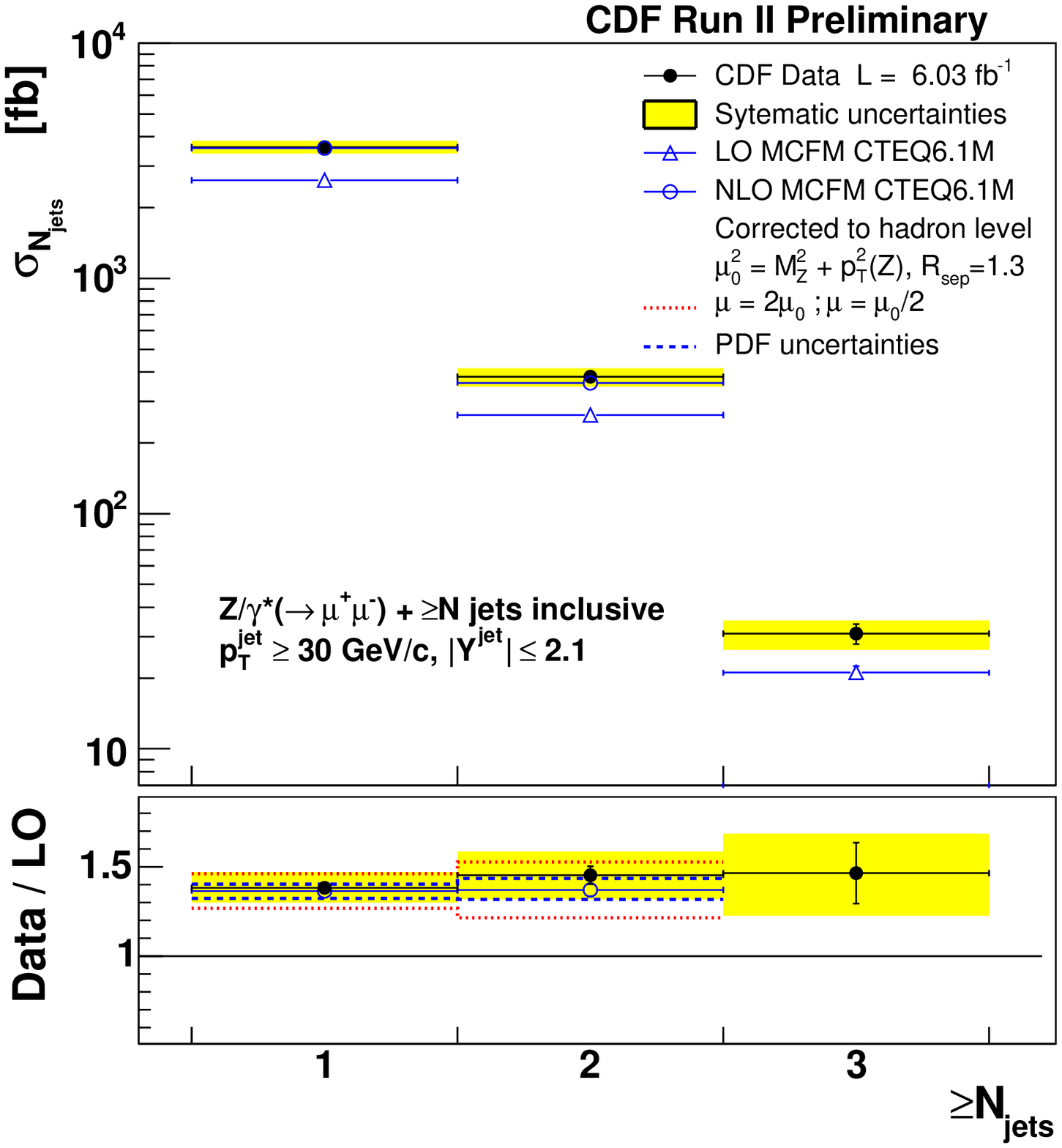}}
\end{center}
\caption{Left-hand frame: (top) Inclusive jet differential cross section measured
by the CDF collaboration as a function of
$p_T^{\rm jet}$ in $Z/\gamma^* + \geq ~1~{\rm jet}$ events (black dots)  compared to NLO pQCD
predictions (open circles). (bottom) Data/Theory versus $p_T^{\rm jet}$. 
 Right-hand frame: (top) Measured total cross section for inclusive jet production
 in $Z/\gamma^* \to \mu^+\mu^-$
events as a function of $N_{\rm jet}$ compared to LO and NLO pQCD predictions.
(bottom) Ratio of data and LO pQCD predictions versus  $N_{\rm jet}$.
 (From ~\cite{CDF-Note-10216}).}
\label{fig:CDF-Z-Jets-pt}
\end{figure}

Inclusive jet production in $p\bar{p}$ and $pp$ collisions in association with a
$Z/\gamma^*/W$ boson provides a stringent test of QCD. As these final states are also
of great importance in the search of the SM Higgs boson arising from the process
$pp(\bar{p}) \to W/Z + H(\to b \bar{b})$, and in the search of supersymmetry in the  missing
${E}_T$ + jets channel, the processes $pp(\bar{p}) \to W/Z/\gamma^* + ~{\rm jets}$ have
received a lot of theoretical and experimental
 attention. In particular, theoretical predictions for vector boson production
 recoiling against a  hadron jet at next-to-leading order were presented
 in ~\cite{Ellis:1981hk,Arnold:1988dp,Arnold:1989ub,Giele:1993dj}. The processes
$p + \bar{p} \to W/Z/\gamma^* + 2~{\rm jets}$ to the same level of theoretical accuracy
were calculated for the Tevatron in ~\cite{Campbell:2002tg} and the corresponding processes
 $p + p \to W/Z/\gamma^* + 2~{\rm jets}$ for the LHC  in~\cite{Campbell:2003hd}. Vector boson 
production in association with $n$-jets for $n \leq 4$ was calculated
 in~\cite{Berends:1989cf,Berends:1990ax}. A parton-level event generator, called
 MCFM~\cite{Campbell:2010ff}, 
which gives theoretical predictions for a large number of processes containing $W$, $Z$ and
$H$ bosons and jets (including heavy quark jets) is available for the Tevatron and the
LHC colliders. Similar theoretical tools have been developed which give predictions for the
transverse momentum distributions of the $Z/\gamma^*/W$ produced in hadron
collisions, based either on fixed order perturbation theory, such as~\cite{Melnikov:2006kv}
and~\cite{Catani:2009sm}, or based on soft gluon resummations valid at
low $p_T$~\cite{Ladinsky:1993zn}, such as RESBOS~\cite{Balazs:1997xd}. 
They have been used in conjunction with the PDFs~\cite{CTEQ} in the
analysis of the Tevatron data~\cite{Abazov:2010kn,CDF-Note-10216}, and we show 
below representative measurements from the CDF Collaboration in Figs.~\ref{fig:CDF-Z-Jets-pt}.
The NLO pQCD MCFM framework describes the data rather well over a large range
of $p_T^{\rm jet}$, as well as the jet-multiplicity.

The production of heavy gauge boson pairs $(WW, WZ, ZZ)$ in $p\bar{p}$ and $pp$ collisions
provides tests of the self-interactions of the gauge bosons and hence deviations from the
SM-based predictions for the production rate could indicate new physics~\cite{Langacker:2010}.
 Since, topologically
diboson production is similar to the associated Higgs boson production $pp(\bar{p}) \to
VH +X$ ($V=W,Z $), the experimental techniques developed in $pp(\bar{p}) \to VV$ are important
for the Higgs boson searches as well. The process $p\bar{p} \to VV$ with both the vector
mesons decaying into lepton pairs ($W^\pm \to \ell^\pm \nu_\ell; Z \to \ell^+\ell^-)$
has been observed at the
Fermilab Tevatron experiments by  CDF~\cite{Acosta:2005mu,Aaltonen:2008mv} and D0~\cite{Abazov:2004kc}.
Diboson production has not been conclusively observed in decay channels involving
only hadrons. However, evidence for diboson decays into a mixed
$\ell \bar{\nu}_\ell q \bar{q}$
final state ($\ell=e,\mu,\tau; q=u,d,s,c,b$) has been presented by  D0~\cite{Abazov:2008yg}
and CDF~\cite{Aaltonen:2009fd}. The experimental analyses involve large transverse
momentum imbalance (due to the escaped neutrino)  and two jets whose invariant mass
 can be reconstructed.
Because of the limited resolution in the dijet invariant mass, decays of $W^\pm \to 2$ jets and
$Z \to 2$ jets are not distinguished separately. The most significant backgrounds
to the diboson signals are $W(\ell \bar{\nu}) +$ jets, $Z (\nu\bar{\nu}) +$ jets and
QCD multijet production.   

\begin{figure}
\center{
\resizebox{0.75\columnwidth}{!}{
\includegraphics{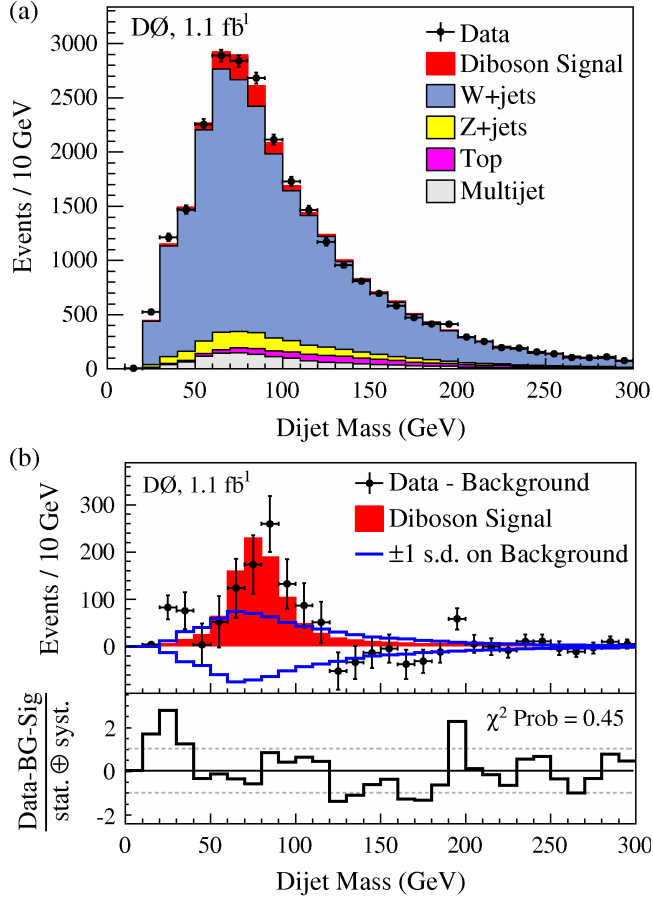}}
}
\caption{\label{fig:Dijetmass-D0} 
(a) The dijet mass distribution from the combined $e\nu q\bar{q}$ and $\mu\nu q \bar{q}$
channels for data from the D0 collaboration at the Tevatron and MC predictions.
(b) A comparison of the extracted signal (filled histogram) to
background-subtracted data (points), along with the $\pm 1 \sigma$ systematic uncertainty on
the background. The residual distance between the data points and the extracted signal,
divided by the total uncertainty, is shown at the bottom [D0~\cite{Abazov:2008yg}].
}
\end{figure}

\begin{figure}
\center{
\resizebox{0.75\columnwidth}{!}{
\includegraphics{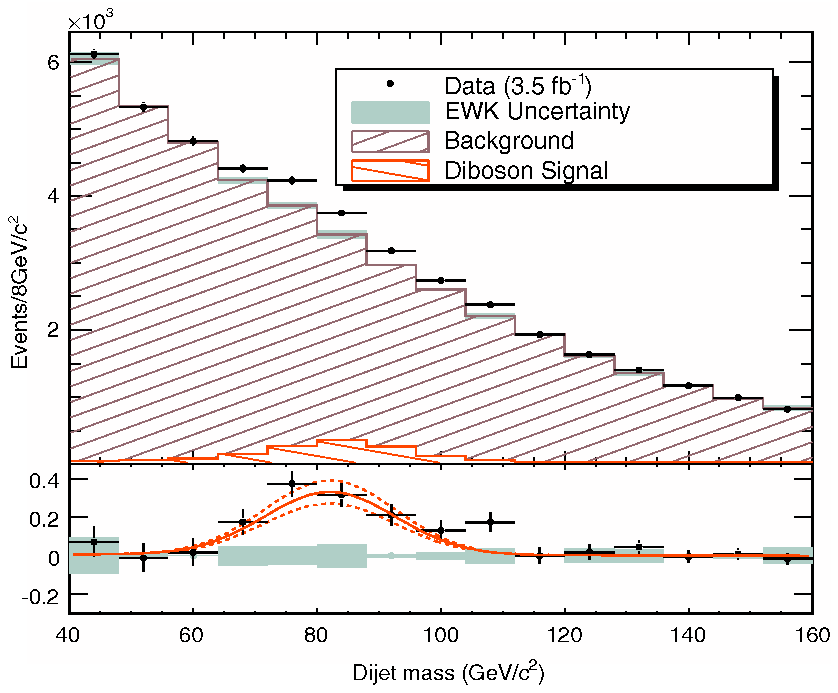}}
}
\caption{\label{fig:Dijetmass-CDF} 
Top: Comparison between data in $p\bar{p} \to VV +X)$ $(V=W^\pm, Z^0)$ from the CDF
collaboration at the Fermilab Tevatron at $\sqrt{s}=1.96$ TeV and the fitted background only
in the Dijet invariant mass.
Bottom: Comparison of the diboson signal (solid line) with the background subtracted data
(points). The dashed lines represent the $\pm 1 \sigma$ statistical variations on the signal
[CDF~\cite{Aaltonen:2009fd}].
}
\end{figure}

In Fig.~\ref{fig:Dijetmass-D0}, we show the dijet mass distribution from the 
$e\nu q \bar{q}$  and $\mu\nu q\bar{q}$ channels for the D0 data~\cite{Abazov:2008yg} and
MC predictions. A clear diboson signal in the dijet invariant mass is seen in the lower frame.
The resulting cross section
$\sigma(WV)=20.2  \pm 4.5$ pb is consistent with the SM prediction
$\sigma(WV)=16.1 \pm 0.9$ pb at $\sqrt{s}=1.96$ TeV~\cite{Campbell:1999ah}.
Fig.~\ref{fig:Dijetmass-CDF} shows the corresponding measurements by 
 CDF~\cite{Aaltonen:2009fd}. This yields a combined $WW+WZ+ZZ$ cross section in
$p\bar{p}$ collisions at $\sqrt{s}= 1.96$ TeV:
  $\sigma (p\bar{p} \to VV)=18.0 \pm 2.8 ({\rm stat})\pm 2.4
({\rm syst})\pm 1.1 ({\rm lumi})$ pb, consistent with the SM prediction.


At the Fermilab Tevatron, top quarks are produced mostly in pairs $p \bar{p} \to t\bar{t} +X$.
In the SM, top quarks decay into a $W$ boson and a $b$ quark almost 100\% of the time. 
The topology of the final states resulting from the $t\bar{t}$ production depends on whether
the $W$ boson decays leptonically $W \to \ell \nu_\ell$, or hadronically
 $W \to q \bar{q}^\prime$
leading to two jets. Following this, $t\bar{t}$ events have been measured in dilepton
$\ell^+\ell^- +X$, single lepton $\ell^\pm +4$ jets and also in the non-leptonic mode with
no energetic  leptons. The non-leptonic $t\bar{t}$ final state has the advantage of
a large branching ratio
$(\simeq 4/9 )$. The major challenge of this channel is the large background from QCD multijet
production. To increase the purity of the candidate sample, methods based on artificial
neural  networks are applied to the data. Further improvement is then obtained from the
requirement of at least one jet identified as originating from a $b$ quark using a 
secondary vertex $b$-tagging algorithm. These techniques have allowed one to measure the
top quark mass and the $t\bar{t}$  cross section in spite of the overwhelming QCD
multijet production.

\begin{figure}
\center{
\resizebox{0.75\columnwidth}{!}{
\includegraphics{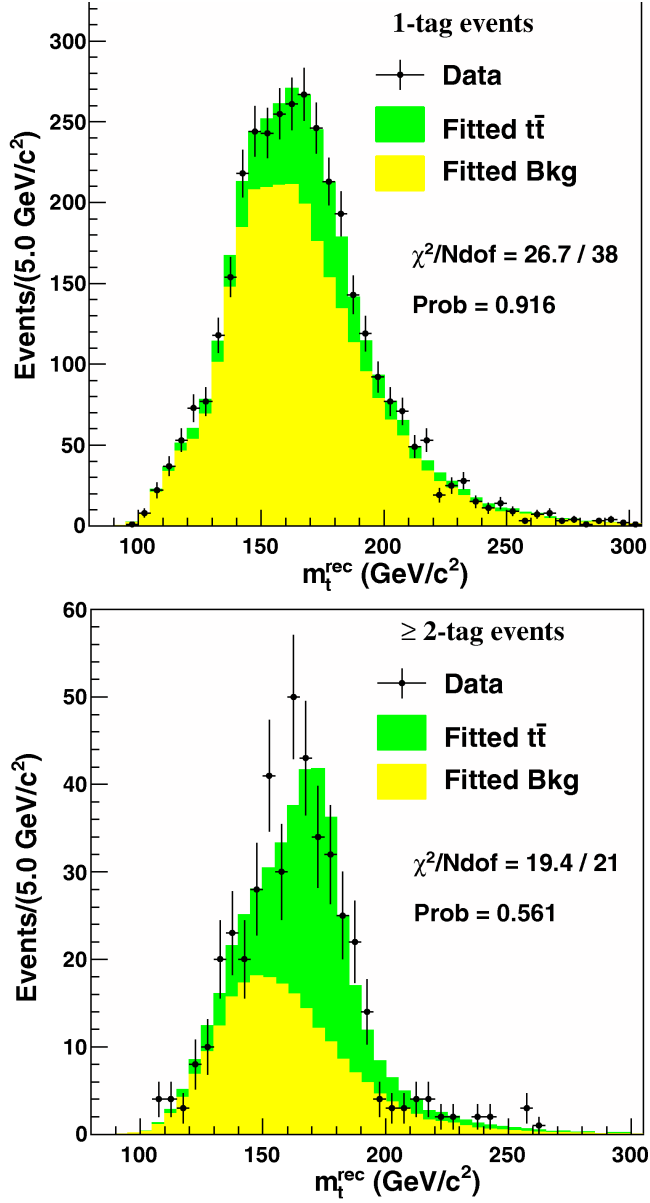}}
}
\caption{\label{fig:ttbar-6jets-CDF} 
Histogram of $m_t^{\rm rec}$ from the CDF data (black points) for 1-tag (upper plot) and
$\geq 2$ tag events (lower plot) compared to the distributions from signal and background 
corresponding to $m_{\rm top}=175$  GeV.
[CDF~\cite{Aaltonen:2010pe}].
}
\end{figure}

To these ends, a reconstructed top quark mass, $m_t^{\rm rec}$, is determined by fitting the
 kinematics of the
six leading jets from the process $p\bar{p} \to t\bar{t} +X \to 6~{\rm jets}$. There exists
a strong correlation between $m_t^{\rm rec}$ and the jet energy scale JES.  However, the JES can
 be calibrated using a selected sample of $t\bar{t}$ candidate events, where a second 
 variable $m_W^{\rm rec}$ is reconstructed from the jets assigned to the $W$ boson.
The variable $m_W^{\rm rec}$ is related to the $W^\pm$ boson mass, which is known accurately.
Relating $m_t^{\rm rec}$ and $m_W^{\rm rec}$ to match the experimental data ({\it in situ}
calibration) reduces significantly the systematic errors. Further improvement comes by using a
multivariate approach taking advantage of the distinctive features of the signal and
background events through a neural network. Fig.~\ref{fig:ttbar-6jets-CDF} shows
the histogram of $m_t^{\rm rec}$ as obtained in the data and compared to the distributions
in the so-called 1-tag and $\geq 2$-tag events
from signal and background corresponding to $M_{\rm top}=175$ GeV.
The best estimates of the top quark mass from this analysis is~\cite{Aaltonen:2010pe}
\begin{equation}
M_{\rm top}= 174.8 \pm 2.4 ({\rm stat + JES}) ~{\rm GeV}~.
\end{equation}
The procedure used to measure the top quark mass also returns the average number of signal 
events expected, given the selected data samples. These results can be turned into a
measurement of the $t\bar{t}$ cross section, and yield
\begin{equation}
\sigma_{t\bar{t}}= 7.2 \pm 0.5 ({\rm stat}) \pm 1.0 ({\rm syst}) \pm 0.4 ({\rm lum})~{\rm pb}~.
\end{equation}

\subsubsection{Searches for the Higgs  particles}
We have discussed 
numerous electroweak processes at the Tevatron in which jets play an essential 
role in the analysis. In particular, $W^\pm$ and $Z$ gauge bosons and top quarks have been
measured using jets. The last and the most-prized on this list is the Higgs boson. 
This is being searched for at the Tevatron feverishly. For $m_H < 135$ GeV, the dominant
decay mode is $H \to b\bar{b}$~\cite{higgs};
analyses of this decay mode open a powerful new Higgs discovery channel \cite{Butter}.
The dominant production modes are
$gg \to H$ and $q\bar{q} \to H$. The $b\bar{b}$ signal in this channel is overwhelmed 
by the QCD $b\bar{b}$ production. A promising production 
and search strategy  is the production of a Higgs boson decaying to a pair of bottom
quarks in association with a vector boson $V$ ($W$ or $Z$) decaying to quarks or leptons,
resulting into a four-jet or a charged lepton + two-jet final states. In either case, two of
the jets are required to have secondary vertices consistent with $B$-hadron decays.  
So far Tevatron Run II searches have used signatures where the $V$ decays to leptons
(see, for example Refs.~\cite{Aaltonen:2008zx,Aaltonen:2008ip}). Recently, also
searches in the four-jet channels have been reported~\cite{:2009bz}.  Using an integrated
luminosity of 2 fb$^{-1}$, Higgs boson searches in this channel provide a weak upper bound.
For example, for $m_H=120$ GeV, CDF is able to exclude a Higgs production cross section
larger than 38 times the SM prediction! Hence, establishing the Higgs signal in this
channel requires much more statistics, but also some fundamental progress in jet algorithms
to be more efficient in Higgs (and other similar particle) searches.

New opportunities are offered by observing $b$ jets in Higgs decays
at the LHC. 
The key technique is the 2-jet splitting of a fat $b \bar{b}$ jet generated in
events in which the Higgs boson is boosted to large transverse momenta in 
the processes $pp \to W^\pm H$ and $ZH$. If the ``fat'' jet 
is characterised by a jet radius $R = R_{bb} \simeq M_H / p_T$, the clustering 
is partially undone by shrinking the radius $R$ until the fat jet $R_{b\bar{b}}$ 
decomposes into two slim $R_b$ subjets with significantly lower mass, 
each containing a $b$ quark. 
Additional criteria will reduce the contamination by standard QCD processes. 
Though the boost will strongly reduce the event rate, the significance will 
nevertheless be raised to such a level that light Higgs events can clearly be 
isolated above background. Extending the method to the channel $pp \to 
t \bar{t} H \to t \bar{t} b \bar{b}$, the crucial $ttH$ coupling, apparently 
not accessible at LHC otherwise, can be measured in the light Higgs sector 
\cite{Plehn:2009rk}.

The concept is useful also for the analysis of other processes, for example, 
the search for supersymmetric particles decaying to jets from the hadronic decays of the
 electroweak and  Higgs bosons \cite{Raklev}, or for detecting strongly interacting $W^\pm,Z$ 
bosons \cite{Cox}, or the search for heavy resonances in decays to 
top-quark jets \cite{Baur}.

\section{Summary}
Quantum Chromodynamics has been established experimentally in the past four 
decades as the microscopic theory of the strong interactions, formulated as
a non-abelian gauge theory for coloured quarks and gluons within the Standard
Model of particle physics. Jet physics has been a crucial instrument for achieving 
this fundamental result. The beginning was made at SPEAR with the observation of
quark jets in $e^+e^-$ annihilation  by the SLAC-LBL collaboration~\cite{EXPquark}.
Subsequent studies undertaken, in particular at DORIS,
PEP and PETRA, involving higher center-of-mass energies largely consolidated the
phenomenological profile of the quark jets (see~\cite{AliSoeding} for a review).
In fact, jets provide an irrefutable
case for the existence of quarks as dynamical entities directly observable in particle
detectors, despite colour confinement, convincing even the most die-hard skeptics about the
reality of quarks. Moreover, making use of the larger masses of the $b$- and $c$-quarks,
relatively long half-lifes of the corresponding hadrons and their characteristic decay patterns, 
one can  efficiently flavour-tag the heavy quark jets. In the meanwhile, these techniques
have been developed to the level of a diagnostic tool to search for new phenomena in which
heavy quarks play a role. The decays $t \to b W$ and $H \to b\bar{b}$ are two good cases
in point. 

 Theoretically, the existence proof of quark jets in fixed order perturbative
QCD was provided by Sterman and Weinberg~\cite{SterW} using a jet-cone definition 
which coincided with 
the actual process of detection of hadrons in finite segments of hadron calorimeters.
Subsequently, Sterman~\cite{Sterman:1979uw}  provided an all-orders argument for the
infra-red safety of jet cross sections.
  Phenomenologically, quark jets
follow from the observation that the transverse momenta of the hadrons produced in
the fragmentation $q^* \to $ hadrons is limited, whereas the longitudinal components
of the hadron momenta scale with energy. A very intuitive and largely accurate 
quark jet fragmentation model was developed along these lines by Field and
Feynman~\cite{FieldF}, which played an important role in the quantitative analysis of jet data. 

 Analysis of
the decays $\Upsilon(9.46) \to $ hadrons measured by the PLUTO
 collaboration~\cite{Berger79} working at 
DORIS were undertaken in terms of the underlying perturbative process $\Upsilon(9.46) \to ggg$.
In particular, the experimental profile of the most energetic parton
($\langle E_1\rangle \simeq 4.1 $ GeV) was close to the phenomenological expectations
of a hadron jet.
  However, a clear three-jet topology using {\it en vogue} jet definitions was not
established in $\Upsilon(9.46)$ decays for lack of energy~\cite{Berger79}. This
 three-jet topology was
established later in the $e^+e^-$ experiments operating at higher energies (typically
$\sqrt{s}=30$ GeV), resulting from the energetic  quark, anti-quark and gluon from the process
$e^+e^- \to q \bar{q} g $. The jet profile of the three jets as well as the inclusive
hadronic measurements undertaken at PETRA in 1979 followed detailed theoretical
 expectations~\cite{THgluon,PHENgluon1,PHENgluon2}. Thus, it is fair to conclude that
the study of the decay $\Upsilon(9.46)\to 3g$, initiated by PLUTO at DORIS~\cite{Berger79}, was
an important step in the confirmation of QCD which served as a prelude to the unambiguous 
discovery of the three-jet topology by the experiments at PETRA. More detailed and
quantitative tests of perturbative QCD in the decays of $\Upsilon(9.46)$ were also presented 
subsequently by PLUTO~\cite{Berger81}.
 Theoretical proofs
of the existence of three-jet topologies, in the Sterman-Weinberg sense, were
provided in 1981, and somewhat later in terms of  the next-to-leading order calculations
 of the three-jet cross sections.
This was done for inclusive jet distributions (such as the Fox-Wolfram shape 
variable~\cite{Ellis:1980nc,Ellis:1980wv}, thrust~\cite{Vermaseren:1980qz,Ellis:1981re}
 and energy-energy correlations\cite{Ali:1982ub,Ali:1983au,Richards:1982te,Richards:1983sr})
 and in terms of topological jet cross
 sections~\cite{Fabricius:1980fg,Fabricius:1981sx,Gutbrod:1983qa,Kramer:1986mc,Gutbrod:1987jt,Gottschalk:1984vy}.
Confirmation of the non-abelian character of QCD in jets~\cite{THchi4,ang34,ang34-2,jet42} in the
four-parton processes $e^+e^- \to q\bar{q} gg$ came from experiments at LEP~\cite{EXPchi4,EXPgroup}.
 In the meanwhile, multijet physics has developed enormously, with the NLO calculation of
$e^+ e^- \to \gamma, Z \to 4$ jets~\cite{Signer:1996bf,Dixon:1997th,Nagy:1998bb} completed around 1996,
and the NLO calculation to five-jet production at LEP reported recently~\cite{Frederix:2010ne}. 
 
 The properties of gluon jets
 have largely been determined by experiments at PETRA and subsequently at LEP. The
fragmentation of gluon jets was initially conceived by treating them as independent
 partons~\cite{PHENgluon1},
or implemented by the perturbative process $g^* \to q \bar{q}$ as the first step followed
by incoherent quark fragmentation~\cite{PHENgluon2} (IJ models). The resulting picture
 could largely account
for the essential properties of the two- and three-jet events seen at PETRA and PEP,
and they helped in the discovery of three-jet events (and hence, of gluons) at PETRA. However,
analysis of the PETRA jets saw the emergence of an alternative fragmentation scheme for
 the $e^+ e^- \to q \bar{q} g$ events - the LUND string model~\cite{Lund} - in which
hadronisation
was implemented in terms of two strings stretched along the colour-anticolour axes
which then fragmented very much like the quark-antiquark string in $e^+e^- \to$ 2-jets.
This model provided a better phenomenological description of data, in particular the particle
flow between the quark, antiquark and gluon
 jets~\cite{Bartel:1981kh,Bartel:1983ij,Aihara:1984du,Althoff:1985wt}.
The LUND-string effect was subsequently
understood in perturbation theory in terms of the antenna radiation pattern 
of QCD~\cite{Azimov:1986sf},
reflecting the colour coherence effect of the non-abelian character of this theory.
Detailed fragmentation models were built along the angle-ordered perturbation
theory, which preserve the colour coherence in QCD, and in which parton showers
were included in the form of cascades. which then finally fragmented as 
hadron clusters according to phase space (cluster hadronisation models)~\cite{Herwig1}.
These Monte Carlo models developed for the PETRA jet analysis have played an
important role in the analysis of all high energy data involving jets. The modern incarnation
of these fragmentation models are PYTHIA~\cite{Pythia}, HERWIG~\cite{Herwig}
 and SHERPA~\cite{Sherpa} , which differ in details on
how the parton showers are matched on to the fixed order perturbative QCD matrix elements
and in the hadronisation schemes. A central role is also played by the jet algorithms,
which starting from the JADE scheme~\cite{JADE} have now evolved as trustworthy tools in
the definition of jets, with the modern versions called the $k_T$~\cite{Catani:1991hj} (mostly
in $e^+e^-$ annihilation processes) and anti-$k_T$ jet algorithms~\cite{Cacciari:2008gp}.

Another large application of QCD is in studying DIS, photoproduction and $\gamma \gamma$
collisions. In  these cases, initial states are not so well known as in
$e^+e^-$ annihilation.  Jets and QCD have played a central role in mapping the
PDFs of the proton and the photon. We have summarised some of the
highlights in this article. In particular, DIS measurements at
 HERA~\cite{Amsler,H1I,Chekanov:2002be,Chekanov:2006xr,:2007pb,ZEUS1} have firmly
established the rise of the structure function $F_2(x, Q^2)$, which is due to the rapid
growth of the gluon density $g(x,Q^2)$ for low values of $x$ as $Q^2$ increases. Likewise,
high energy $p\bar{p}$ collisions at the Tevatron, in particular the Tevatron Run II data on
inclusive jet production~\cite{Tevjet,Aaltonen:2008eq,Abulencia:2007ez},
 have led to greatly firming up the PDFs of the proton.
On the theoretical
side, the complete next-to-next-to leading order (3-loop) parton splitting functions
 have been derived
by Moch {\it et al.}~\cite{Moch:2004pa,Vogt:2004mw}. They have been used in
working out the  proton PDFs by the CTEQ~\cite{CTEQ} and the MSTW~\cite{MSTW} collaborations.
 Thus, the  HERA and the
Tevatron measurements and the progress in the QCD splitting functions 
 will prove to be an asset in understanding the forthcoming
LHC data. 

In the meanwhile, a fundamental change of paradigm has taken place concerning ``Jets and QCD''. 
The theory (QCD) is so well controlled (in particular, $\alpha_s(M_Z)$ is known to an
accuracy of better than 1\% and the crucial property of asymptotic freedom is now fully
established) that jet-physics can serve as a tool to chart out
new territories in high energy physics. We have reviewed here some applications of
jet techniques in quantifying the properties of the top quark and the electroweak gauge
bosons $(W,Z)$. They have already played a significant role in determining the properties of
the SM particles in experiments at the Tevatron and they will play an even more 
important role in the analysis of data from the experiments at the LHC. For example,
jets are now increasingly used in developing search strategies for the Higgs
boson,  and even particles suggested in theories beyond the Standard
Model.  New jet techniques, such as the 2-jet splitting of a fat $b\bar{b}$ jet
will be required to disentangle the decay $H \to b\bar{b}$ from an overwhelming
QCD background in hadron colliders.  
 
One problem in jet physics however remains unsolved up to now. While,
due to asymptotic freedom, the dynamics of quarks and gluons can
theoretically be described with high accuracy at short distances,
matched by numerical lattice calculations for static properties
of hadrons, the transition from small to large distances in the evolution 
of jets is theoretically not understood. However, a bridge is built,
at the present time, by intuitively formulated models, which are constrained
experimentally so stringently that hadron jets can be exploited to draw
a valid picture of quarks and gluons and their interactions at
short distances. New theoretical methods may help solve this 
multi-scale problem rigorously in the future.   

{\bf Acknowledgements} We thank Peter Zerwas for the collaboration in the early stages
of this work, for numerous helpful discussions that we had with him all along, 
and for his valuable input in good parts of this manuscript. Helpful discussions with
Hans-J\"urgen Meyer, Hinrich Meyer and Bruno Stella on the PLUTO analysis of the
$\Upsilon(9.46)$ data are thankfully acknowledged.
 We also thank a large
number of our colleagues and collaborators whose dedicated and painstaking work over decades
has contributed decisively to the development of QCD and jet physics.
This article is dedicated collectively to all of them. 
\end{document}